\documentclass{SCIS2024}
\usepackage{graphicx}
\usepackage{hyperref}
\usepackage{float}
\usepackage{tabularx}
\usepackage{multirow}
\usepackage{longtable}
\usepackage[normalem]{ulem}

\begin{document}
\ArticleType{RESEARCH PAPER}
\Year{2024}
\Month{}
\Vol{}
\No{}
\DOI{}
\ArtNo{}
\ReceiveDate{}
\ReviseDate{}
\AcceptDate{}
\OnlineDate{}
\title{Overview of AI and Communication for 6G Network: Fundamentals, Challenges, and Future Research Opportunities}{Title keyword 5 for citation Title for citation Title for citation}

\author[1]{Qimei Cui}{}
\author[2]{Xiaohu You}{{xhyou@seu.edu.cn}}
\author[3]{Ni Wei}{} 
\author[1]{Guoshun Nan}{{nanguo2021@bupt.edu.cn}}
\author[1]{Xuefei Zhang}{} 
\author[4]{Jianhua Zhang}{} 
\author[1]{\\Xinchen Lyu}{} 
\author[5]{Ming Ai}{} 
\author[1]{Xiaofeng Tao}{} 
\author[4]{Zhiyong Feng}{} 
\author[4]{Ping Zhang}{}
\author[6]{Qingqing Wu}{} 
\author[7]{\\Meixia Tao}{} 
\author[2]{Yongming Huang}{} 
\author[8]{Chongwen Huang}{} 
\author[9]{Guangyi Liu}{} 
\author[10]{Chenghui Peng}{}
\author[2]{\\Zhiwen Pan}{}
\author[9]{Tao Sun}{} 
\author[11]{Dusit Niyato}{}
\author[12]{Tao Chen}{}
\author[13]{Muhammad Khurram Khan}{}
\author[14]{\\Abbas Jamalipour}{}
\author[15]{Mohsen Guizani}{}
\author[16]{Chau Yuen}{}

\AuthorMark{Author Cui Q}



\address[1]{National Engineering Research Center of Mobile Network Technologies, \\Beijing University of Posts and Telecommunications, Beijing {\rm 100876}, China}
\address[2]{National Mobile Communications Research Laboratory,
 Southeast University, Nanjing {\rm 210096}, China}
\address[3]{Fudan University, Shanghai {\rm 200433}, China}
\address[4]{School of Information and Communication Engineering, Beijing University of Posts and Telecommunications, \\Beijing {\rm 100876}, China}
\address[5]{CICT Mobile Communication Technology Co., Ltd., Beijing {\rm 100020}, China }
\address[6]{Department of Electronic Engineering, Shanghai Jiao Tong University, Shanghai {\rm200240}, China}
\address[7]{Department of Electronic Engineering and the Cooperative Medianet Innovation Center (CMIC), \\Shanghai Jiao Tong University, Shanghai {\rm 200240}, China}
\address[8]{College of Information Science and Electronic Engineering, Zhejiang University, Hangzhou {\rm310058}, China}
\address[9]{~China Mobile Research Institute, Beijing {\rm 100053}, China}
\address[10]{~Huawei Technologies, Shanghai {\rm 201206}, China}
\address[11]{~College of Computing and Data Science, Nanyang Technological University, Singapore {\rm 117583} }
\address[12]{~VTT Technical Research Centre of Finland Ltd., Espoo FI-{\rm 02044} VTT, Finland}
\address[13]{~Center of Excellence in Information Assurance, King Saud University, Riyadh {\rm 11362}, Saudi Arabia}
\address [14]{~School of Electrical and Information
 Engineering, University of Sydney, NSW {\rm 2006}, Australia}
\address[15]{~Mohamed bin Zayed University of Artificial Intelligence (MBZUAl), Abu Dhabi UAE}
\address[16]{~School of Electrical and Electronics Engineering,
 Nanyang Technological University, Singapore {\rm639798}}

\abstract{
With the growing demand for seamless connectivity and intelligent communication, the integration of artificial intelligence (AI) and sixth-generation (6G) communication networks has emerged as a transformative paradigm.
By embedding AI capabilities across various network layers, this integration enables optimized resource allocation, improved efficiency, and enhanced system robust performance.
This paper presents a comprehensive overview of AI and communication for 6G networks, with a focus on their foundational principles, inherent challenges, and future research opportunities. We first review the integration of AI and communications in the context of 6G, exploring the driving factors behind incorporating AI into wireless communications, as well as the vision for the convergence of AI and 6G.
The discourse then transitions to a detailed exposition of the envisioned integration of AI within 6G networks, divided into three progressive stages. 
The first stage, \textit{AI for Network}, focuses on employing AI to augment network performance, optimize efficiency, and enhance user service experiences.    
The second stage, \textit{Network for AI}, highlights the role of the network in facilitating and buttressing AI operations and presents key enabling technologies.
We compare \textit{wireless network large models} with conventional large language models (LLMs), and identify key design principles and components for building wireless network architectures.
In the final stage, \textit{AI as a Service}, it is anticipated that future 6G networks will innately provide AI functions as services, supporting application scenarios like immersive communication and intelligent industrial robots. Specifically, we define the quality of AI service, which refers to a framework for measuring AI services within the network.
We further summarize the standardization process of AI for wireless networks, highlighting key milestones and ongoing efforts. In addition, we analyze the critical challenges faced by the integration of AI and communications in 6G. Finally, we outline promising future research opportunities that are expected to drive the development and refinement of AI and 6G communications.
}
 
\keywords{6G, AI, AI and communication, AI for network, AI as a service, LLMs, network for AI}
\maketitle

\section{Introduction}
\label{Section_Intro} 
In recent years, the rapid development of wireless communication technology has profoundly reshaped various aspects of our society, driving unprecedented connectivity and enabling innovative applications \cite{7414384}. Following the widespread deployment and success of the fifth-generation (5G) networks, attention has shifted towards the sixth-generation (6G) wireless communication systems. With its enhanced capabilities, it is anticipated that 6G will bring transformative changes, including ultra-low latency, significantly higher data transmission rates, increased reliability, and ubiquitous connectivity \cite{9335927}. Among these advancements, integrating artificial intelligence (AI) into 6G networks is expected to be a game-changer, providing new paradigms and opportunities across multiple fields \cite{9237460}.

{{AI technology has advanced rapidly over the past decade, particularly in machine learning (ML) \cite{2017A}, deep learning (DL) \cite{GUO201627}, and natural language processing (NLP) \cite{chowdhary2020natural}. These advancements have enabled AI to play a crucial role across various industries. By virtue of its powerful data analysis and learning capabilities, AI can excavate massive data in wireless communication networks, realizing intelligent management and optimization of the network. In the 5G era, AI has been successfully applied to the aspects of wireless networks, e.g., network optimization, traffic prediction, and fault detection, significantly enhancing network performance and user experiences \cite{9023918}. However, many unresolved issues need to be addressed to achieve AI-native support in 6G networks.}}

The introduction of AI can enhance coding efficiency by utilizing compressed semantic information (SI) to transmit more data with less bandwidth, which helps alleviate network congestion and improve data transmission rate \cite{10208153}. However, AI algorithms generate additional data that needs to be transmitted, such as model parameters, training data, and real-time feedback. This raises the question of whether the overall data volume in future networks will decrease or increase. This change in data volume will directly impact network design and architecture. Moreover, both base stations (BS) and user devices will employ AI algorithms for resource allocation to reduce energy consumption and improve resource utilization~\cite{10118940}. The operation of AI itself may increase power consumption, leading to another question: Will energy consumption in future networks increase or decrease? In the pursuit of high efficiency, it is crucial to consider optimizing energy efficiency in AI applications to ensure the sustainable development of networks. Additionally, while networks can leverage AI algorithms to enhance transmission reliability and better respond to changing network conditions and user demands, the inherent uncertainty of AI algorithms—especially in complex and dynamic network environments—raises concerns about whether future networks will operate more reliably. These uncertainties may lead to decision-making errors, affecting user experience and overall network performance.

It is essential to consider integrating AI into the architecture, network elements, and functional processes of the new 6G system from a closer and deeper perspective to address these issues. Specifically, we need to re-examine the network's design philosophy to ensure a genuine synergy is formed between AI and the 6G network. This comprehensive review explores the intricate relationship between 6G and AI, delving into the fundamental principles, key technologies, and potential applications this integration brings forth. We discuss the critical technical enablers of 6G, including advanced wireless communication techniques, spectrum management, and network architectures. Additionally, we examine the role of AI in optimizing network operations, enhancing security, and enabling intelligent decision-making processes. The review also highlights the open challenges to fully realizing the potential of 6G and AI integration.  This review can be a valuable resource for researchers, practitioners, and policymakers involved in developing and deploying 6G and AI technologies by providing a holistic overview of the state-of-the-art and future directions. Through this collaborative effort, we can harness the full potential of these cutting-edge technologies to build a brighter, more connected, and more sustainable future. 
\begin{figure}[ht] 
    \centering
    \includegraphics[width=1\textwidth]{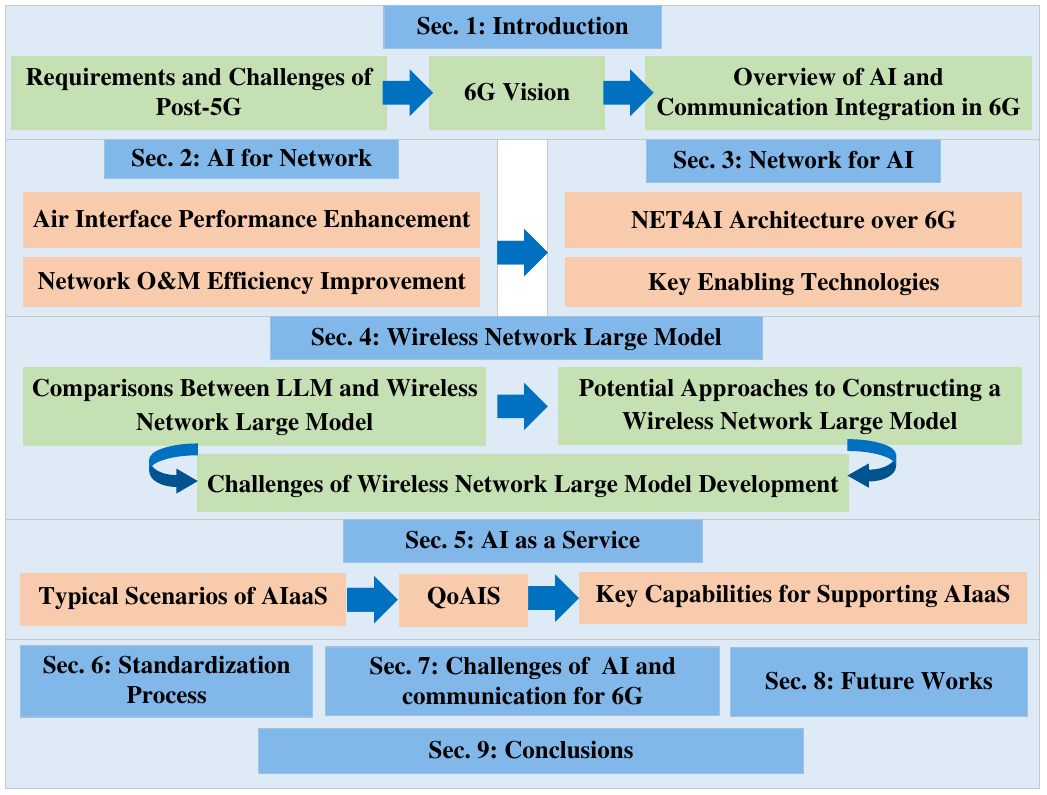}
    \caption{Organization of this article}
    \label{sturcture}
\end{figure}

The organization of this paper is illustrated in Figure \ref{sturcture}, and we summarize the main abbreviations used throughout this work in Appendix A.  

\subsection{Requirements and Challenges in the Post-5G Era}
{{\subsubsection{Better Utilization of Spectrum Resources}
The volume of network traffic data has reached unprecedented levels \cite{wong2024real}. Networks face the challenges of explosive growth in data traffic and the demand for massive devices to connect everything. 
As an essential mobile communication resource, spectrum remains the most critical element in improving network capacity \cite{9023920}. Whether the limited spectrum resources can be better managed is the key to ensuring that the network can provide high-quality services to users. However, existing wireless communication systems still adopt a static spectrum management model, where authorized users have exclusive access. This model lacks the ability of dynamic adjustment and is unable to flexibly allocate resources like spectrum and power in accordance with the real-time requirements of services. This likely leads to resource waste or insufficient resources for some services, which impacts network performance and user experiences.}}

{{\subsubsection{Lower-Carbon Wireless Coverage}
The energy consumption of 5G systems is about three times that of the fourth-generation (4G) system due to its wider bandwidth, more channels, and more complex equipment architecture \cite{10269761}.
Moreover, due to the use of higher-frequency bands in 5G technology, there is a further reduction in the coverage area per BS. To achieve the same coverage targets, the number of 5G BS required would be three to four times that of 4G, significantly increasing the cost of network deployment for 5G BSs. The current operational management systems of actual networks lack flexibility and intelligence, as a BS cannot be dynamically adjusted in precise accordance with real-time changes in regional user traffic. During periods and in areas with low traffic demand, the BSs may maintain a high workload, leading to energy waste.}}

{{\subsubsection{More Efficient and Cost-effective Network O\&M}
The operation and maintenance (O\&M) management of current 4G/5G networks relies primarily on manual on-site troubleshooting, resulting in low levels of automation and high maintenance costs. This approach is rather reactive, addressing issues as they arise, as opposed to proactively preventing them. Providing rapid response reports and swift emergency handling capabilities is challenging. At the same time, 5G networks introduce new technologies, such as virtualization and mobile edge computing (MEC) \cite{9652477,9059015}. Edge sites are expanding, characterized by their large numbers, wide distribution, and heterogeneity, resulting in expansion of maintenance teams and uncontrolled labor costs \cite{8663994}.

Compared to conventional core network (CN) equipment's typical millisecond-level fault detection efficiency, network resource virtualization introduces new potential fault points. It is difficult for operators to delineate the responsibility boundaries among suppliers when faults occur. Using software mechanisms for fault detection, virtual network functions at the edge site typically experience longer response times, which complicates timely network maintenance and management. Moreover, the incessantly emerging novel applications and services have presented elevated requisites with respect to network bandwidth, latency, reliability, and security.}}

{{\subsubsection{More Personalized and Customized On-demand Service Capability}
In the post-5G era, a mobile communication network, as a critical infrastructure for digital social transformation, is no longer confined to the business domain of conventional mobile communication. Instead, it places greater emphasis on the vastly diverse new demands for digital transformation across a multitude of industries. Specifically, the application scenarios of mobile communication networks are rich and diverse, covering fields such as mobile communication, mobile internet, internet of things (IoT), smart cities, and satellite internet. Different application scenarios have significant differences in requirements for network user experience, data rate, service latency, reliability, etc. The network needs to possess the ability to provide personalized and customized services to meet the user needs of different scenarios \cite{wang2022high}.}}
 
\subsubsection{More Secure and Reliable Transmissions}
From 2G to 5G, the design goal of mobile communication networks in terms of security and reliability has been to ensure the authenticity of user and network identities, prevent data interception and tampering during transmission, and primarily employ security mechanisms, such as network authentication, data encryption, and integrity checks, to safeguard communications. Moreover, these security measures have undergone several ``plug-in'' enhancements and refinements. Establishing security measures is ``post-hoc", meaning that security mechanisms are added to the system after the design of the network communication functions is complete. Furthermore, some security vulnerabilities are only addressed in subsequent generations of mobile communication networks—for example, 5G remedied the capture attacks on user identity tags in 4G \cite{Security,9999559}. The defense systems are deployed primarily based on a perimeter-centric defense architecture, most often erected on the premise of ``known risks". Protective devices, such as firewalls and intrusion detection systems, are positioned at the physical boundaries of mobile communication networks. Internally, these systems adopt a trusting stance, while externally, they use a ``patching" approach to ward off security risks.

Additionally, the currently released 5G network security protocol standards target primarily the enhanced mobile broadband (eMBB) scenario. The standardization progress of security protocols for two other typical cases: massive machine type communications (mMTC) and ultra-reliable low latency communications (uRLLC)—lags. In the post-5G era, mobile communication networks will increasingly focus on empowering vertical industry services. They will do so by opening up specific network capabilities to third parties through upper-layer interfaces or by providing customized security services tailored to the differentiated needs of various industries \cite{CustomizedSecurity}. This approach is intended to effectively adapt to the new characteristics of mobile communication networks, which are more open, feature heterogeneous integration, include many terminals, and exhibit diverse connection types.

\subsection{6G Vision}
\subsubsection{Grand Vision}
As new-generation information and communication technologies and applications such as big data, cloud computing, the IoT, and AI develop, and as these intertwine deeply with information, communication, data, and technology, the technology landscape transitions into an era of ``digital twin (DT) and ubiquitous intelligence" in the 6G era \cite{9239911}. 6G networks will build a closely interconnected network space by providing communication services that integrate the physical and virtual worlds, enabling seamless interaction between the human society, physical world, and virtual world. This integration will create new value through the digital world, realizing the promising vision of ``6G changing the world."

In 2030 and beyond, mobile communication application scenarios will exhibit entirely new characteristics within the context of the DT world and pervasive intelligence. These scenarios will support ubiquitous wireless connectivity, big data, and new technologies such as AI, giving rise to three major application areas: intelligent living, intelligent production, and intelligent society. These areas will encompass integrated air-space-ground-sea networks, communication and sensing interconnection, and intelligent interaction. 6G networks will not be confined to providing communication functions alone. Following the major trend of integrating sensing, communication, computing, AI, and security, 6G will expand from conventional single-function capabilities to new network capabilities such as sensing, computing, AI, and security.

\subsubsection{Usage Scenarios of 6G}
\begin{figure}[h!]
    \centering
    \subfloat[Usage scenarios]{
    \includegraphics[width=0.44\linewidth]{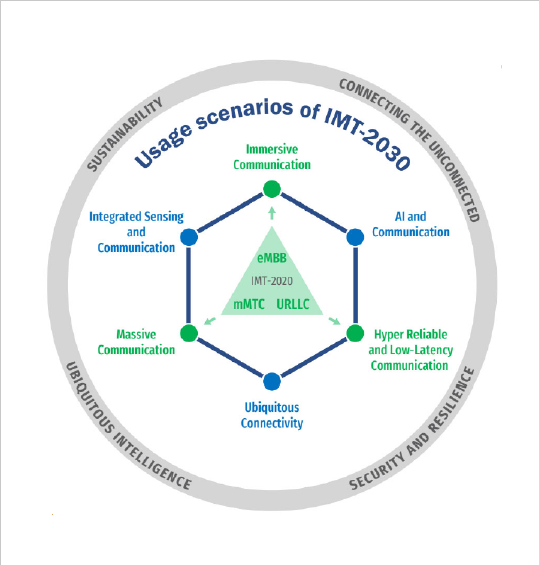}}\quad
    \subfloat[Capabilities]{
    \includegraphics[width=0.44\linewidth]{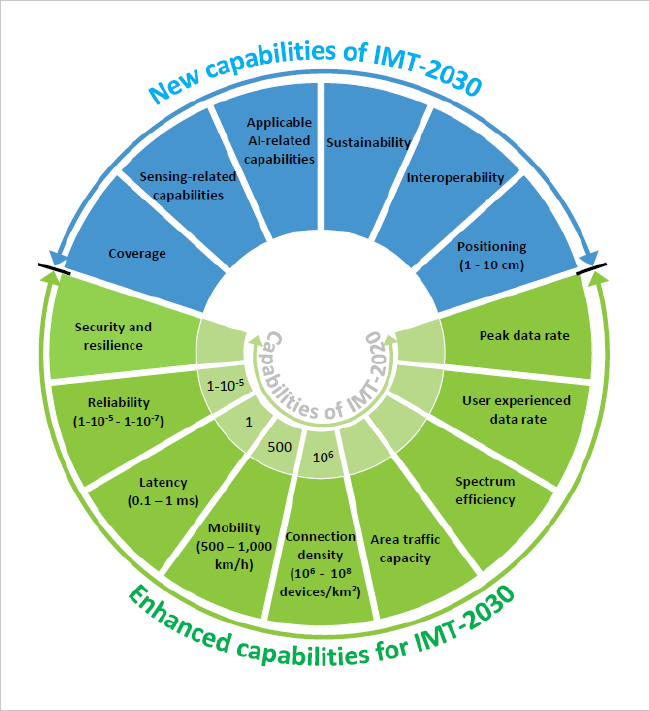}}
    \caption{Usage scenarios and new capabilities of 6G proposed by IMT-2030 \cite{ITU-R}}
    \label{imt2030}
\end{figure}

In June 2023, the International Telecommunication Union (ITU) radio sector (ITU-R) approved the ``Framework and Overall Objectives Recommendation for the Development of International Mobile Communications (IMT) for 2030 and Beyond" \cite{ITU-R} at the 44th ITU-R WP 5D meeting. This document outlines the development goals, typical scenarios, and capability indicators for 6G, providing essential guidance for subsequent 6G technology and standards research. IMT-2030 (6G) has defined six significant scenarios. As shown in Figure \ref{imt2030}, building on the ``iron triangle" of IMT-2020 (5G), IMT-2030 (6G) extends outward to form a hexagon. On the outermost circle of the hexagon, four design principles applicable to all scenarios are listed: sustainability, ubiquitous intelligence, security/privacy/resilience, and connecting the unconnected.

\textbullet\textbf{ \emph{Immersive Communication.}} The immersive communication scenario extends the eMBB of 5G. It includes use cases that provide users with rich interactive video (immersive) experiences, including interactions with machine interfaces. Typical use cases in this scenario include immersive extensive reality (XR) communication, remote multi-sensory presentation, and holographic communication. In immersive communication, supporting time-synchronized mixed traffic of video, audio, and other environmental data is essential, including independent support for voice. Moreover, the ability to improve spectrum efficiency, provide a consistent service experience, and balance higher data rates and enhanced mobility in various environments is crucial. Some immersive communication use cases may also require high reliability and low latency support to enable responsive and accurate interactions with real and virtual objects and greater system capacity to connect numerous devices simultaneously.

\textbullet \textbf{ \emph{Hyper-Reliable and Low-Latency Communication.}} The hyper-reliable and low-latency communication scenario extends the uRLLC capabilities of 5G, covering specialized use cases that are expected to have more stringent requirements for reliability and latency. Typical use cases include communications for comprehensive automation, control, and operations in industrial environments, such as robotic interactions, emergency services, remote medical care, and power transmission and distribution monitoring.

\textbullet \textbf{ \emph{Massive Communication.}} The massive communication scenario expands upon the mMTC capabilities of 5G, focusing on connecting many devices or sensors. Typical use cases include smart cities, transportation, logistics, healthcare, energy, environmental monitoring, agriculture, and many other fields with expanded and new applications. This scenario involves various IoT devices that may not have batteries or have long-life batteries. This scenario requires support for high connection density. Depending on the use case, the scenario requires different data rates, low power consumption, mobility, extended coverage range, and high security and reliability.

\textbullet \textbf{ \emph{Ubiquitous Connectivity.}} The ubiquitous connectivity scenario aims to enhance connectivity to bridge the digital division. One of the critical focuses of this use case is to address areas that currently lack coverage or have minimal coverage, especially in rural, remote, and sparsely populated areas \cite{10540053,10001188}. Typical use cases include, but are not limited to, IoT and mobile broadband communication in these underserved regions.

\textbullet \textbf{ \emph{AI and Communication.}} The AI and communication integration scenario will support distributed computing and AI-driven applications. It will enable unprecedented use cases by leveraging data collection, local or distributed computation offloading, and distributed training and inference of AI models across intelligent nodes. Typical use cases include assisted autonomous driving, autonomous collaboration between medical devices, offloading intensive computations across devices and networks, creating and predicting DT, and assisting collaborative robots. This scenario requires support for high regional traffic capacity, user experience data rates, low latency, and high reliability tailored to specific use cases. In addition to communication aspects, this usage scenario is expected to include a range of new functionalities integrating AI and computing capabilities into IMT-2030. These include data collection, preparation, and processing from diverse sources, distributed AI model training, model sharing, and distributed inference across IMT systems, as well as coordination and linking of computing resources.

\textbullet \textbf{ \emph{Integrated Sensing and Communication (ISAC).}} This scenario contributes to new applications and services that require sensing capabilities. It leverages IMT-2030 to provide wide-area multidimensional sensing, offering spatial information about unconnected devices and spatial information about connected devices, movement, and the surrounding environment. Typical use cases include IMT-2030 assisted navigation, activity detection, and motion tracking (e.g., posture/gesture recognition, fall detection, vehicle/pedestrian detection), environmental monitoring (e.g., rainwater/pollution detection), and providing sensory data/information about the surrounding environment for AI, XR, and DT applications. Besides the provided communication functions, this scenario also requires support for high-precision positioning and related sensing capabilities, including distance/velocity/angle estimation, object and presence detection, localization, imaging, and mapping.

\subsubsection{Capabilities of 6G}
Regarding development goals, 6G aims to achieve seven significant objectives: inclusivity, ubiquitous connectivity, sustainability, innovation, security/privacy/resilience, standardization and interoperability, and accessibility. It may become a new digital infrastructure that better connects the physical and virtual worlds, supports new users, and empowers new applications. Regarding performance indicators, the Recommendation \cite{ITU-R} specifies 15 key capability metrics for 6G, divided into two categories. The first category focuses on enhanced capabilities for IMT-2020, including peak rate, user experience rate, spectrum efficiency, regional traffic density, connection density, mobility, latency, reliability, and security/privacy/resilience performance—totaling nine indicators. The second category supports new functionalities for extending IMT-2030 use cases, encompassing coverage, perception, AI, sustainability, interoperability, and positioning—totaling six indicators. Each capability may exhibit varying relevance and applicability across different usage scenarios. The ITU's vision for 6G encompasses a range of services and scenarios to advance communication capabilities beyond what the current 5G achieves.

{{Among the above-mentioned newly added capabilities, the applicable AI-related capabilities have drawn the most attention. The possession of AI-related capabilities by 6G networks can be understood from several aspects. Firstly, the network can optimize its own performance through AI. Secondly, the network is capable of providing support for the operation of AI. Eventually, the network can offer AI services to users and equipment just as it provides communication connections. Such capabilities can only be achieved through the in-depth fusion of wireless communication and AI.}}

\subsection{Overview of AI and Communication Integration in 6G}
{{This section elaborates on why and how to integrate AI and wireless communication. Generally speaking, AI can endow wireless networks with ``autonomy" and ``intelligence", while wireless networks provide AI with a broad application space. These two inseparable strategic development fields will continuously promote the convergent innovation and technological innovation of information, communication, and computing, and realize people's visionary prospects in the spatiotemporal domain, information interaction types, and cross-industry convergent and innovative applications.}}

\subsubsection{Drivers of Integrating AI into Wireless Communication}
{{Future wireless network systems will evolve into integrated wireless infrastructure platforms deeply integrating ``communication", ``perception", ``computing", ``intelligence", and ``storage." These platforms will be capable of providing customized or personalized services on demand, surpassing the capabilities of conventional communication systems that rely on fixed architectures and predefined rules. Such systems will need to leverage AI technologies. They will utilize big data and knowledge bases for inference and decision-making, employing models' generalization abilities to adapt to various environments and scenarios and providing optimal resource allocation and management solutions. This section analyzes the driving forces of the integration of wireless communication and AI from the following two perspectives.}}

\begin{table}[h!]
\centering
\caption{Data information}
\begin{tabularx}{\textwidth}{m{0.18\textwidth}XX}
\toprule
\textbf{Data Type} & \textbf{Data Name} & \textbf{Application Direction} \\
\toprule
Air interface data & Channel state information, intra/inter-cell interference, multipath delay & Channel state information prediction, wireless channel modeling, power control, interference cancellation \\  \hline

Terminal data & Measurement reports, minimization of drive tests, block error rate, port flow & Network traffic detection and congestion control, traffic prediction and scheduling optimization, personalized service \\  \hline

Network data & KPI, extended detection and response, running log & Dynamic load balancing and interference avoidance in wireless networks, network energy conservation, intelligent routing \\  \hline

Business data & IP information, uniform resource locator information, subscriber tariff data, internet access time & Business identification and awareness, business anomaly detection, traffic management \\ 
\bottomrule
\end{tabularx}
\label{data in network}
\end{table}
{{\textbf{(1) AI to push limits of wireless communication systems.} AI possesses remarkable capabilities in handling big data. Mobile communication networks generate vast and diverse data every moment\cite{CuiBigData,8473689}. The international data corporation has predicted that by 2025, the global daily data generation will reach 491 exabytes \cite{reinsel2018data}. These data typically include terminal, wireless air interface, network, and service data in various formats, such as text, images, extensible markup language, hyperText markup language, graphics, and audio/video information, as shown in Table \ref{data in network}. Network operators can use AI to train and make inference decisions on different types of mobile big data, enhancing network performance and optimizing across various dimensions and objectives.}}

{{AI has predictive capabilities absent in conventional algorithms. AI can learn from historical data via ML algorithms and construct models to forecast future network requirements and possible issues, including network congestion, device failures, or alterations in user behavior. For instance, AI can anticipate potential traffic peaks during specific holidays or large-scale events and take resource allocation measures in advance, such as increasing the capacity of temporary BS or optimizing routing strategies, thereby effectively averting network breakdowns and enhancing user experience.}}

{{Adaptivity is a prominent characteristic of AI. The wireless communication environment is highly dynamic and affected by various factors, such as weather, terrain, buildings, and changes in the location of mobile devices. The adaptive ability of AI enables it to make rapid adjustments according to environmental changes and fluctuations in network conditions. For example, when a user moves from indoors to outdoors or is in a high-speed vehicle, the signal strength and interference situation will change. AI can sense these changes in real time and automatically adjust communication parameters to ensure the stability of communication quality, adapt to different communication scenarios, and improve the reliability and flexibility of wireless communication systems.}}
 
{{In contrast to the hierarchical management of different levels in conventional communication systems, in theory, AI can learn hidden structures and parameters to fit arbitrarily complex functions. This provides a more effective way to sense the complex and variable wireless environment and characterize the network state space. On the other hand, the designs of different communication system modules might have conflicting goals, and there might be performance constraints between modules. There are trade-offs between performance metrics, such as channel capacity and interference, transmission reliability, and system energy consumption. Optimizing each module individually often fails to achieve overall optimal performance. In such cases, AI can facilitate joint optimization design among modules.}}
 
{\textbf{(2) AI for resolving challenges in wireless communications.} The core reason for the challenges summarized in Section 1.1 is that the current wireless network is not intelligent enough. Future wireless network systems will evolve into an integrated wireless infrastructure platform that deeply integrates ``communication", ``sensing", ``computation", ``intelligence", and ``storage".

On the wireless radio access network (RAN) side, AI can be combined to conduct data analysis and decision-making for the control unit plane of 5G new radio (NR), realizing the immediate dynamic performance optimization of wireless resource scheduling and management. The next generation node B (gNB) BS can achieve intelligent on/off handoff based on AI to achieve dynamic energy savings. On the CN side, with the network data analytics function (NWDAF) network elements achieving comprehensive interconnection with surrounding CN elements and realizing data collection, AI will also participate in the control and decision-making of the CN, including quality of service (QoS) guarantee, traffic handing, 5G edge computing, and load balancing of network functions. International communication standard organizations, such as ITU, 3rd Generation Partnership Project (3GPP), and IMT-2030, have initiated the standardization of AI applications in communication systems, and made significant progress. In the future, AI is anticipated to further empower the infrastructure within networks, network management, operation systems, and service systems, fully unleashing the potential of integrating wireless communication and AI in the communication ecosystem and applications.
}

\subsubsection{Vision of AI and 6G Integration}
\begin{figure}[ht]
    \centering
    \includegraphics[width=1\linewidth]{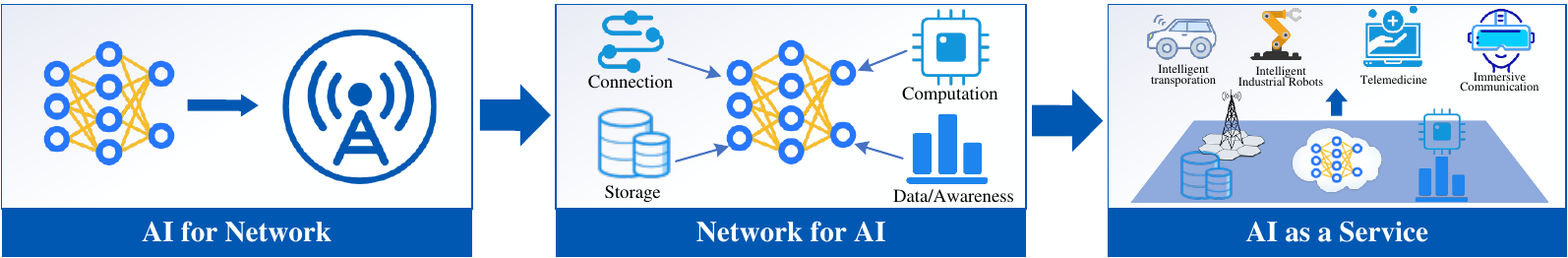}
    \caption{The extent of 6G and AI integration: AI for Network, Network for AI, and AI as a Service.}
    \label{6G and AI 1}
\end{figure}

The advancement of AI technology has significantly empowered the development of wireless networks. Figure \ref{6G and AI 1} illustrates the degree of integration between 6G and AI from three perspectives: AI for Network (AI4NET), Network for AI (NET4AI), and AI as a Service (AIaaS). The application of AI to enhance wireless networks is generally referred to as AI4NET. Its essence lies in using AI to improve wireless networks' performance, efficiency, and user experience.  AI4NET's application in 5G networks has significantly facilitated the intelligent development of mobile communication networks and vertical industries. Its role primarily manifests in O\&M intelligence and network element intelligence \cite{zhinengfenlei}. The former emphasizes utilizing AI to optimize conventional algorithms and automate and intellectualize tasks across network layers and operations. The latter focuses on proposing AI-driven intelligence for individual network elements and functional entities, breaking down proactive network responses into predictions of multiple performance metrics. This approach ultimately aims to optimize network resource allocation and reduce network latency.

To meet the transformative and developmental demands of AI in wireless networks, 6G will no longer be merely a communication network that serves as a connectivity enabler for intelligent services. Instead, it will evolve into an integrated information network that combines communication, sensing, and computing functionalities. While AI empowers 6G networks, 6G networks will also empower AI, achieving NET4AI, which provides comprehensive support for the deployment and application of AI in wireless environments \cite{tong2022nine, tao2023wireless}. The essence of NET4AI is to provide AI with various support capabilities, enabling more efficient and real-time AI training/inference and enhancing data security and privacy protection. The IMT-2030 (6G) Promotion Group in China envisioned the 6G network architecture in \cite{IMT2030-6gnetwork}, as an open and innovative platform for information services, offering capabilities that transcend mere connectivity. These capabilities encompass computing power networks, trusted security, sensing, and data services, which are essential for the operation of AI in networks.  The authors of \cite{yang2022kubeedge} proposed a novel network architecture for providing native support for AI, called an AI-oriented network, which is represented as a network management framework with distributed AI computational capabilities and multi-party participation built in 6G networks.

Moreover, to effectively support ``inherent intelligence" and achieve ``Native AI", 6G networks will treat AIaaS for provision and processing. This concept gives rise to the idea of 6G AIaaS. Specifically, 6G AIaaS utilizes resources and functionalities within the network (including 6G CN, wireless access networks, and terminals), such as connectivity, computing, data, and models. It aims to construct a distributed, efficient, energy-efficient, and secure AI service ecosystem, which includes AI model training, inference, deployment, and other functionalities in a low-carbon open environment \cite{UserExperience}. Not only can it redefine the ecosystem of edge cloud, but also build new business models through 6G mobile networks, transitioning from past connectivity-oriented networks to service-oriented networks, ultimately achieving ubiquitous intelligence \cite{ZJU-6g, LLMFanhua}. AIaaS's typical scenarios include but are not limited to, smart cities, smart agriculture, universal education, and smart industry, etc. Typical applications include unmanned taxi services, smart grid inspections, home health monitoring, and virtual classrooms.

\begin{figure}[ht]
    \centering
    \includegraphics[width=0.9\linewidth]{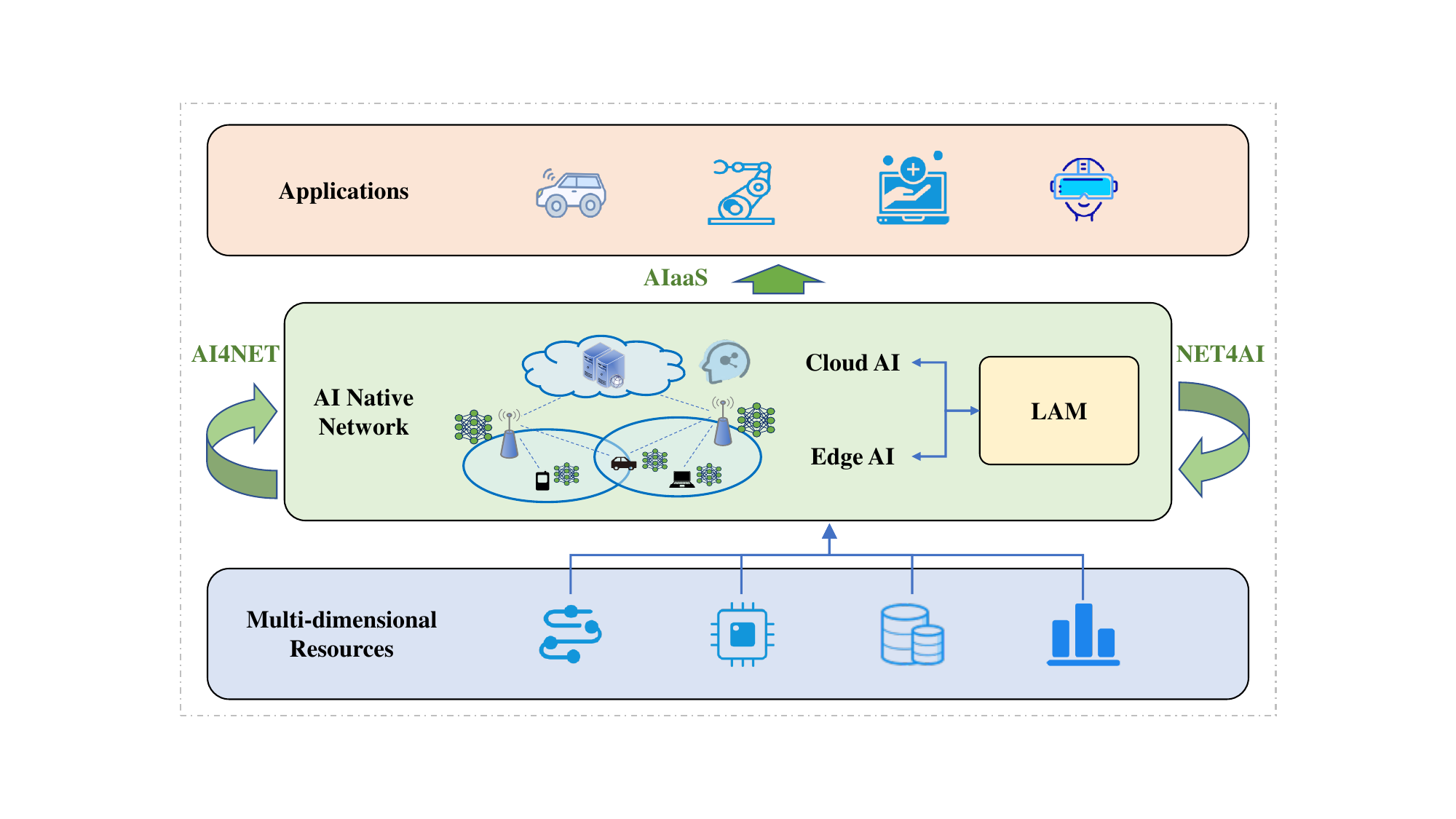}
    \caption{The design of 6G AI integration.}
    \label{6G and AI}
\end{figure}

{{
Large language models (LLMs) \cite{cao2023comprehensive} have represented a significant breakthrough in NLP and can potentially contribute to the development of 6G AI. Compared to traditional smaller parameter models, LLMs exhibit strong contextual understanding, coherent text generation, logical reasoning, and generalization capabilities. Existing general-purpose large models, such as the closed-source GPT model developed by OpenAI and the open-source Llama model developed by Meta (formerly Facebook), can address a wide range of domain-specific problems
\cite{6GLLM,10558819}. Recently, the open-source DeepSeek LLM developed by the DeepSeek team in China has potentially become a game-changer, demonstrating the feasibility of training large-scale AI models (LAMs) at unprecedentedly low cost \cite{deepseekai2025}. LLMs can potentially play a critical role in communication systems, e.g., in task processing and intelligent services.

The emergence of LLMs can bring a new paradigm for the integration of 6G and Al. The intrinsic Al architecture in 6G can provide linkage, computation, model decomposition, and model distribution services for the training and inference of LLMs. LLMs can empower various domains of 6G networks (e.g., air interface, network side, network security, network O\&M, etc.), enabling intelligent services and closed-loop control. In 6G networks, customized specialized large models can be deployed across different network layers. These specialized large models can collaboratively address network issues through task decomposition and composition, cross-layer collaboration, and cloud-edge collaboration, enhancing the intelligence level of the network.

In addition to using customized AI models tailored for specific functionalities and tasks, large models demonstrate superior generalization performance across multiple tasks and breakthrough capabilities on complex tasks that small models cannot achieve. The functionalities of large models increasingly align with the multi-scenario, multi-task characteristics of 6G, effectively alleviating the workload in model perception and performance metric setting within 6G networks and showing broad application prospects. A possible form of integrating 6G with Al is illustrated in Figure \ref{6G and AI}. Leveraging data from diverse heterogeneous networks within 6G, these large models can synthesize different 6G scenarios and services, laying a crucial foundation for Al empowerment in 6G networks. For example, the proposal of 6GANA in the 6G NETGPT white paper has built an LLM similar to ChatGPT, which provides a new paradigm for advanced operations and management of 6G networks \cite{LLMFanhua}.}}
 
\section{AI for Network} \label{Section_AI4Net}
{{The most fundamental goal of integrating wireless communication and AI lies in enhancing the performance, and efficiency of the network and the service experience of users by virtue of AI, which is known as AI4NET \cite{AI4Net}. At this stage, the focus is on how to utilize AI algorithms to optimize communication performance and network functions. For example, AI can optimize the signal modulation and demodulation process, making signal transmission more accurate and efficient; with the help of AI, network resources can be intelligently allocated to achieve load balancing; and an automated O\&M management mode can be constructed to improve O\&M efficiency and reduce costs. This introduction of AI is not anticipated to have a significant impact on the original network architecture. Instead, it is to improve the communication problems in specific recognition by training AI algorithm models. This process is analogous to precisely implanting intelligent patches in the network. It can not only maintain the stability of the network architecture but also gradually enhance the intelligence level of the network.
\begin{figure}[H]
    \centerline{ \includegraphics[width=1\textwidth]{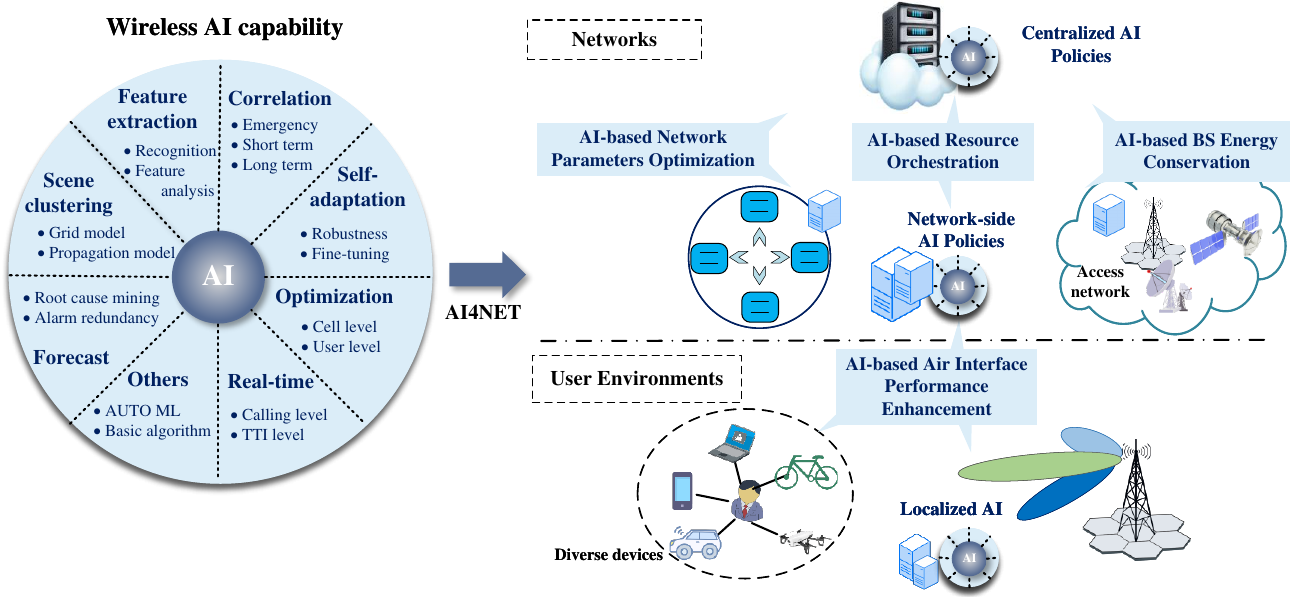}}
     \caption{Schematic of AI's capabilities for network}
    \label{Schematic of AI's capabilities for Network}	
 \end{figure}
 
The concept of AI4NET is presented in Figure \ref{Schematic of AI's capabilities for Network}, where the wireless AI possesses capabilities such as feature extraction, prediction, adaptation, optimization, real-time processing, correlation, and scene clustering. These capabilities form the foundation for AI-enabled wireless networks and can be directly deployed on the BS side and the CN to empower wireless communication. AI optimizes conventional methods to enhance network performance on various terminal devices. On the network side, AI takes advantage of its capabilities to improve the quality of end-to-end services, optimize the functions of network elements, conduct automated management of the upper-layer network, and boost network O\&M efficiency. The advent of AI has also given rise to a series of new services aimed at improving and serving users in a better manner.

AI4NET has initiated relevant research and applications in 5G. In 6G, with the more mature AI technologies represented by DL and the emergence of new infrastructures integrating connectivity and computing power, the relevant applications will become more abundant and sophisticated and may further evolve in depth. This section elaborates on some cases from two aspects, namely air interface performance enhancement and network O\&M improvement, to illustrate the concept of AI4NET.

\subsection{Air Interface Performance Enhancement}

{
In this subsection, we show how AI technology can improve transmission in channel state information (CSI) feedback, orthogonal frequency division multiplexing (OFDM) receivers, beam management, and wireless localization.}

\subsubsection{AI-Based CSI Feedback Algorithm}
Within the 6G framework, the feedback overhead is expected to experience a sharp increase along with a significant rise in the number of antennas. Consequently, ensuring the accuracy of channel reconstruction and minimizing the CSI feedback cost is one of the bottleneck problems that need to be surmounted.
The utilization of AI technology in CSI feedback can profoundly dissect these elaborate channel characteristics. Through training with a substantial quantity of labeled samples, the AI model can acquire the characteristic patterns of the channel within diverse fading environments, interference scenarios, and multipath circumstances. In contrast to conventional CSI estimation approaches relying on statistical theory \cite{10247266} or simple linear models \cite{8308193}, AI-based methods exhibit stronger generalization competencies and can estimate CSI with greater precision in complex real-world communication settings.

The massive CSI data and the inherent random characteristics of CSI make AI potentially useful for designing a new CSI feedback mechanism.
For instance, the nonlinear characteristics of DL can be utilized to efficiently extract the features of CSI, and the original CSI data can be transformed into a more compact feature representation, thereby reducing the communication overhead in the CSI feedback process.
An example of its implementation is the autoencoder architecture, which employs neural networks to extract and compress the features of CSI. The authors of \cite{8934725} were the first to apply DL to CSI feedback. The proposed CsiNet scheme utilizes convolution to extract channel features, and compress and reconstruct channels, and it is superior to the CSI feedback scheme based on compressed sensing in terms of feedback accuracy and computational complexity. The encoder of CsiNet compiles the high-dimensional CSI into code words, while the decoder is responsible for decoding the codewords back into the original CSI. The internal structures of encoders and decoders of different algorithms vary, resulting in differences in channel reconstruction performance.
As shown in Figure \ref{CSI Feedback}, Huawei demonstrated the huge advantages of AI-based CSI feedback over conventional codebook approaches through system-level simulations. AI with additional information can effectively improve the performance on the square of the generalized cosine similarity (SGCS) and throughput \cite{Huawei}.
\begin{figure}[h!]
    \centering
    \subfloat[SGCS versus bits]{
    \includegraphics[width=0.495\textwidth,height=0.37\textwidth]{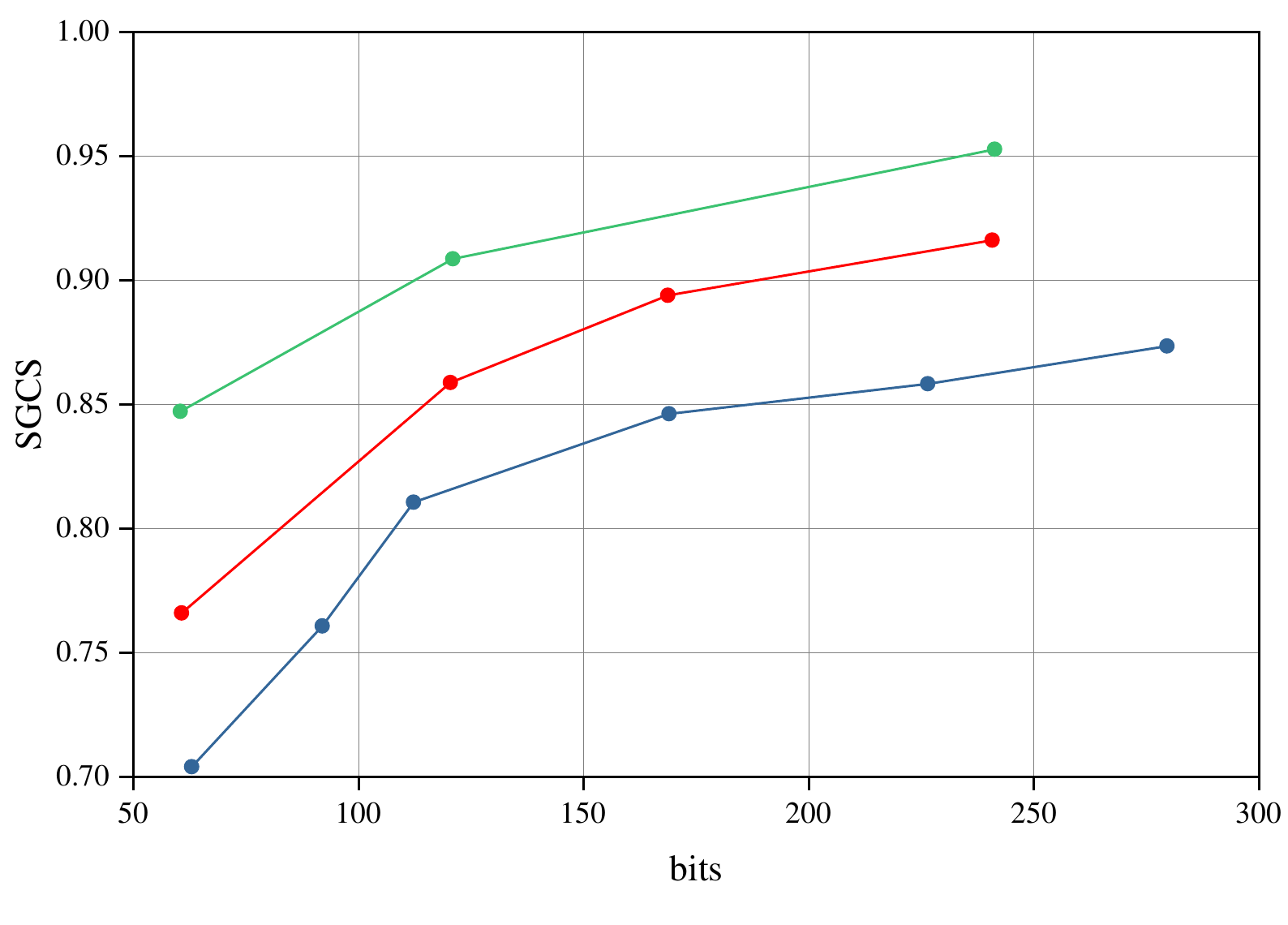}}
    \subfloat[Throughput gain versus bits]{
    \includegraphics[width=0.495\linewidth,height=0.37\textwidth]{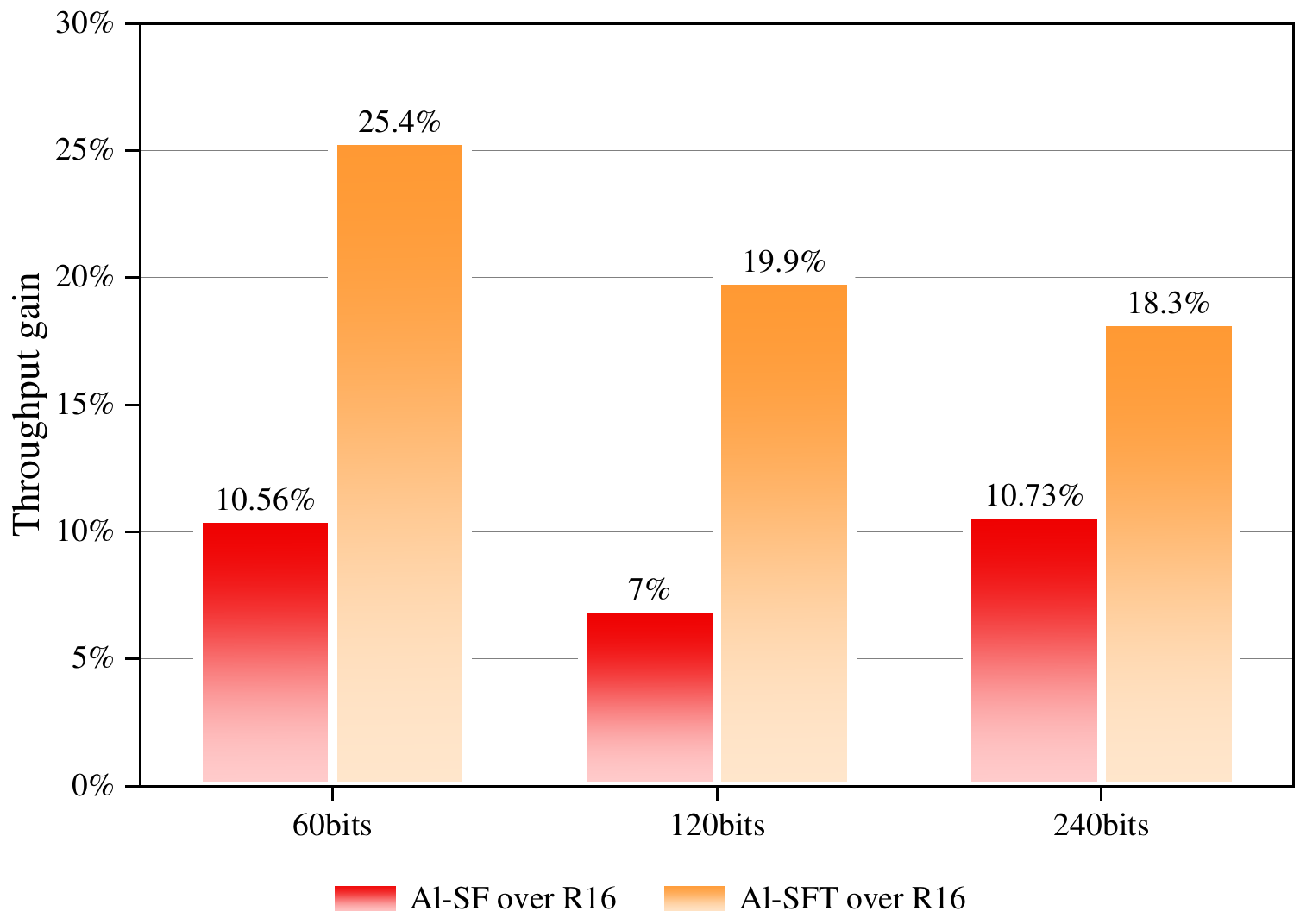}}
    \caption{The performance of R16 Type II codebook, AI-SF, and AI-SFT on SGCS and throughput gain. AI-SF means the AI/ML-based spatial-frequency compression method, which transformer is used as the backbone of both the encoder and decoder. AI-SFT is AI-SF with additional past CSI.}
    \label{CSI Feedback}
\end{figure}

The studies in \cite{8972904} increased the size of the convolutional kernel based on CsiNet to improve the perceptual field of view of the convolutional layer, which is conducive to extracting the sparsity of the channel and enhancing the accuracy of channel reconstruction. Although a larger perceptive field of view could effectively extract the sparsity of CSI, a smaller convolution kernel could extract finer features from CSI. A DL-based CSI compression and quantization method was developed in \cite{9466243} considering high compression ratios. A multi-rate CSI compression framework was designed in \cite{9625585} to improve the generalization of the model in the field of transfer learning. In \cite{9178295}, a non-local neural network was introduced based on the CsiNet network to capture a wide range of dependencies, and the accuracy of channel recovery was improved compared to CsiNet. The CRNet was proposed in \cite{9149229}, which used convolution kernels of different sizes for channel feature extraction and recovery, reducing the computation and improving the accuracy of channel reconstruction.
In order to better deploy AI-based CSI feedback models in real-world wireless environments, model monitoring, model updating and AI-based signaling process management are hot issues that require our attention in future studies.

\subsubsection{AI-Based OFDM Receiver}
AI method can extract essential information in sparse and time-varying pilot frequencies. The complex mapping relationship between the input and the output is effectively constructed, thus AI-based receiver can learn and adapt to channel characteristics in complex environments. Each module of the conventional receiver requires accurate signal modeling and calculation, leading to the complexity and calculation burden of the system. AI has the potential to extract features directly from the original signal and demodulate it in a data-driven manner, simplifying the complex process of conventional receivers.

\begin{figure}[H]
    \centerline{ \includegraphics[width=0.9\textwidth]{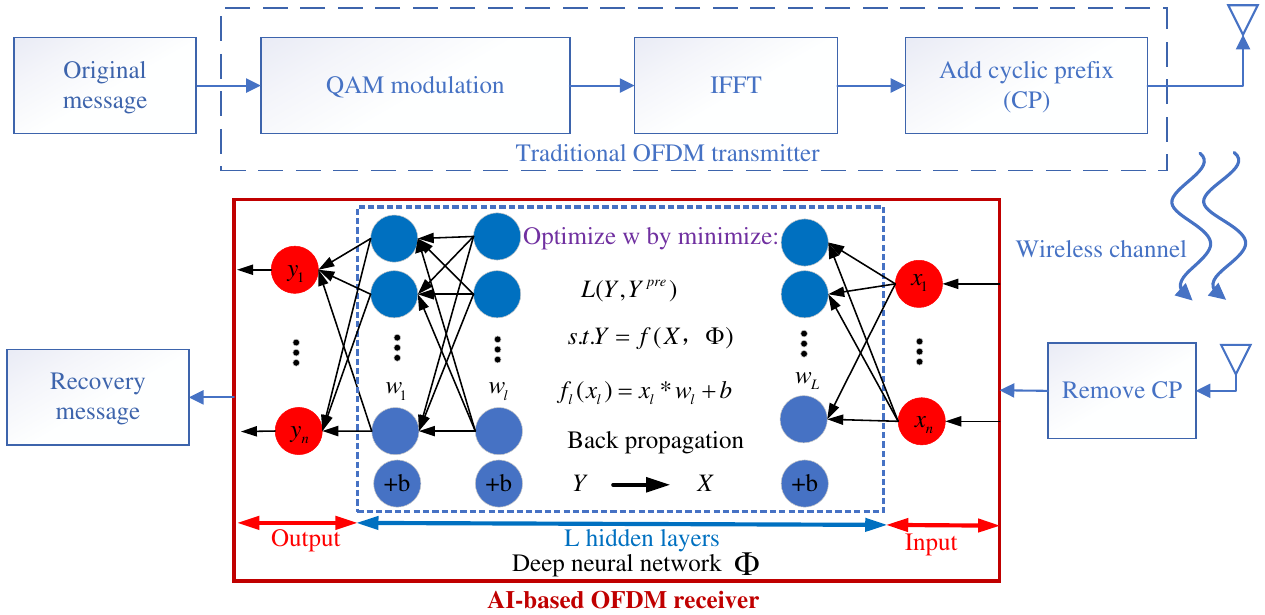}}
     \caption{The AI-based OFDM receivers. Using AI modules to replace conventional physical layer processing, AI's capabilities can compensate for signal distortion.}
    \label{OFDM receiver}	
 \end{figure}
Regarding wireless OFDM receivers, some researchers have used a fully connected deep neural network (DNN) to improve the existing modular OFDM receiver. A typical framework of an AI-based receiver is given in Figure \ref{OFDM receiver}. We use DNN parameters $\Phi $ to construct the relationship between the input $X$ and output $Y$ as $Y:f(X,\Phi )$. With the assumption of $L$ hidden layers, we denote $w_{l}$ as the parameter weights of the $l$-th layer, where $l$ ranges from $0$ to $L+1$. $w_{0}$ and $w_{L+1}$ represent the weights of the input layer and the output layer, respectively. $b$ denotes the bias of each layer. In the training stage, the optimization direction of the model minimizes the loss between the output $Y$ and true label $Y^{true}$. Thus, the network parameters $\Phi$ are updated by changing $w_{l}$ during the process of back propagation. Hence, the AI model can encode the information of actual scenes into neural networks and optimize the performance of algorithms by adjusting the structure and parameters of neural networks. 

In \cite{8509622}, a model-driven DL approach was proposed, which combines DL with expert knowledge to replace existing OFDM receivers in wireless communications. An ML-assisted physical layer receiver technique was proposed to demodulate the OFDM signals, subject to very high Doppler effects and corresponding distortions in the received signal \cite{9723316}. To strike a balance between full-size cyclic prefix (CP) and non-existent CP, the authors of \cite{9287725} investigated the redundancy problem and proposed a minimum redundant OFDM receiver using DL tools. In \cite{9446039}, the receivers were designed based on DNN, consisting of a layer of DNNs and soft decisions. The problems of channel estimation error, delay, and decoding limits between users with conventional detection methods were solved. The symbols of all users were recovered at one go to jointly perform channel estimation and signal detection. A novel generative supervised DNN was designed in \cite{8663458}, which could be trained using a reasonable number of pilots. After channel estimation, the neural network-based receiver jointly learns the pre-encoder and decoder for data symbol detection. An intelligent receiver for OFDM communication systems was designed in \cite{9637770} based on the DNN structure, and realized by optimizing the DNN structure. This method can recover information on the receiving side and avoid complicated pilot operations and signal error accumulation. Moreover, a convolutional neural network (CNN) was used to reduce the bit error rate using the mathematical calculation function of discrete Fourier transform and the training of OFDM signal samples \cite{9526047}. 

\subsubsection{AI-Based Beam Management}
In the existing literature, the beam selection methods of multi-antenna systems can be divided into beam selection based on beam scanning \cite{8023460}, beam selection based on position prediction \cite{8444984}, and beam selection based on hierarchical search \cite{7460513}. Conventional beam management methods have some limitations when applied to millimeter wave massive multiple-input multiple-output (MIMO) systems \cite{9665388}. The primary cause is that the real-world wireless propagation environment often has various characteristics. After a beam is narrowed, the signal transmission is more susceptible to occlusion. The increase in the number of beams significantly increases the number of beam search operations in beam alignment. It is difficult to accurately select the optimal beam instantaneously.

\begin{figure}[H]
    \center
    \includegraphics[width=0.7\textwidth]{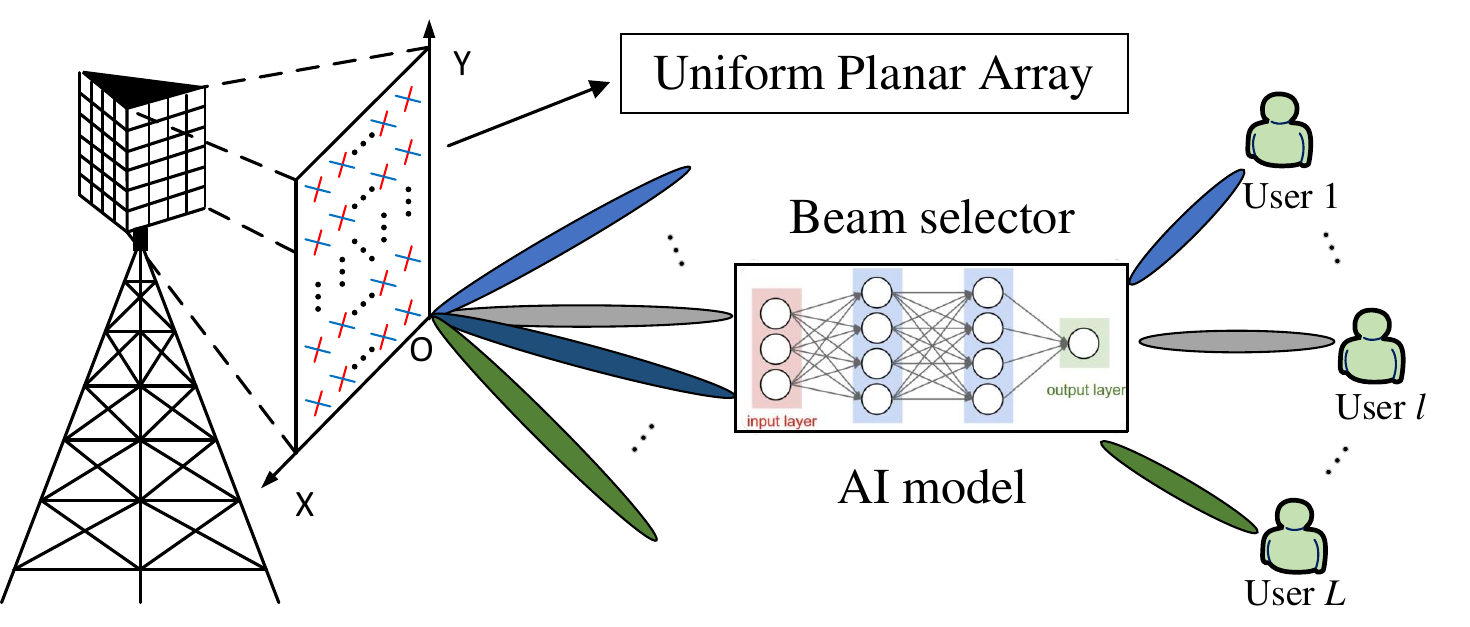}
     \caption{The structure diagram of AI-based beam management. AI selects the optimal beam by looking for the complex mapping relationship between input and output, which can significantly reduce the search cost of the beam while ensuring prediction accuracy.}
    \label{beam managment}	
 \end{figure}
With the application of AI, the optimal beam can be selected without scanning all beam pairs, thus alleviating the scanning overhead existing in the conventional beam management method, as shown in Figure \ref{beam managment}. In \cite{heng2021machine}, a DNN was utilized for beam alignment based on the contextual information of the UE location. This method can reduce the search time by four times for AP selection and over tenfold for beam selection. By taking advantage of the mmWave channel structure, a novel hierarchical beam alignment framework was proposed in \cite{yang2023hierarchical} to leverage DL techniques to seek the optimal beamformer with learnable probing codebooks. This novel framework learns two tiers of probing codebooks and uses their measurements to predict the optimal beam in a coarse-to-fine search manner. A DL-based beam alignment method was proposed in \cite{9690703} by jointly training the probing codebook and the beam predictor.
Simulation results demonstrate that the method can achieve high beam alignment accuracy while reducing the beam sweeping complexity by ten times.

The above literature aims to use DL methods to improve the accuracy and robustness of beam alignment, and has demonstrated the superiority of AI. For practical deployment, we should further consider the acquisition of high-quality tags in complex environments. Considering the training overhead, improving the generalization ability of AI models among different BS environments also becomes an important future work.

\subsubsection{AI-Based Wireless Localization}
The 6G network is mandated to furnish high-precision wireless location capabilities for diverse application scenarios. 
This location information is usually in the form of the mobile user's geographic coordinates relative to some reference points \cite{sayed2003wireless}.
Many applications need high-precision positioning, such as industrial automated guided vehicle and asset tracking, especially indoor precision positioning, but the global positioning system (GPS) cannot be used indoors. Despite having a specific positioning reference signal, LTE cannot meet the requirements for high-precision positioning due to its relatively low positioning accuracy and the large distance between BS (about 100 meters even with a 20 MHz bandwidth). Bluetooth, Wi-Fi, and other wireless location-based technologies can have high deployment costs and are difficult to become a universal positioning technology \cite{luan20226g}.

The application of AI technology has been introduced to achieve high-precision positioning in the 5G era. One approach is to utilize the random forest technique \cite{maung2020enhanced}, which can create a classification model to split a vast area into many small grids and then predict the grid where the user is located. It can also make a regression model to predict the user's position coordinates. In another method, multi-layer perceptron has a similar application to random forest and can establish regression and classification models to locate targets \cite{shan2018outdoor}. AI algorithms can also be trained to recognize the fingerprint of a wireless signal in a specific location and fuse data from different sensors to achieve high-precision positioning.

With the widespread use of massive MIMO technology, AI has shown more significant advantages in high-precision positioning. 
Using the location in the line-of-sight (LoS) and non-line-of-sight (NLoS) channel co-existence scenario based on the neural network as an example, ML and a significant volume of channel data can help effectively map the relationship between channel response \cite{zhang2022indoor} and position coordinates, resolving the location problem in complex environments and increasing location accuracy. Very high positioning accuracy is achieved without significant additional cost, even about 20 mm accuracy under indoor LoS and NLoS conditions.
Due to the diversity of data modes in complex environments, the further development and practical application of AI-based localization requires us to consider the robustness and reliability of the AI model to enable mobility management.

\subsection{Network O\&M Efficiency Improvement}
{
AI can analyze a large amount of data and make real-time decisions, which is especially useful in wireless networks where many variables and parameters must be continuously monitored and adjusted to optimize system performance \cite{10505907}. It can dynamically allocate resources such as bandwidth and power in wireless networks to optimize system performance.
In addition, the integration of AI enables the network to better understand and predict the complexity and dynamics of the network. Based on the prediction results, proactive and timely network maintenance can be carried out. By leveraging AI's DL and pattern recognition technologies, a more in-depth understanding of network behavior can be achieved, realizing more optimal resource management and scheduling strategies. Subsequently, the role of AI in network O\&M management will be introduced through specific cases.}

\subsubsection{AI-Based Traffic Prediction}
The progressive deployment of 6G networks \cite{2020A} heralds the increasing prevalence of new application scenarios like virtual reality (VR) and immersive communication. In such applications, the network traffic exhibits high variability.
By predicting network traffic, it is possible to identify traffic fluctuations and peak periods in advance, thereby enabling dynamic allocation and scheduling of network resources.
For instance, the authors of \cite{2017Spatiotemporal} employed a recurrent neural network (RNN) to achieve joint spatiotemporal prediction, aiming to improve the accurate modeling of network traffic variations. The method proposed in \cite{2017Network}, based on deep belief networks, effectively predicts the long-term variation trends of network traffic. The result in \cite{2020Cellular} demonstrated that, compared to conventional ‌autoregressive integrated moving average models, long short-term memory outperforms in traffic prediction tasks, particularly in handling nonlinear and complex time-series data.

In centralized traffic prediction methods, a BS typically needs to collect traffic data from various geographical locations, which inevitably introduces additional communication overhead and potential security risks. As a result, federated learning (FL)-based traffic prediction methods have emerged. The FedDA method was proposed in \cite{zhang2021dual}, which effectively reduces the delay and bandwidth overhead associated with data transmission by only transmitting model parameters, instead of raw traffic data. This approach not only ensures privacy but also improves the efficiency of data processing. The authors of \cite{zhang2022efficient} proposed a distance-weighted federated traffic prediction method that better captures the spatiotemporal characteristics of traffic and improves the accuracy of traffic forecasting.

To reduce the model training time, the authors of \cite{2019Deep} proposed a transfer learning-based approach, which accelerates the training process by uncovering the potential correlations between traffic patterns in different regions. Other studies have employed techniques such as data sharing \cite{zhao2018federated} and meta-learning \cite{2017Model}, to further optimize the performance of traffic prediction models. In the future, network traffic prediction methods will extract spatiotemporal features and long-term dependencies, but also enhance the lightweight design and privacy protection capabilities of the models.

\subsubsection{AI-Based BS Energy Conservation}

The information and communication technology (ICT) industry directly contributes approximately 4\% of global greenhouse gas emissions, with mobile communication networks accounting for over 10\% of this figure~\cite{9215362}. In response to the challenges posed by climate change, the ITU released Recommendation 1470, titled ``Greenhouse Gas Emission Trajectories for the ICT Industry in Line with the United Nations Framework Convention on Climate Change Paris Agreement", which calls for mobile operators to achieve a 45\% reduction in carbon emissions from 2020 to 2030, aiming to drive the industry's transition towards sustainability. 

The energy optimization within 6G networks exhibits a close correlation with user latency. A key focus of academic research has been how to achieve network energy savings by effectively shutting down BS while ensuring the maintenance of basic service quality. The authors of \cite{Jinlin2014Stochastic} proposed a BS shutdown strategy based on traffic thresholds. During peak network periods, this approach may result in frequent state handoffs, leading to service interruptions. The authors of \cite{2015AE} classified BS states into four different sleep modes based on the activation time scale of the BS. While deeper sleep modes can significantly reduce energy consumption, they also result in increased wake-up delays, affecting service quality. 

\begin{figure}[H]
    \centerline{ \includegraphics[width=0.65\textwidth]{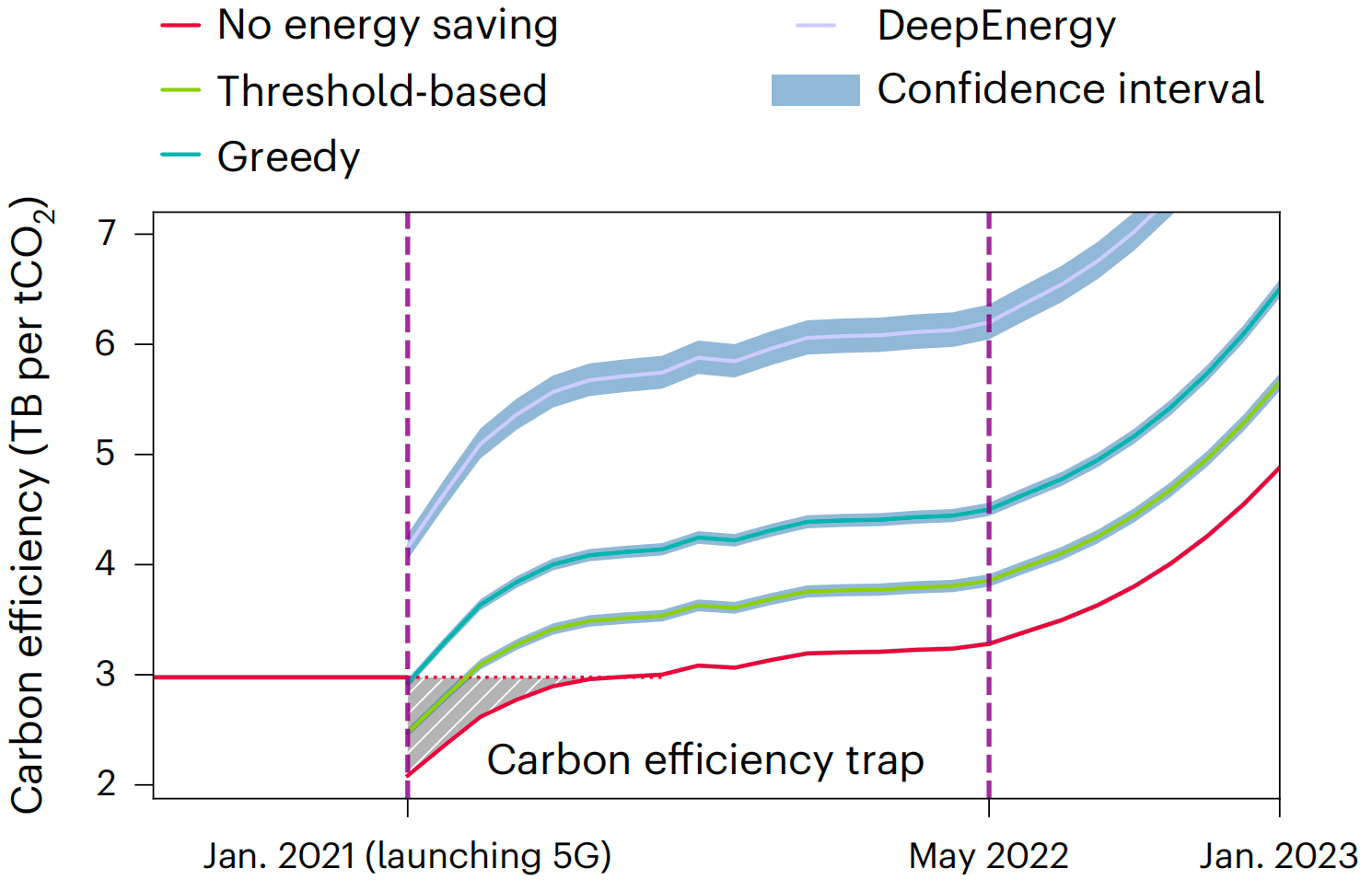}}
     \caption{Carbon emissions of different strategies \cite {li2023carbon}}
    \label{Energy_Conservation}	
 \end{figure}
The energy-saving strategy for BS predicated on reinforcement learning (RL), via the incessant interactive learning between the agent and the operational milieu, dynamically modifies the parameters to accommodate the variations in service load and the oscillations in channel conditions. In \cite{2019Reinforcement}, a BS state management scheme was proposed based on Q-learning. It dynamically adjusts the energy consumption of the BS in response to load changes by exploring the optimal duration of different sleep modes. The authors of \cite{masoudi2020reinforcement} considered the interference between multiple BS, and designed an energy optimization method based on collaborative strategies. By enabling cooperation between BS, this method reduces the overall system energy consumption. A cascading RL method was proposed in \cite{lin2022dades}, which optimizes the decision process of user management and BS states in multi-BS scenarios, further improving network efficiency. 

Introducing renewable energy, especially clean energy sources, e.g., solar power, as the energy supply for networks can significantly improve network energy efficiency and reduce dependence on conventional power grids. As depicted in Figure \ref {Energy_Conservation}, the authors of \cite {li2023carbon} integrate RL with graph neural networks and accomplish the optimization of energy conservation and carbon emissions in wireless networks by capitalizing on photovoltaic solar technology. Energy-saving efforts in wireless networks should be deeply integrated with the use of renewable energy, providing more sustainable solutions. 

\subsubsection{AI-Based Network Parameter Optimization}
The rational setting of network parameters is crucial, as these parameters determine the design, energy consumption, and performance of the 6G network, and must be aligned with the real-time distribution of users and the electromagnetic environment. 
Wireless network parameters can be categorized into two types: BS parameters and threshold parameters. 
The BS parameters involve basic handoff functions such as paging and slicing. 
The threshold parameters configure the thresholds for cell reselection, event decision-making, and neighbor cell handoff. Moreover, spectrum resource planning in communication systems is a critical component of parameter optimization. Studies have shown that spectrum is often underutilized \cite{Mchenry2007XG}, and AI technologies have demonstrated great potential in cognitive radio \cite{kaur2022comprehensive}. AI can monitor and mitigate interference \cite{alabi2023artificial}, predict spectrum usage trends \cite{data2020machine}, and significantly improve the utilization of spectrum resources.

Conventional parameter optimization relies mainly on drive tests and expert experience. 
When network operators identify significant deficiencies in communication service quality, specific operators will deploy drive test vehicles to collect and report network performance metrics, such as network throughput and audio/video transmission delay. Subsequently, operators create scatter plots to further characterize network performance. Finally, expert experience is used to analyze and optimize the network parameters \cite{6666502}. 
Due to the significant increase in the number of 6G antennas, the channel is not a simple one-dimensional circuit loss structure but has expanded into an $N$-dimensional angle power spectrum. The efficiency of expert-based parameter tuning is low, making it difficult to modify billions of 6G network parameters. Furthermore, conventional experience-based tuning methods often rely on inherent rules, which makes it difficult to fully capture and adapt to the complex relationships in the network \cite{10155734}. Table \ref{Various parameters in the current network} summarizes various parameters in the current networks.

\begin{table}[H]
\centering
\caption{Various parameters in the current network}
\label{Various parameters in the current network}
\begin{tabularx}{\textwidth}{XXX}
\toprule
\multicolumn{1}{l}{\textbf{Parameter Type}} & \multicolumn{1}{l}{\textbf{Parameter Name }} & 
\multicolumn{1}{l}{\textbf{Application Area }}\\
\toprule
\multirow{3}{*}{\centering Terminal parameter} 
& Block error rate  & Scheduling optimization  \\
& Traffic information & Traffic monitoring\\
& Measurement report  & Adaptive Coding \\
\hline
\multirow{3}{*}{\centering Air interface parameter} 
& Multi-path time delay & Channel modeling \\
& Inter-cell interference & Interference cancellation \\
& Channel state information & Channel state prediction \\
\hline
\multirow{3}{*}{\centering Core network parameter} 
& Operation log  & Fault handling  \\
& Network topology  &  Load balancing \\
& Network energy consumption & Energy saving \\
\hline
\multirow{3}{*}{\centering Service parameter} 
& Online duration  & Personalized service \\
& Transport protocol  & Transmission control  \\
& Consumption history  & Service monitoring \\
\toprule
\end{tabularx}
\end{table}

AI models have demonstrated unprecedented advantages in large-scale parameter tuning, nonlinear fitting, and real-time optimization \cite{8382166}. 
In the context of the highly intricate architecture and nonlinear traits of the 6G networks \cite {you2014development}, AI technology can gain a more profound understanding of the network's nonlinear features by learning from and dissecting extensive network data, and optimizing network parameters based on real-time data and scenario.
Deep RL (DRL) has been deployed for the optimization of network parameters \cite{10332666}. Specifically, each performance metric within the current communication network is considered a distinct RL state, and the adjustment operations carried out on network parameters are designated as diverse actions. In this procedure, by incessantly adapting the behavior of the RL model across different states, the model will receive feedback information from the environment to assess the merit of each action. RL models typically reinforce those actions that can yield higher reward returns.

\subsubsection{AI-Based Mobility Management}
As early as 2015, 3GPP began researching how AI technology could be integrated into the 5G network. For example, the 3GPP SA2 and SA5 working groups (WG) launched projects on ``Enablers for Network Automation" and ``Enhancement of Management Data Analytics Services", which made good progress \cite{wang2021research}. Using technologies such as DL, AI is capable of predicting user behaviors, including moving speed and direction, thereby permitting the proactive selection and handoff between cells to diminish signal overhead. In ultra-dense network scenarios, frequent cell exchanges can lead to increased overhead and connection disruptions, adversely affecting QoS. The authors of \cite{shubyn2020deep} proposed the use of a GRU-based RNN to adaptively handle handoff decisions, significantly reducing handoff latency and improving network throughput. In \cite{liu2020proactive}, the authors suggested a method for predicting the next position of vehicles using AI in ultra-dense network scenarios, which allows for BS selection based on predicted positions, thereby reducing handoff frequency and failure rates, and saving handoff signaling overhead. This demonstrates that AI-based mobility management is feasible and practical.

On the other hand, through RL and other techniques, AI can dynamically optimize mobility management strategies based on network environment and user behavior, better addressing complex network environments. As wireless networks in higher frequencies require more dense deployment of BS, the authors of \cite{prado2023enabling} addressed the issue of reduced network capacity due to frequent handoffs, proposing a centralized and multi-agent deep Q-network (DQN) algorithm. Both algorithms can find near-optimal user-to-BS allocations. As shown in Figure \ref{Mobility_Management}, the DQN algorithm significantly reduces or even completely avoids ping-pong handoffs compared to the baseline (conventional signal-to-interference-plus-noise ratio (SINR)-based handoff algorithm). The authors of \cite{yan2019machine} proposed an ML-based framework to optimize the handoff between sub-6 GHz and mmWave bands in vehicular networks, using RL to enable vehicles to predict and discover the optimal handoff target in real-time. The method reduces handoff latency, improves connectivity in vehicular networks, achieves seamless handoff even in high mobility scenarios, and enhances user experience.
\begin{figure}[H]
    \centerline{ \includegraphics[width=0.6\textwidth]{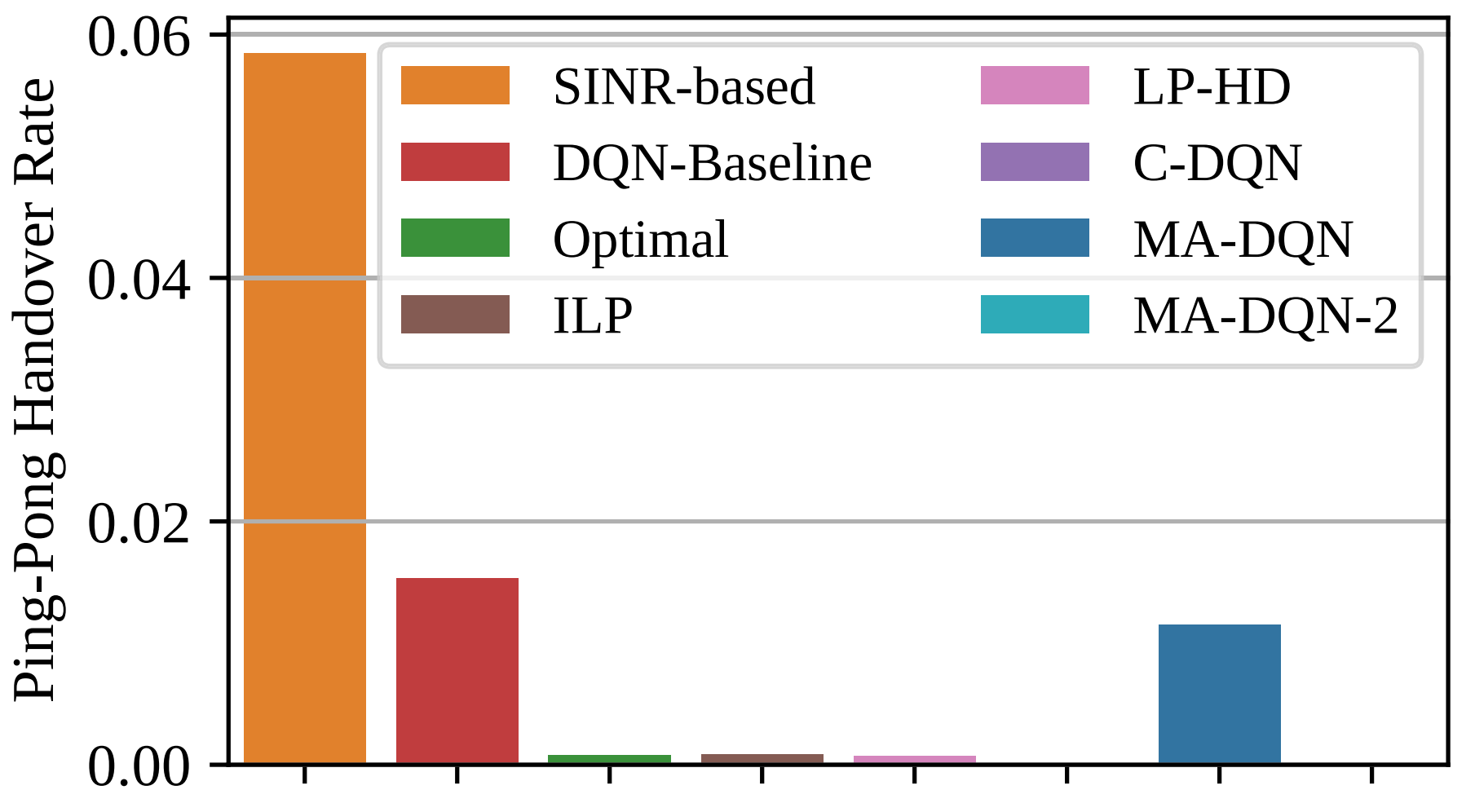}}
     \caption{Ping-pong handoff rate performance comparison in \cite{prado2023enabling}}
    \label{Mobility_Management}	
 \end{figure}
 
\subsubsection{AI-Based Resource Orchestration}
With hundreds of billions of devices being connected, the generated data is undergoing continuous explosive growth \cite{10121779}. Merely strengthening the communication capabilities is insufficient to meet the real-time requirements for data processing in control application scenarios \cite{9998551}. By harnessing the communication and computing capabilities at the network edge and on the device side, the entire process of data collection, processing, analysis, and decision-making can be accomplished closer to the device side, avoiding the latency drawbacks caused by congestion in the CN. 

The joint scheduling of communication and computing resources founded on RL attains the collaborative and optimal allocation of communication and computing resources via the interaction and learning between the agent and the 6G network environment \cite{10003251}.
Confronted with the resource allocation difficulties within the space-air-ground integrated network in the 6G era, the authors of \cite{9858867} put forward an RL-assisted bandwidth-aware virtual network resource allocation algorithm. This algorithm uses RL with a policy network for node embedding, preferentially handling high-bandwidth requests to meet strict user bandwidth demands.
Given the relevant problems that exist within 6G mobile devices and communication systems, motivated by the concept of virtual network embedding, the authors of \cite{10323248} pioneered a multi-objective aware dynamic resource scheduling algorithm designed for multi-layer computing networks, aiming to augment resource flexibility. They established a scheduling network underpinned by DRL and refined the learning process, thus proffering a sustainable resolution for resource scheduling strategies.
Regarding the circumstance that the applicability of the open radio access networks (O-RAN) in governing and optimizing the functions of the wireless access network within the 6G network has not been extensively explored, the authors of \cite{10330565} developed low-complexity algorithms based on methods such as RL to address the problems of jointly optimizing traffic splitting and allocation, congestion control, and scheduling across different time scales. This effort provides insights for realizing a fully automated network with enhanced control and flexibility.

To address the challenges of computing resource demands and intelligent resource allocation in 6G network IoT services, the authors of \cite{9380677} proposed an effective DRL-based solution for IoT resource expansion and service placement, and verified the effectiveness of its multi-application autonomous resource allocation via dataset simulation.
In \cite{9655323}, the authors proposed a multi-task DRL method based on graph convolutional networks by introducing a joint network slicing and routing mechanism, achieving robustness in various network environments.
In the study of \cite{9459763}, the authors established a model-free DRL framework by a hierarchical structure integrating modified deep deterministic policy gradient (DDPG) and double DQN to actualize an intelligent RAN slicing strategy with two-layer control granularity for maximizing QoS and slice spectrum efficiency.
Moreover, a dynamic spectrum allocation scheme was proposed in \cite{9810019}, which utilized the advantages of the DQN and reduced the search state explosion by a reward and punishment framework to dynamically allocate unallocated resource blocks to mobile units and obtained the Pareto optimal solution of sub-problems via Chebyshev decomposition.

The AI-based resource orchestration still faces numerous challenges and opportunities in the future. 
For instance, the improvement of DRL algorithms is urgent to cope with complex and changeable environments. 
It is necessary to strive to enhance the adaptability to complex scenarios, meet the diverse resource requirements in various special scenarios of different industries, and improve performance evaluation and enhance algorithm interpretability.

\subsubsection{AI-based Situational Awareness and Fault Detection}
Network situational awareness technology endows mobile communication networks with real-time monitoring, prediction, and response capabilities. Its applications include network management and optimization, security monitoring, as well as fault detection and prediction \cite{9075574}.
AI-based situational awareness collects, stores, and analyzes vast amounts of network data via big data platforms and DL techniques, thereby providing precise threat intelligence and full traffic inspection \cite{7587350}. This technology employs DL to precisely identify the states of various elements including network traffic, logs, and network key performance indicator (KPI), predict network development trends, and formulate accurate response strategies through the analysis of perceived information and predicted states \cite{9775698}, realizing the shift from “passive defense” to “active defense”. The AI-driven threat representation and causal reasoning can identify and infer the origins of attacks, preventing the recurrence of similar security events. 

The network fault analysis constitutes one of the crucial elements in the O\&M of mobile communication networks. As the number of mobile devices increases and the network scale expands, the quantity of network alarms is growing explosively, and the relationships among them are highly intricate \cite{8330210}.
Employing AI to build a fault detection analysis model endows the model with the capacity for high-dimensional data analysis, enabling it to serve as a more intelligent and efficient tool for fault analysis~\cite{10541561}. 
By conducting a fusion analysis of these data, the system can effectively identify common issues within the network and perform accurate fault diagnoses.
The results obtained after fault handling will be utilized as feedback to update and optimize the fault diagnosis model, allowing the system to operate efficiently under constantly changing network environments and business requirements \cite{10543904}. The specific operational workflow is illustrated in Figure \ref{SA_1}.
\begin{figure}[H]
    \centerline{ \includegraphics[width=1\textwidth]{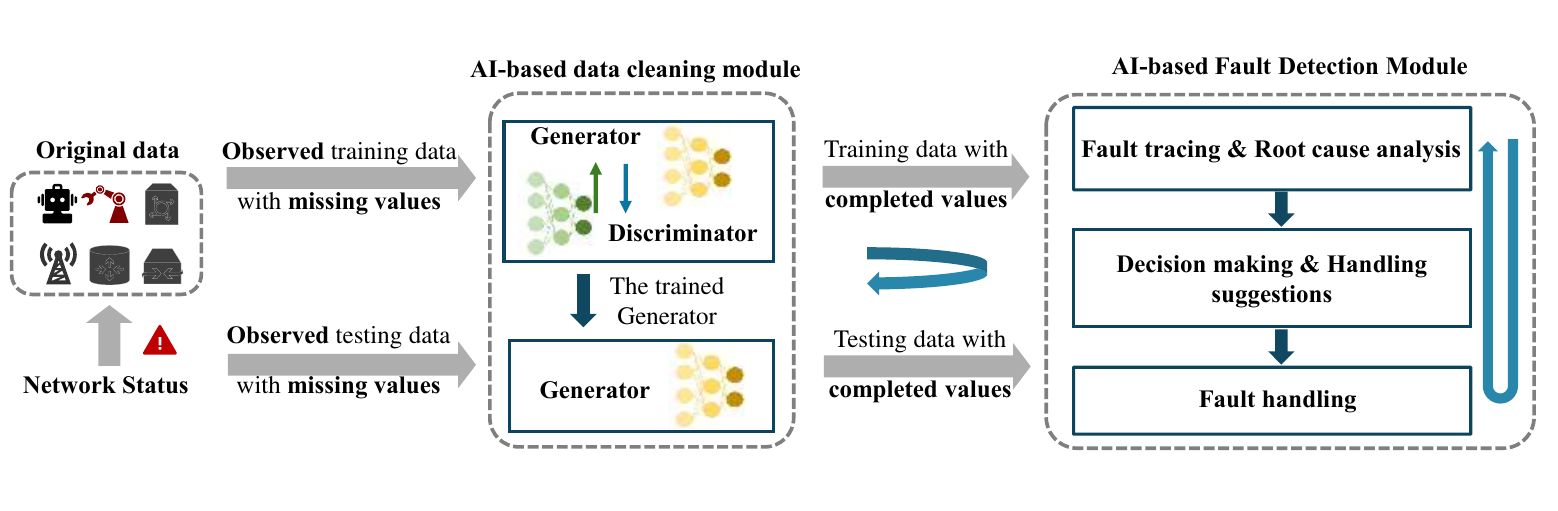}}
     \caption{AI-based fault detection}
    \label{SA_1}	
 \end{figure}

By integrating AI technology, we can establish a systematic and intelligent network fault analysis and tracing back scheme, which can provide automated, accurate, and flexible fault detection and tracing back capabilities and continuously enhance the efficiency of fault diagnosis and handling. Meanwhile, it can also adapt to the changes in network environment and business requirements.}}

\section{Network for AI}
\label{Section_Net4AI}
{
A series of AI use cases and scenarios have been developed, and part of the standardization work has been accomplished under the leadership of 3GPP. This has led to the creation of various module-level ``plug-in'' AI functions in 5G networks. Although this approach is convenient, it still lacks considerations of systematicity and interpretability, limiting the scalability of AI capability and the ability to model generalization. 

Key technologies such as cloud-native architectures, software-defined information technology, and service virtualization have advanced significantly in 5G (e.g., CN supporting service-based architecture and cloud-based deployment of CN functions). 3GPP has also standardized the NWDAF as an AI-enabled component within the CN. This includes defining relevant mechanisms and processes while introducing standardized AI operational paradigms.
However, these advancements remain constrained by several factors, including technological maturity, security concerns, and the complexity of system O\&M. Consequently, the current access network architecture largely retains the traditional ``siloed communication-technicalized BS design." In this architecture, BSs are highly tailored to communication services, lacking the flexibility to interface with emerging AI-driven services. 
From an end-to-end AI service perspective, there is no clear definition of AI functional components on the access network side. The integration and coordination of AI service functions between the CN and the access network remain unaddressed. This disconnect hinders the current network architecture from supporting the intelligent vision envisioned for 6G.

To support ``native intelligence'' and achieve ubiquitous general intelligence, the 6G network requires groundbreaking innovations in the network architecture to enable the endogenous embedding of AI capabilities within the network. This represents the core concept of Phase II (i.e., NET4AI) in the deep integration of 6G and AI, signifying a shift from a communication network solely providing connectivity services to a new network paradigm that offers a diverse array of services, including connectivity, computing power, data, and algorithms. This new paradigm would span CN, access networks, and terminals, to provide comprehensive support of AI. 
} 

\subsection{NET4AI Architecture over 6G}\label{Section_NET4AI}

{
China Mobile \cite{ChinaMobile-6g} proposed a systematic 6G network architecture design that transcends connectivity in the industry, centered on the core concept of ``three layers and six planes''. The ``three layers'' refer to the physical resource layer, the network function layer, and the application\&service layer. The ``six planes'' comprise the communication plane, the data plane, the computing plane, the intelligence plane, the security plane, and the management\&orchestration plane. The first five planes are independent functional planes formed by the decoupling of the 6G network from end to end, providing corresponding capabilities and services. The management\&orchestration plane realizes the flexible combination of functions across layers and planes, realizes the management and scheduling of multiple resources, and forms a complete service chain. This architecture focuses on multi-dimensional capability elements within the network. In \cite{ChinaTelecom-6g}, China Telecom proposed a ubiquitous and ultra-converge 6G network architecture, built upon the core concept of ``three layers and three sectors''. The ``three layers'' consist of the infrastructure layer, the network function layer, and the network operating system layer, while the ``three sectors'' embody data integration, intelligent simplicity, and trustworthiness. This architecture leverages the intelligent simplicity sector to integrate the intelligent brain, connections, and facilities across the three layers, thereby enabling intelligent inclusiveness. China Unicom \cite{ChinaUnicom-6g} envisioned a future 6G network as intelligent, converged, green, and trustworthy. Its architecture is structured vertically into the land-based, air-based, and space-based communication layers and horizontally into three domains: perception resource domain, function control domain, and service application domain. These layers and domains are interconnected by two chains: intelligent native and secure\&trustworthy. 

Huawei \cite{Huawei-task} unveiled a task-centric AI architecture of 6G networks, asserting that 6G networks will introduce new resource dimensions and support the coordination of multi-dimensional resources. This design allows 6G networks to natively support AI and achieve a transformation from session-centric to task-centric operations. CICT \cite{CICT-6gvision, CICT-6gnetwork} put forward the concept of multi-network integration and ubiquitous intelligence, envisioning that 6G will realize global three-dimensional deep coverage, develop body area networks centered on human beings, form multi-layer coverage with both breadth and depth and learn the ubiquitous intelligence system with multi-network integration. OPPO \cite{OPPO-aicube} suggested that 6G would redefine human interaction and AI, by making AI a ubiquitous infrastructure. The 6G system architecture will emphasize the coordination across intelligence, performance, and flexibility, and integrate deeply AI capabilities into the user plane, control plane, and function plane. More authoritative and official, The IMT-2030 (6G) Promotion Group in China envisioned the 6G network architecture in \cite{IMT2030-6gnetwork}, as an open and innovative platform for information services, offering capabilities that transcend mere connectivity. These capabilities encompass computing power networks, trusted security, sensing, data services, and AI-based network intelligent autonomy. 

Globally, Ericsson \cite{Ericsson-6g} has focused on cognitive networks by leveraging AI to achieve data-driven security and highly automated network operations. Qualcomm advocated that AI and ML would fundamentally transform the design and deployment of wireless communication and networking systems. As the leader of the EU's 6G project Hexa-X, Nokia \cite{Nokia-6g} stated that AI would be the primary driving force behind the technological transformation of the NR interface for 6G. The company is creating unique AI use cases and scenarios and developing foundational AI technologies. LG Electronics is focusing on the research and development of 6G AI technology, anticipating that 6G systems will usher in the era of the ``Internet of Everything and Environment'' powered by AI. Japan's ``Beyond 5G Promotion Strategy'' project\cite{B5GPC-6g} aimed to introduce AI into network management, enabling network autonomy and driving network evolution towards intelligence. 

Academic research has also proposed various innovative ideas for shaping the future 6G network architecture. The authors of \cite{Zhang-6g} put forth a 6G network design philosophy of ``Human-Machine-Thing-Spirit'', emphasizing the inevitable trend of AI integration in 6G, leveraging cognitive enhancement and decision-making simulation to intelligently define network requirements, ensuring secure and reliable network transmission, and realizing intent-driven networks. In \cite{Cui-6gran}, a fusion of 6G RAN and AI was advocated, identifying four key characteristics: green, multi-dimensional, stereo, and full-scenario service. 6G would integrate communication, computation, control/caching, and AI to support full-application scenarios, and would be inherently secure\cite{Southeast-6g}. The authors of \cite{UESTC-6g} demonstrated a vision of integrating 6G and AI and studied implementation examples in wireless communications based on two classic AI algorithms: DL and RL. It was suggested in \cite{ZJU-6g} that AI would arguably become the cornerstone of 6G with ``Intelligent inclusion'' serving as an essential feature of 6G, enabling the network to provide intelligent services to all types of end users. The authors of \cite{THU-6g} promoted that AI would be used in wireless communications and change the top-level architecture of wireless networks. The authors of \cite{SJTU-6g} emphasized that the NET4AI paradigm should leverage the capabilities of 6G networks, such as precise wireless sensing, high data-rate transmission, and ubiquitous connectivity, for intelligence distillation at the network edge, hereby promoting edge AI and enabling ubiquitous and trustworthy AI applications in wireless networks. Among them, federated edge learning is expected to be an important learning architecture over 6G. 6G networks and AI will be highly integrated, and customized AI services will be provided on the wireless network side\cite{CUHK-6g}. 6G would enhance its progress in intelligence, especially in edge intelligence\cite{Oslo-6g}. Among them, integrating DT and edge networks is a promising attempt. The 6G Flagship research program was initiated in \cite{Oulu-6g}, advocating for realizing context-aware intelligent services and applications for human and non-human users through AI. }

The industry has reached a core understanding of the intelligent design of 6G network architectures, emphasizing ``distributed structure'', ``endogenous intelligence'', and ``integrated simplicity''. Future networks will evolve into advanced, integrated platforms that surpass mere communication connectivity services, offering multi-dimensional services and corresponding network capabilities beyond communication. They will cater to diverse businesses by offering end-to-end lifecycle management services throughout their entire process. 
Additionally, the network’s AI capability can be opened to the outside world, while external AI capability can be introduced into the network, enabling the crowdsourcing of AI capability. 

\begin{figure}[H]
    \centering
    \includegraphics[width=0.95\linewidth]{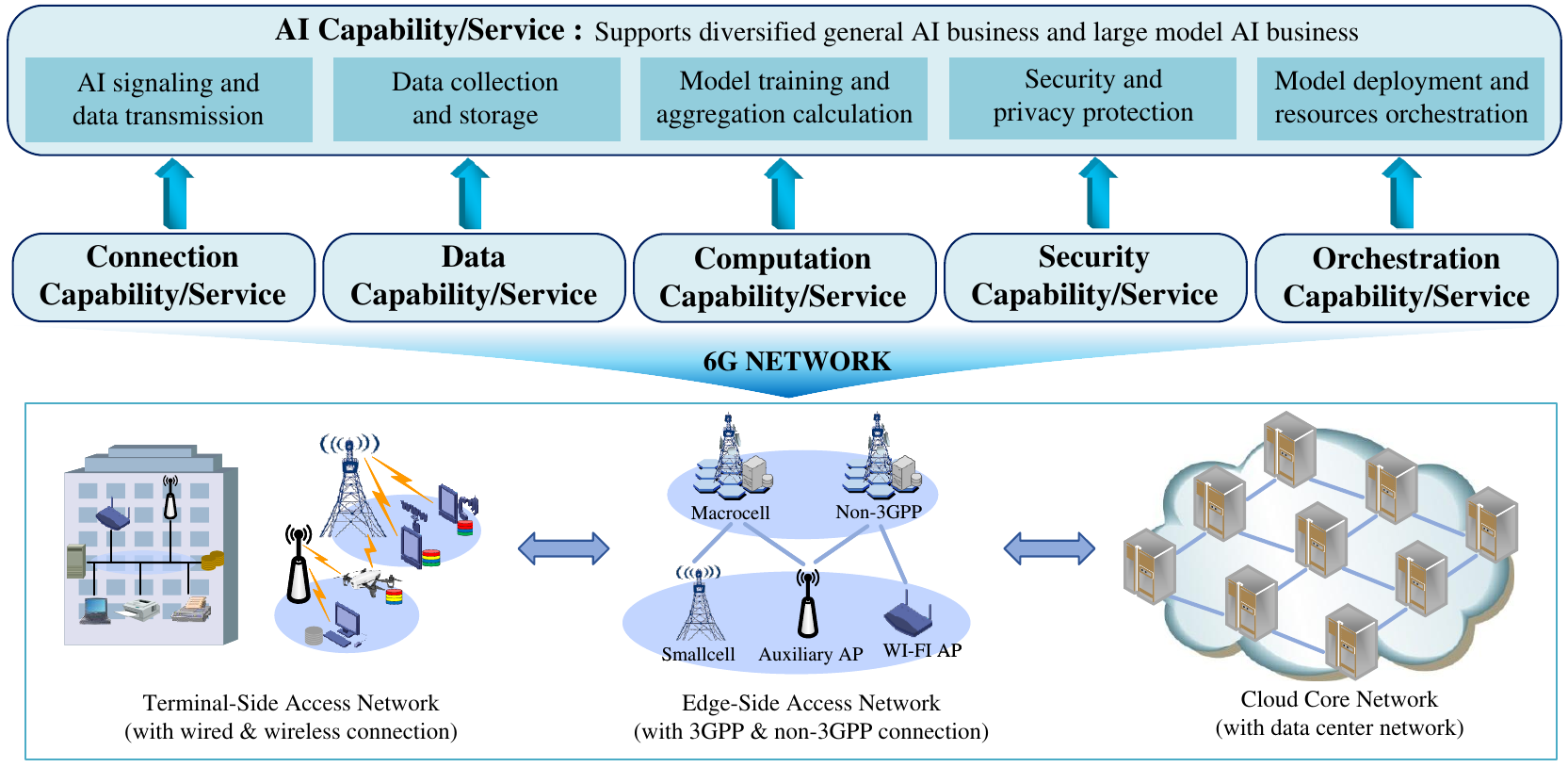}
    \caption{Architecture diagram of NET4AI. The 6G network physically comprises three entity networks: the cloud CN (including centralized data/computation centers), the edge-side access network (including 3GPP and non-3GPP connection methods), and the terminal-side access network (including wired and wireless connection methods), as depicted in the lower half of the figure. NET4AI represents the mapping and supporting role of communication connectivity, data, computing, security, and management orchestration within AI, among the six capabilities and services (i.e., communication connectivity, data, computation, security, AI, and management \& orchestration) abstracted from the 6G network's functional services, as indicated by the arrows in the upper half of the figure.}
    \label{NET4AI_Architecture}
\end{figure}

We elaborate on the NET4AI architecture, i.e., the support provided by the other five capabilities/services in the network (i.e., communication connectivity, data, computation, security, and management \& orchestration) for the AI capabilities/services, with an emphasis on their interaction with AI, as illustrated in Figure \ref{NET4AI_Architecture}. NET4AI will provide architectural support for the implementation of AI-native 6G, achieving a deep integration of the network and AI.  

\subsubsection{Connection for AI}
In 6G, the control plane and user plane will be further enhanced, aiming to deepen the service-oriented and open nature of network functions and provide support for emerging AI services. On the one hand, 6G needs to further enhance and optimize the service management and interface of the existing control plane functions to reduce the complexity of network functions and signaling interaction processes\cite{liu2021-6gnative}. Additionally, 6G will extend the conventional control plane by integrating space-terrestrial ubiquitous access control management, enabling new functionalities and service-oriented offerings for session management and policy control across communication, computing, data, and other essential elements. On the other hand, the user plane will be enhanced by augmenting or introducing new, more generalized, and flexible user plane forwarding protocols that can perform agile service forwarding and processing based on communication, computing, and other elements. Leveraging microservice technologies and next-generation hardware support, e.g., high-performance programmable chips, cloud-native technologies, and the P4 (programming protocol-independent packet processors) advanced language, user plane functions (UPF) will be restructured into service-oriented modules\cite{Huawei-task,liu2022-nativeai}. This flexible invocation and efficient collaboration among functional modules will break the constraints of conventional layered protocols, enhancing data forwarding efficiency and bandwidth utilization. 


Considering the control plane and user plane separation principle \cite{Cui-6gran}, the types of AI service-related data carried by communication connections can be naturally categorized into two major groups: AI signaling and AI data. Connection for AI embodies providing connectivity support for AI signaling and data. 

\textbullet \textbf{  \emph {AI Signaling.} }AI Signaling is used for transmitting AI-related control messages, e.g., AI service request/response messages; request/response messages related to computing power required for AI analysis, etc. It can also include auxiliary information for collaborative AI analysis among multiple entities.

\textbullet \textbf{  \emph {AI Data.} }AI data is used to transfer AI input data, generally referred to as input vectors for invoking an AI model or represented as inputs in an AI algorithm. It is also used for transmitting AI outputs, generally called the output of the invoked AI model/algorithm. It can be used for transmitting AI models, e.g., transmitting a neural network model usually means transmitting its parameter weights (which can be represented as vectors) and its network structure (which can be represented as scalars). It can be further used for transmitting AI training data. The data can often be usually huge and dedicated to the training phase of intelligent models.

For AI business scenarios and related connectivity performance requirements, 3GPP TS 22.261\cite{22.261} (based on the conclusions of TR 22.874\cite{22.874}) also has some KPIs requirement level analysis and results for 5G systems, mainly on the impact of conventional connectivity performance metrics, such as throughput, latency, reliability, etc.; in other words, the current 3GPP standard only analyzes the impact of the 5G system (mainly the CN portion) in terms of how it supports different AI services/applications from the functional and performance perspectives. From an overall end-to-end process analysis perspective, AI signaling and AI data involved in AI services/applications necessitate transmission via the air interface. However, in the current standards and architectures, no relevant modules guide the transmission process or standardize the transmission content. 

\subsubsection{Data for AI}
6G networks necessitate a novel data capability that differs from the conventional user plane. This capability must adhere to data regulation and supervision requirements, while enhancing data analysis and processing efficiency, addressing data management challenges, and supporting AI services. This data capability supports the realization of information collection, cleaning, and storage, and enables the complete lifecycle management of data service processes while ensuring data privacy and security. By leveraging a hybrid data processing and storage model that combines centralized and distributed approaches, the 6G network can flexibly address diverse scenario requirements: 1) For sensitive or high-privacy-level data, local processing and storage can be performed at the RAN/CN and terminal sides to ensure no data leakage; 2) Global and comprehensively analyzed data, on the other hand, is processed and stored centrally in the data center. Through tight cooperation between the central data center and edge nodes, 6G can efficiently manage massive amounts of data originating from various technological and business domains while safeguarding privacy. 

{Data capability provides a solid data foundation for 6G's endogenous intelligent AI services \cite{6gdata}. Several publicly available datasets have been widely utilized, such as DeepMIMO \cite{alkhateeb2019deepmimo} and RadioML \cite{oshea2016}. The DeepMIMO dataset, based on ray-tracing simulations of real wireless communication environments, provides MIMO channel data and is extensively applied in scenarios like channel modeling, beamforming, and user location in 6G networks. By contrast, the RadioML dataset encompasses radio signal data with multiple modulation schemes and is mainly used for signal modulation recognition, interference detection, and spectrum management. Moreover, the ``AI/ML in 5G Challenge" dataset \cite{AI/MLITU} released by ITU covers a variety of communication scenarios in 5G networks, offering crucial support for research on network optimization, traffic prediction, and resource allocation. These datasets not only lay a solid foundation for AI research in the communications field but also provide unified benchmarks for validating and comparing algorithm performances. }

Data for AI aims to efficiently support end-to-end data collection, transmission, storage, and sharing, addressing how to facilitate, expedite, and securely provide data to AI functions within or external to the 6G network. Depending on the potential functional scope, the data support provided by data for AI should encompass five components: 
\textit{1) Data Collection/Distribution: }This component provides a basic publish-subscribe mechanism for data producers and consumers, enhancing data collection/distribution efficiency and supplying samples for large-scale training required by AI models to achieve optimal results. 
\textit{2) Data Security and Privacy: }This component leverages security and privacy protection technologies to provide high-quality, trusted data services tailored to user and network needs, ensuring user and network privacy protection and data security, immutability, and traceability. 
\textit{3) Data Analysis: }This component utilizes AI models, algorithms, and knowledge to provide statistical information, predictive insights, network anomaly analysis, and optimization suggestions, enhancing the data consumption experience for internal and external network functions and related intelligent applications. 
\textit{4) Data Preprocessing: }This component applies generic tools for format conversion, noise reduction, feature extraction, and other preprocessing tasks to collected data, fulfilling AI model training requirements for input data and facilitating the management of multi-dimensional complex data. 
\textit{5) Data Storage: }This component stores and retrieves collected data, providing storage support for data services related to data security and privacy, data analysis, or data preprocessing, and further establishing a data foundation to support AI services. 

{Through standardized data service functions, a comprehensive data service process can be established within the 6G network. 
As researchers often need to construct dedicated datasets for specific application scenarios, Data for AI can provide valuable guidance. Take the construction of communication datasets as an example. Firstly, data sources can be selected from collected data in real-world communication networks or generated simulation data using high-fidelity simulation tools (such as NS-3 or OMNeT++). Real-world data can reflect actual scenarios but may involve privacy and security issues; whereas simulation data offer higher flexibility and can be customized to specific scenarios according to particular requirements. Secondly, data preprocessing and labeling are two essential steps. Communication data usually contain a large amount of noise, thus requiring preprocessing operations, such as cleaning and normalization, and accurate labeling based on research objectives (e.g., channel state prediction or interference detection). 
When using real-world user data, strict compliance with data privacy protection regulations is mandatory. Techniques such as anonymization or differential privacy should be adopted. Last but not least, researchers are encouraged to make datasets publicly available under the premise of protected privacy, to promote collaborative research and algorithm comparison within the field. It is recommended to use a unified storage format and standard to facilitate data sharing and reproducibility of research results. }

\subsubsection{Computation for AI}
{
The evolution of 6G will bring significant advancements in computation, including computational sensing, control, and execution. \textit{Computational sensing} will gather information about the ubiquitous computing resources distributed across the spectrum, from central clouds to terminal devices. \textit{Computational control} will cater to computing service demands by providing intelligent strategies for pervasive computing power control and resource scheduling. \textit{Computational execution}, guided by these strategies, will deliver services such as session management, traffic control, and policy execution.

Computational capabilities must synergize with communication connectivity to optimize the efficiency and energy consumption of emerging services like AI. This integration necessitates the development of a unified environment that ensures continuous operation of computing services across terminals, edge nodes, networks, and clouds. Such an environment will enable the dynamic selection of optimal computing nodes for computational tasks, considering factors like latency, bandwidth, computing power, and energy efficiency. This unified framework will allow computational tasks to transition among terminals, networks, edge nodes, and clouds\cite{9050836}. This computational power, along with the AI algorithms or functionalities, serves the network or devices to improve network operations but is also potentially exploited through unified interfaces to serve upper-layer applications. 
Computational power can be categorized into three types: dedicated computing power for network elements, distributed external computing power, and endogenous computing power within the network.} 

\textbf{(1) Network Element Dedicated Computing Power for AI. }This type of computing power typically consists of computing and storage units composed of specialized processors and programmable devices. It serves network elements in mobile communication networks (BS or CN). It is primarily used for customized AI application services (i.e., AI4NET) to enhance communication performance or optimize network operations, like CSI feedback, channel estimation, and beamforming. Due to resource restrictions, it cannot support AI services and applications requiring large-scale computations and training. It is challenging to support third-party applications.

\textbf{(2) Distributed External Computing Power for AI. }This type of computing power typically exists in the form of distributed edge computing/MEC, primarily utilizing general-purpose central processing units (CPU), high-performance graphics processing units (GPU), and programmable acceleration cards. External computing power shifts computations from centralized data centers to the edge of the access network, enabling network optimization and supporting industry applications with high computational demands and stringent latency requirements, such as video acceleration and augmented reality (AR)/VR scenarios. Due to its external nature, it requires management through unified network functions, introducing some latency that may impact extremely latency-sensitive AI services and applications. Moreover, this external overlay approach may not optimize resource utilization, hindering efficient construction, deployment, and usage of AI services.

\textbf{(3) Distributed Network-Native Computing Power for AI. }In the vision of future networks, each network element will possess control, forwarding, and computing capabilities, with numerous computing nodes deployed throughout the network. This computational power, known as endogenous computing power, promotes the development and deployment of endogenous AI, enabling large-scale intelligent distributed collaborative services. It compensates for the shortcomings of external computing power, promptly responding to changes in mobility and networks while fostering the emergence and development of future AI applications, such as immersive cloud XR, holographic communication, sensory interconnection, communication-sensing integration, and DT.

\subsubsection{Security for AI}
Some security vulnerabilities have remained after the design of 2G-5G systems, and conventional security measures have often failed to ensure safety. 6G networks necessitate establishing an inherently trustworthy and self-driving security system at the core of the system. Rapid advancements in cloud computing, big data, and AI technologies offer technical support to build a 6G network security system. In recent years, trusted endogenous security has emerged as a new security approach for 6G, characterized by four main features: collaboration, intelligent proactive defense, trustworthiness, and privacy protection. The concept of trusted endogenous security also applies to ensuring the security of AI. 

\textbf{(1) Ubiquitous Collaboration. }Collaboration has become a technological approach to enhancing communication capabilities. Intelligent ubiquitous collaboration will enable NET4AI to adapt to various complex and dynamic environments and scenarios while improving the robustness of the AI plane to identify potential malicious attacks or abnormal behaviors accurately. Future security defenses will shift from isolated to highly efficient collaboration, achieving individual collaborative defenses between entities or intelligent agents, inter-layer defenses between protocol layers or architecture layers, and inter-domain defenses between network space domains. 

\textbf{(2) Intelligent Active Defense. }The rapid development of 6G networks and AI technology presents opportunities for intelligent security defenses. The intelligent, proactive defense will enable NET4AI to continuously learn and adapt while automatically adjusting defense strategies based on changes in the network environment and AI security threats to counteract attacks promptly. This transition represents a shift from passive protection to proactive sensing and defense \cite{security5}. 

\textbf{(3) Trustworthiness. }6G will further evolve towards trustworthiness capabilities, achieving a comprehensive endogenous security system encompassing security (i.e., confidentiality, integrity, availability), reliability, resilience, and safety \cite{security6}. This can provide a trusted ecosystem for NET4AI. Furthermore, flexible and reliable access control \cite{cui2021edge} and effective identity authentication\cite{li2022lightweight} can, to some extent, prevent unauthorized access from impacting the endogenous trusted environment of NET4AI. 

\textbf{(4) Privacy Protection. }The deep integration of AI, big data, and 6G networks will exacerbate the challenges of data privacy protection. NET4AI must handle massive business and personal data while preventing privacy leakage. Emerging distributed ML, while achieving distributed ubiquitous intelligence and partially avoiding privacy leakage caused by AI model training, still has security risks \cite{9446488}. Therefore, NET4AI will integrate distributed ML with existing privacy protection methods at multiple points to enhance data privacy protection and build an efficient and secure data ecosystem.

\subsubsection{Orchestration for AI}
6G is considered to possess an extremely high level of network autonomy, which means that the management \& orchestration capability of the network can conduct unified and dynamic orchestration and scheduling for various functional components of service requirements, thus enabling the self-optimization, self-evolution, and self-healing of the network. The 6G system will introduce external applications and service demands while exposing internal diversified resources and functions. This requires flexible cross-functional flow and scheduling of multi-dimensional resources, such as computing, communication, and storage within the network, offering customized and personalized application services based on virtualization and microservice technologies. For instance, AI capabilities and analyzed data within the network can be opened to third parties to provide services and necessary support, a crowdsourcing behavior that necessitates the support of management and orchestration functions. Furthermore, through joint analysis and prediction of computing resources and network traffic, containerized gNBs can automatically scale up or down based on predicted network traffic loads, enabling dynamic resource allocation and network energy savings. Therefore, management and orchestration capabilities are closely related to all other capabilities, orchestrating and managing resources and functions efficiently and flexibly across all functional layers of the 6G network, achieving joint optimization. 

Orchestration for AI of the intelligence plane signifies efficient support for deploying, operating, and optimizing AI services within the network through orchestration. It automatically configures and orchestrates network resources based on the specific requirements of AI services, ensuring that AI applications obtain the necessary computational power, storage, network bandwidth, and other resources. Additionally, this orchestration capability enables real-time monitoring of AI service performance, dynamically adjusting resource allocation to accommodate fluctuations in service loads, thereby guaranteeing the stability and efficiency of AI services \cite{6gorchestration}. This seamless integration and support facilitate smoother AI service operation within the network, providing robust network guarantees for various intelligent applications. Specifically, this manifests in the following three aspects:

\textbf{(1) Automated Deployment and Resource Allocation. }The orchestration capability swiftly responds to AI services' network resource demands through automation. Upon AI service deployment, the orchestrator intelligently analyzes the current network resource utilization, automatically configuring network devices, establishing links, and deploying services. This automated process enhances deployment efficiency, mitigates human errors, and ensures rapid AI service launch and stable operation.

\textbf{(2) Dynamic Optimization and Elastic Scaling. }As AI services operate, their resource requirements may fluctuate. The orchestration capability features dynamic optimization and continuous monitoring of AI service performance, including CPU usage, memory consumption, network bandwidth, and other key metrics. Upon detecting resource shortages or surpluses, the orchestration system automatically adjusts resource allocation, such as adding compute nodes, expanding storage capacity, or optimizing network bandwidth, to meet AI services' dynamic demands. Furthermore, the orchestration system supports elastic scaling, automatically resizing resources based on AI service loads, ensuring service stability and efficiency.

\textbf{(3) Fault Self-Healing and Intelligent O}\&\textbf{M. }By integrating monitoring, alerting, and troubleshooting functions, the orchestrator promptly identifies and locates faults within the network, automatically triggering recovery mechanisms. This fault self-healing capability minimizes service disruptions for AI services, enhancing service availability and reliability. Additionally, the orchestration facilitates intelligent network O\&M, empowering network administrators to manage network resources and AI services more efficiently, reducing O\&M costs, and improving efficiency.

\subsection{Key enabling technologies}
\begin{figure}[H]
    \centerline{ \includegraphics[width=0.9\linewidth]{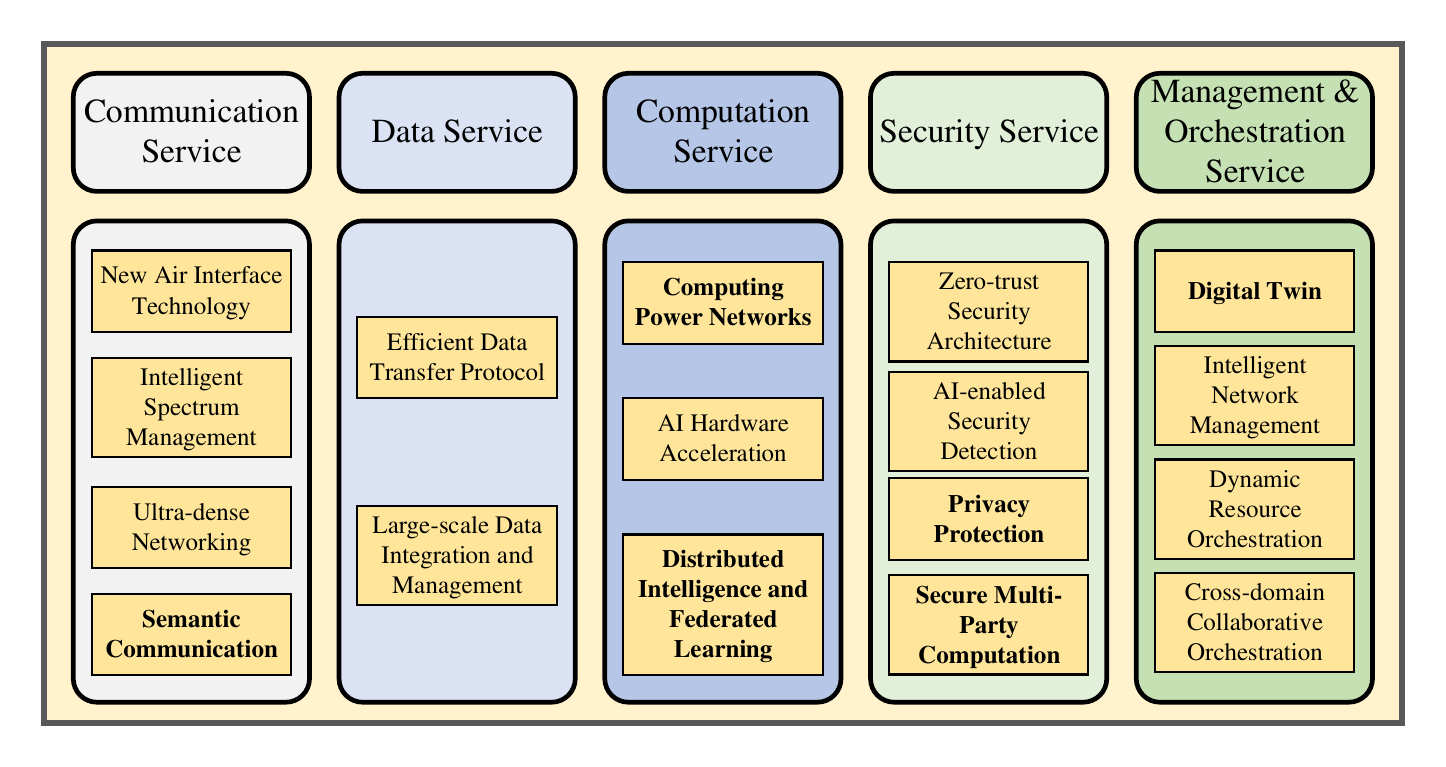}}
     \caption{NET4AI Enabling Technology}
    \label{3.2-v4}	
 \end{figure}

{This section describes the key enabling technologies for NET4AI. Each of the five capability/service planes in the future 6G network architecture, as presented in Section \ref{Section_NET4AI} shall contain its own enabling technologies, as categorized in Figure \ref{3.2-v4}. In the following, we focus on those closely related to the needs of AI, introducing their research background, significance, and progress. }
 
\subsubsection{Distributed Intelligence and Federated Learning}
{The NET4AI architecture must address the dual demands of low latency and high reliability for communication and computing, as well as growing concerns about privacy.} Distributed intelligence and FL are the key technologies that can solve these challenges. Conventional centralized architectures in wireless networks can no longer meet low-latency, high-reliability communication and computation demands. Additionally, conventional centralized architectures are inefficient in supporting the ubiquitous intelligence required by future wireless networks. This trend towards deploying intelligent decentralized elements has become apparent, with intelligence gradually shifting from centralized to distributed systems. Therefore, introducing distributed intelligent computing architectures is necessary to fully utilize the multi-dimensional data and computing resources held by user terminals and nodes, enabling intelligent connectivity.

Wireless distributed intelligence refers to organizing AI tasks in a distributed manner within wireless networks and incorporating collaborative AI and ML without uploading all raw data to a central cloud. This approach alleviates the network bandwidth pressure and reduces the central cloud's computational burden. Moreover, as the computing power of smartphones and IoT devices increases and users become more concerned about data privacy, mobile devices can handle more AI computations and training. This proximity of AI to users and data allows for faster iteration and upgrade cycles, efficiently supporting the pervasive intrinsic intelligence of 6G networks.

Distributed intelligence is a cornerstone technology for realizing the intrinsic intelligence envisioned for 6G networks. By adopting a decentralized architecture, it disperses network functions and resources across multiple nodes or devices, thereby eliminating reliance on a single central node to manage the entire network. Each node in a distributed network operates with a degree of autonomy, enabling independent decision-making and task execution without the need for constant communication or coordination with other nodes. This autonomy enhances network flexibility, making it more adaptable to dynamic environmental conditions and evolving demand patterns.

\begin{figure}[h]
    \centering
    \includegraphics[width=1\linewidth]{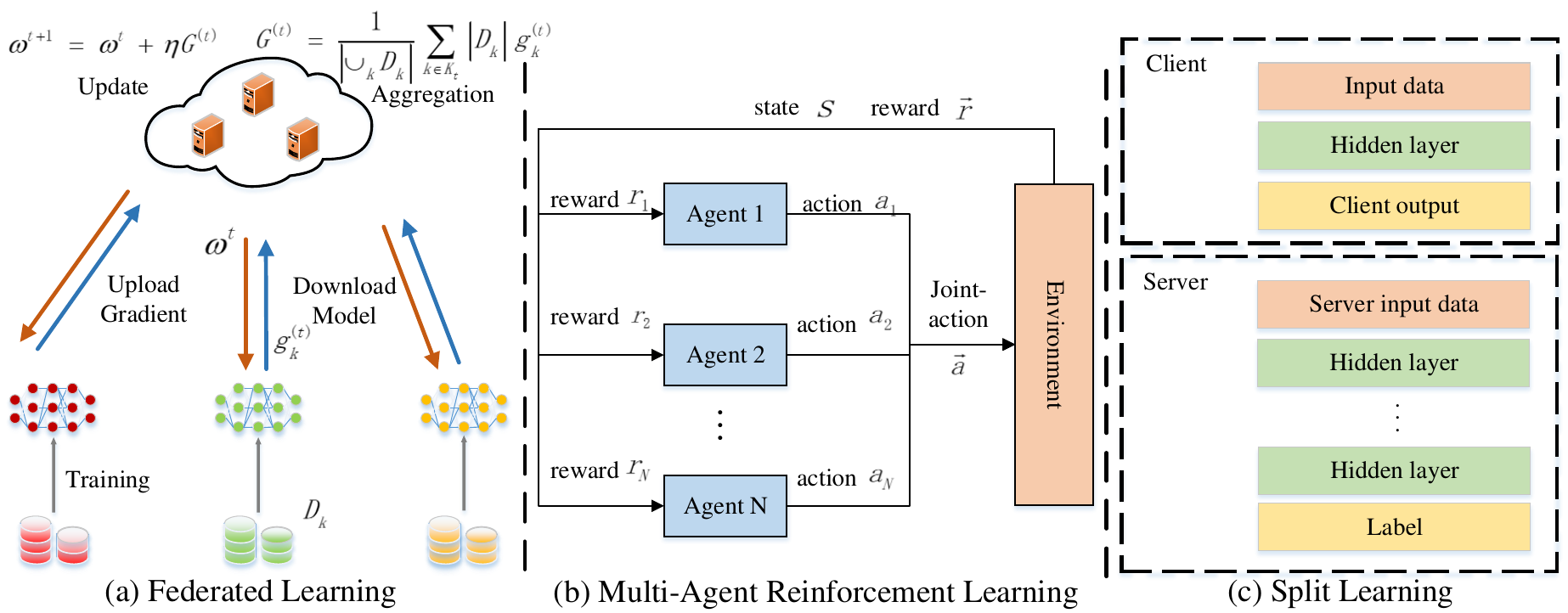}
    \caption{An illustration of examples of FL, MARL, and SL in distributed intelligence.}
    \label{3.2.1-v0}
\end{figure}

In distributed networks, data is stored across multiple nodes, and computational tasks are processed collaboratively among them. This decentralized approach to data storage and computation improves network scalability, accelerates the training and inference of LAMs, and enhances overall system efficiency. Moreover, distributed intelligence facilitates online learning and continuous optimization by enabling real-time data collection, processing, and model updates, which improves the performance and accuracy of AI systems.
Distributed AI technology addresses the limitations of centralized AI systems, such as communication bottlenecks and data privacy concerns. It reduces communication overhead and ensures data security by keeping sensitive information closer to its source. Distributed AI is poised to create a new intelligent ecosystem for 6G, fostering a more scalable, efficient, and privacy-preserving network paradigm.
As shown in Figure \ref{3.2.1-v0}, the wireless distributed intelligence architectures include:

\textbf{(1) Federated Learning.}
{
FL is a typical paradigm of distributed intelligence, which enhances privacy by training AI models on local devices and only sharing model parameters instead of raw data~\cite{10073536}. During training, each client node trains the model locally and uploads the model weights regularly. A central node aggregates these weights and feeds the aggregated weights back to the clients for the next round of training or local inference.
A typical example of FL in wireless networks is illustrated in Figure \ref{3.2.1-v0}a. The process begins with downloading the global model \( \omega^t \) for the current round \( t \) to local devices. Each device \( k \) trains the model on its private dataset \( D_k \), generating a local gradient update \( \mathbf{g}^{(t)}_k \), which is then uploaded to the parameter server. The server applies the federated averaging (FedAvg) algorithm to aggregate these updates into a global gradient \( G^{(t)} \), subsequently updating the global model to \( \omega^{t+1} \) for the next round. This process continues until model convergence.
FL can be divided into horizontal FL and vertical FL. 
Horizontal FL emphasizes inter-source cooperation and data sharing, whereas vertical FL facilitates collaboration and information exchange across distinct levels within the same data source. }

\textbullet \textbf{\textit{Horizontal FL.}}
Also known as sample-based FL, horizontal can be applied in scenarios where the datasets of the various participants in FL have the same feature space but different sample spaces. Through horizontal FL, different participants can jointly train a model based on local data sets while protecting privacy and improving the accuracy and generalization of the model. Horizontal FL enables real-time learning on distributed edge devices, allowing for timely model parameter updates to constantly changing environments and data. It supports collaboration across devices with varying computational capabilities and data distributions, catering to the diverse or personalized needs of 6G network terminals~\cite{10838599}. It also supports over-the-air commputations to achieve excellent scalability \cite{XIAO2024,10145450}, tolerates to imperfect wireless transmissions~\cite{10649032}, and can be potentially implemented in a fully decentralized fashion~\cite{10542235}.

{The authors of \cite{li2019convergence} considered limited resources in wireless networks, and analyzed the convergence of FedAvg in horizontal FL under full and partial device participation, providing important theoretical support for the application of horizontal FL in wireless networks. In \cite{9796935,10229070}, the authors represented the partial device participation as a probability distribution and eliminated the device selection bias by modifying the aggregation rule in FedAvg to ensure the unbiasedness of gradient updates. This allows for a more stringent convergence bound and is committed to the efficient implementation of horizontal FL. The authors of \cite{9745426,10750008} provided a variety of solutions to guide the efficient application of FL in wireless networks, such as layering, aggregation frequency optimization, device selection, and communication compression, in support of low-carbon green economics. In \cite{fallah2020personalized}, the authors focused on the shortcomings of horizontal FL in dealing with the heterogeneity of device data, introduced a personalized federated learning (PFL) method, and provided a theoretical basis for the application of FL in multiple fields to realize personalized models. The authors of \cite{tan2022towards} categorized PFL techniques based on key challenges and strategies, highlighting key ideas and future research directions in PFL architecture, realistic benchmarks, and trustworthy methods. In \cite{passerat2020blockchain}, the authors considered the application of horizontal FL in the medical field to solve the problem of collaborative research and patient health testing among multiple hospitals, and proposed a blockchain-based orchestration framework to enhance medical data privacy. The authors of \cite{polap2021agent} focused on IoT healthcare and proposed an intelligent medical system agent architecture that integrated horizontal FL and blockchain technology to achieve efficient and secure processing of medical data.} The authors of \cite{kevin2021federated} proposed a new FL framework for cross-domain prediction for intelligent manufacturing, designed to quickly adapt to new products, processes, and applications with limited training data.
{Horizontal FL has important application prospects in providing efficient and personalized AI services in wireless networks, coordinating research with multiple hospitals in healthcare, and realizing cross-application in intelligent manufacturing. However, it also has defects such as slow convergence speed and susceptibility to data heterogeneity and poisoning attacks~\cite{10423783,10419367}.}

\textbullet \textbf{\textit{Vertical FL.}}
Vertical FL refers to a setup where datasets have different features but the same samples/users. It is suitable for scenarios where there is a high overlap among participants but low or no overlap of their features. During the training process, participant data must first be aligned through sampling to ensure overlapping datasets across users. Subsequently, the model is trained locally, and the resulting parameters are uploaded to a central server \cite{liu2024vertical}. 
Vertical FL leverages correlations between diverse data sources to achieve data augmentation, enabling complementary and enriched datasets that enhance the generalization and robustness of AI models. By sharing model parameters, vertical FL facilitates cross-level learning and knowledge transfer, allowing data sources to exchange learning experiences and improve overall model effectiveness.
{ 
To tackle the problem of user diversity, a novel approach for a vertical collaborative recommendation system based on clustering algorithms and latent factor models is put forward \cite{zhang2021vertical}. This method effectively enhances the accuracy of the recommendation results.
In \cite{tang2023ihvfl}, the authors developed the idea of intent-hidden vertical FL for multi-party collaborative training of disease prediction models in medical diagnosis, and built a security protocol using homomorphic encryption for data privacy protection. The authors of \cite{yuan2022fedstn} proposed a vertical federated DL based on the spatial-temporal long- and short-term networks for intelligent city traffic in MEC environments. 
Leveraging diverse sample features (e.g., different regions) within the same sample space (e.g., a specific traffic scene and time range) enables the construction of a federated model for traffic flow prediction. 
Vertical FL can uncover relationships between different features of data samples, making it highly valuable for applications such as recommendation systems and traffic prediction. 
Research on vertical FL remains in its infancy, with limited theoretical advancements to date.
}

\textbf{(2) Multi-Agent Reinforcement Learning (MARL).}
{ In a shared environment, multiple learning agents pursue their own rewards. 6G networks, supporting diverse AI applications like smart cities and intelligent transportation, can benefit from MARL. With its adaptive decision-making, MARL enables agents to autonomously optimize and adjust, enhancing user experience across various applications.}
{
The authors of \cite{chu2019multi} introduced a decentralized MARL algorithm based on advantage actor-critic to address large-scale traffic signal control, effectively tackling congestion in intelligent transportation systems. In \cite{wu2020multi}, a multi-agent recurrent DDPG algorithm was proposed to manage traffic light phases in real-time using Internet of Vehicles data, mitigating congestion in complex and partially observable road environments. In \cite{palanisamy2020multi}, a partially observable Markov game was used to model connected autonomous driving, resulting in a multi-agent learning platform for interconnected autonomous vehicles. The authors of \cite{10754633} focused on collaborative computing in low earth orbit (LEO) satellites, developing a real-time offloading decision model optimized through MARL with a penalty function. Additionally, an MARL-based online production scheduling method was proposed in \cite{zhou2021multi} for smart factories, while multi-agent DRL was applied in \cite{cao2020multiagent} for multi-channel access and task offloading in MEC-enabled Industry~4.0.}

\textbf{(3) Split Learning (SL).}
{ Due to the limitations of resource-constrained devices that cannot support complex DL models or FL collaboration \cite{duan2022combined}, SL has emerged as a viable solution~\cite{10646623}. SL divides ML models, distributing different parts between the client and server, ensuring that original data remains secure \cite{10529950}. While high-computing tasks are handled by powerful central nodes, lighter data processing layers are retained on the terminal where the data are stored. As an alternative or complement to FL, SL reduces the processing burden on resource-limited devices and holds the potential for advancing distributed intelligence in 6G.}
{
The authors of \cite{vepakomma2018split,poirot2019split} introduced SL in healthcare applications, enabling multiple entities to collaboratively train DL models without sharing sensitive raw data, achieving promising performance results on medical datasets. In \cite{li2024split}, SL was explored for multi-institutional collaborative training, demonstrating improved client efficiency and more reliable privacy protection. To address issues such as overfitting and slow convergence in the original sequential training method, the authors of \cite{9016486} proposed a parallel SL method. The method mitigated overfitting through adaptive training batch adjustments and node layer synchronization, improving training efficiency while maintaining privacy. In \cite{10040976}, SL was applied to wireless networks to leverage terminal privacy data in IoT. A cluster-based parallel training method was proposed, optimizing joint cutting layer selection, device clustering, and spectrum allocation to reduce training delays effectively. However, SL remains vulnerable to data heterogeneity, complex training methods, overfitting, slow convergence, and potential failure to converge.}

\textbf{(4) Summary and lessons learned.}
{The above-mentioned technologies can all contribute to distributed intelligence. 
FL is suitable for AI services in scenarios with data privacy requirements, and applicable to various emerging intelligent demand applications of 6G owing to its compatibility with other intelligent learning paradigms. 
MARL is in line with the needs of online learning in complex environments and completes AI tasks in an observable and interactive environment. 
SL is suitable for collaborative learning of DL models in resource-constrained devices, and can adapt to different levels of resource constraints through multi-level model splitting.
{
Distributed intelligence may expand into additional research directions. These include: 1) the adoption of error correction and retransmission mechanisms, such as advanced error correction codes and adaptive retransmission strategies, to ensure reliability in unreliable communication environments; 2) the implementation of secure protocols and privacy protection measures, including data encryption, identity authentication, and access control, to safeguard user data from unauthorized access and tampering; 3) the efficiency of data/model parameter exchange to enhance the energy efficiency of learning algorithms, and 4) fairness through incentive mechanisms for participating devices \cite{10637271}.}
}

\subsubsection{Digital Twins for 6G Network Native AI}
\begin{figure}[htbp]
    \centerline{ \includegraphics[width=0.9\textwidth]{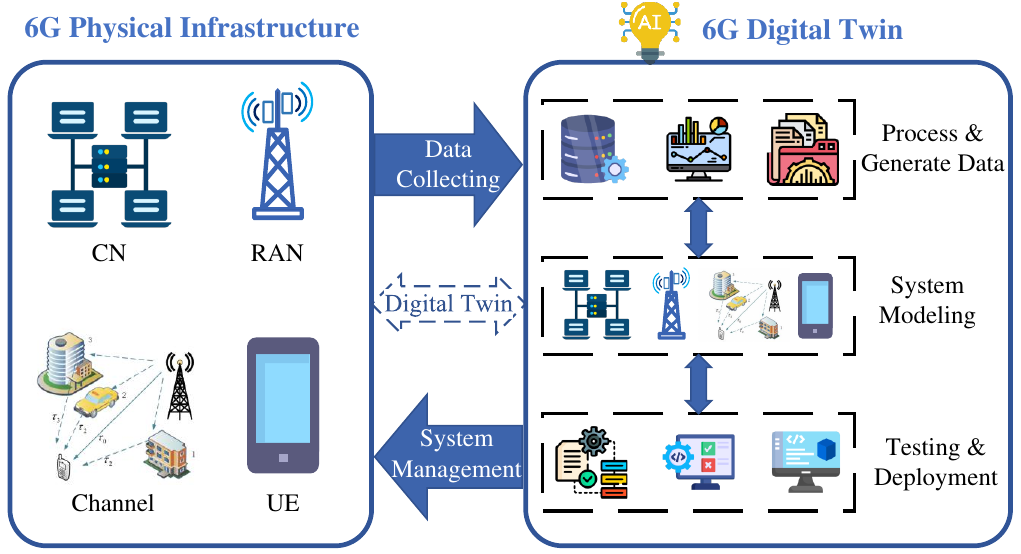}}
     \caption{Digital Twin for AI}
    \label{Digital Twin for AI}	
 \end{figure}

Conventional communication systems relying on historical data and static models struggle to adapt swiftly to dynamic network changes, making real-time adjustments difficult. These limitations hinder the adaptability, reliability, and flexibility of the systems. DT technology has been introduced into the communication domain to address these challenges, offering a flexible, efficient, and precise approach to resource optimization. Adopting DT not only enhances system agility but also introduces innovative models for network operation, significantly reducing research and development costs and risks \cite{nguyen2021digital}. Figure \ref{Digital Twin for AI} depicts the relationship between 6G physical infrastructure and DT for~AI.

The concept of the DT was first proposed in \cite{grieves2005product}. By creating a virtual replica of a physical object, DTs establish a bidirectional connection between the physical and digital worlds, enabling real-time reflection of the physical entity's operational state and the simulation of its future behavior~\cite{cao2024channeltwinningenablernextgeneration}. 
In 6G communication systems, DT technology facilitates the construction of comprehensive network digital replicas to monitor network devices in real time, predict network states, and optimize resource allocation~\cite{tao2018digital}. 

The implementation of DT relies on the synergy of several key technologies, including high-precision data collection, real-time modeling and simulation, bidirectional data interaction, and intelligent analysis and optimization. In communication systems, distributed sensors continuously gather diverse data about network device statuses, traffic dynamics, environmental parameters, and user behavior patterns. These data are transmitted to the DT platform via high-speed, low-latency communication links. 
The DT platform employs physical models and data-driven ML algorithms to construct high-fidelity virtual replicas that comprehensively simulate physical entities' operational state and dynamic changes~\cite{10234427}. 
With bidirectional data interaction, physical systems and virtual replicas achieve real-time data synchronization. 
When a network traffic surge is anticipated, the DT can proactively adjust load distribution strategies, optimize spectrum and power resources, and alleviate stress on network devices.
The integration of edge computing further enhances the efficiency of this technology by distributing modeling, simulation, and optimization tasks to edge nodes closer to physical entities. 
DT enables intelligent management and dynamic optimization of complex networks in 6G communication systems, providing robust support for efficient resource utilization, network stability, and intelligent upgrades \cite{barricelli2019survey}.

{The 6G network aims to achieve ultra-high capacity, ultra-short-range communication, beyond-best-effort service, high-precision communication, and the integration of multiple communication types \cite{li2022novel}.} 
However, this vision introduces security, spectrum efficiency, intelligence, energy efficiency, and affordability challenges. The emergence of DT technology offers promising opportunities to address these challenges.
DT provides a virtualized counterpart for 6G networks, enabling comprehensive network traffic monitoring and analysis. 
Leveraging feedback from the virtualized network, 6G systems can enhance their security by preparing for potential threats in advance. Additionally, DT enables the automation of demand identification and service provisioning by analyzing communication data to discern usage patterns and rules. Predictive insights into communication demands allow 6G networks to reserve resources, such as spectrum, in anticipation of future needs.
Moreover, the integration of DT technology empowers 6G to support innovative services, including AR/VR, and autonomous driving. By addressing key challenges in security, spectrum efficiency, intelligence, energy efficiency, and customization, DT technology redefines and accelerates the development of 6G networks \cite{qin2024machine}.

Significant research has integrated DT technology with 6G networks to enhance network performance. For instance, DT has been proposed for managing metasurface reflectors in 6G terahertz communications \cite{wu2022digital}. By modeling, predicting, and controlling the propagation characteristics of indoor signals, DT maximizes the system’s terahertz signal-to-noise ratio (SNR). Additionally, Lu et al. \cite{cui2023human} introduced the concept of DT wireless networks by integrating DT into wireless networks. This approach shifts real-time data processing and computation to the edge plane and leverages DT to mitigate the unreliability of long-distance communications between end users and edge servers in 6G networks. 

One of the critical capabilities of 6G networks will be supporting massive-scale terminal devices that generate extensive data traffic. 
For example, the authors of \cite{Oslo-6g} designed a DT model for edge computing in 6G wireless networks. By incorporating mobile users' dynamic network state and DT, the study addressed edge association problems using DRL and transfer learning, successfully reducing average system latency and improving resource utilization. Similarly, the authors of \cite{liu2021digital} proposed a DT-assisted task offloading method, creating DTs for all device states to achieve lower latency and power consumption in the network. Furthermore, the authors of \cite{huang2023collective} investigated dynamic resource allocation mechanisms for DT-based services in the 6G IoT. The study improved resource allocation efficiency by establishing service function chains with DTs and employing a collective RL method.

In addition, the DT channel (DTC) has been proposed as a digital virtual mapping of a wireless channel that reflects the entire process of channel fading states and variations in the physical world \cite{wang2023towards}. DTC can be applied in various typical 6G usage scenarios, offering significant performance enhancements in data rate, latency, spectral efficiency, energy efficiency, and link reliability \cite{wang2024digital}. 
A novel DTC implementation platform was proposed, including data acquisition, environment sensing, and reconstruction module to build the relationship between environment and channel for channel prediction, communication decision, and interaction \cite{nie2022predictive,miao2023demo}. A cluster-nuclei based channel model was proposed to analyze the mapping relationship between clusters and scatterers in the propagation environment, enabling channel reconstruction across various scenarios \cite{zhang2016interdisciplinary}. In addition, a radio environment knowledge (REK) pool was proposed to serve as a specific enabler for DTC. 
An electromagnetic wave property-inspired REK construction method was proposed, bridging the gap in the interpretable mathematical representation of the relationship between environmental information and the channel \cite{wang2024electromagnetic}.

The ITU documentation defines the requirements and architecture of DT networks DTNs\cite{ITU-T-DT}, emphasizing their core role as virtual representations of physical networks. 
DTNs enable physical network analysis, diagnosis, simulation, and control. The ITU proposal introduces a ``three-layer, three-domain, double closed-loop" architecture design, highlighting applications in complex network operations, optimization, innovation, and security strategy drills. 
For instance, DT optimizes computation offloading in industrial IoT in uRLLC links. Integrating DT with blockchain and FL technologies in wireless communication strengthens system security and privacy protection~\cite{10.1145/3659099}.  
In industrial IoT environments, the authors of \cite{van2022urllc} minimized computation offloading delays in uRLLC links by leveraging DT-enabled wireless edge networks. 
To enable wireless communication systems based on proactive online learning, a DT design framework was introduced in \cite{khan2022digital}. This framework considers aspects such as twin object access, security and privacy, and air interface design. Additionally, in the work of \cite{lu2020communication}, a blockchain-empowered FL framework was proposed to enhance communication security and data privacy within DT-enabled edge networks. The study incorporates asynchronous aggregation mechanisms and DT-driven RL to optimize user scheduling and spectrum resource allocation.

An innovative vehicular edge computing network combining DT and multi-agent learning was proposed in \cite{zhang2021adaptive}, which uncovers potential matches for edge services among large-scale vehicles, reducing the complexity of service management. The authors of \cite{hu2021digital} presented a DT-assisted real-time traffic data prediction method capable of delivering accurate short-term forecasts for traffic flow and speed. A DT-based load-balancing prediction model for autonomous vehicles was introduced in \cite{chen2022artificial}, which accurately predicts road network conditions while ensuring high data transmission security. In \cite{liu2023cooperative}, the authors explored the use of DT in vehicular edge computing, developing models that reflect the real-time state of the vehicular environment. The study introduced two metrics, Quantum DT and Cognitive DT, to evaluate the quality and cost of DT models for edge nodes.

\subsubsection{Computing Power Network and AI}
The NET4AI architecture requires powerful computing service capabilities because wireless networks show great interest in computing tasks related to high-performance computing, such as ultra-large-scale data processing and DL. The computing power network is a key enabling technology for computing services\cite{IMT2030-6gkeytechnology}. It achieves ubiquitous computing interconnection through mutual sense and collaboration between the network and computing resources\cite{yukun2024computing}, harmonizes cloud, edge, and terminal resources, and serves all computing tasks in the network, especially AI-related computing services. 

 \begin{figure}[H]
    \centerline{ \includegraphics[width=0.9\linewidth]{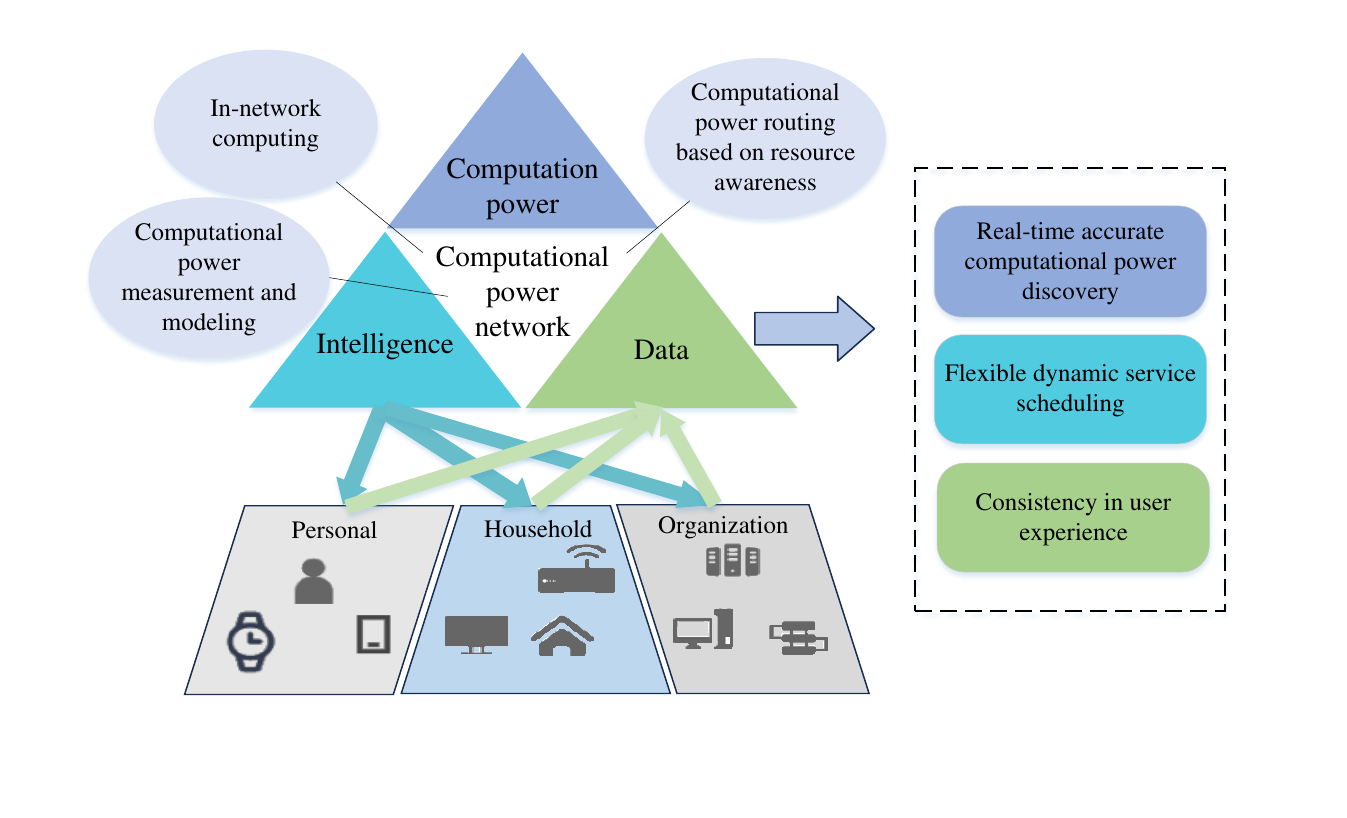}}
     \caption{Computing power network and AI}
    \label{Computing Power Network and AI}	
 \end{figure}

The computing power network is a new type of information infrastructure. As shown in Figure \ref{Computing Power Network and AI}, it supports computing power discovery, computing power management, task offloading, and network computing\cite{qingmin2022design,yukun2024computing}. 
It connects geographically distributed computing power center nodes through new network technologies\cite{IMT2030-6gkeytechnology}, 
and computing power, data, and application resources are gathered and shared\cite{yao2019trend}. 

The technical principles of the computing power network have three aspects: 1) Real-time and accurate computing power discovery; 2) Flexible and dynamic service scheduling; and 3) Consistency of user experience. 
Currently, the computing power network has gained wide recognition in both academia and industry and is classified into the following three categories. 

\textbf{(1) Computational power measurement and modeling. }This is a foundational technology for providing computational power services\cite{tang2021computing,yukun2024computing}. In the future, computational power providers in computational power networks will not be confined to a specific data center or computing cluster. They will include ubiquitous computational power from cloud, edge, and end devices. Efficient sharing of this ubiquitous computational power through network connections requires accurate sensing of the computational capacity of these heterogeneous chips, the business types suitable for different chips, and their locations in the network, as well as effective management and supervision~\cite{10146517}.

\textbf{(2) Computational power routing based on resource awareness. }In computational power networks, after measuring and modeling computational resources, the information is encoded and loaded into network control layer packets for sharing\cite{wang2020net,IMT2030-6gkeytechnology}. The network control layer makes network decisions based on shared computational resource information, guiding business routing to different computational resource pools or through collaboration between computational resource pools for business processing, thus enabling network awareness of computational resources to guide global routing.

\textbf{(3) In-network computing (INC). }Leveraging the deployment of programmable network technologies, INC processes packets within the network\cite{tokusashi2019case}. Sharing INC power using open and programmable heterogeneous resources accelerates data processing close to the source without altering the original business operating mode, reduces application response delays, and simplifies application deployment processes. 
Computing power networks support users to dynamically adjust resource scale according to demand to adapt to AI applications of different scales and complexities\cite{wang2020net}. 
In addition, the computing power network can also be combined with other novel technologies. For example, by adopting blockchain technology\cite{ren2022ai}, the computing power network can provide AI with a secure and privacy-protected computing environment, ensuring the transparency and traceability of the computing process~\cite{10194241}. The introduction of technologies, such as smart contracts, can further protect user privacy and data security, allowing users to safely use the computing power provided by the computing power network to process sensitive data and tasks and improving overall security and credibility~\cite{10130487}. 

In \cite{liu2022computing}, a Kubernete-based prototype testing platform for the computing power network using a microservices architecture was implemented, achieving key enabling technologies for the computing power network, including computational modeling, sense, announcement, and offloading. The authors of \cite{mingxuan2020research} described a lightweight multi-cluster hierarchical edge resource scheduling scheme based on a cloud-native resource scheduling mechanism, successfully managing and deploying many heterogeneous edge devices within a unified framework using a lightweight cloud-native platform. An integrated ICT network architecture of ``connectivity + computing + intelligence'' was proposed in \cite{yao2019trend}, which can sense computational resources and perform related management and control. Numerical results in \cite{chen2024two} indicate that on-demand compute resource scheduling scheme significantly improves the efficiency and stability of task offloading and compute resource scheduling in edge computing networks. Detailed power analysis of several INC use cases in \cite{tokusashi2019case} showed that INC becomes more energy-efficient at very low processing loads, with increased processing loads having little impact on INC power consumption. 
In \cite{sodhro2020ai}, the fog computing network offers powerful computing power support for AI model training, reasoning, and other tasks. This helps shorten model training cycles, improve training efficiency, and enable AI models to iterate and optimize faster. The study presented in \cite{deng2022uav} takes DNNs as a typical AI application, and formulates optimization problems under the constraints of energy consumption, delay, computing, and communication resources to efficiently utilize computing resources and avoid resource waste.

\subsubsection{Secure Multi-Party Computation}
\begin{figure}[h!]
    \centerline{ \includegraphics[width=0.7\textwidth]{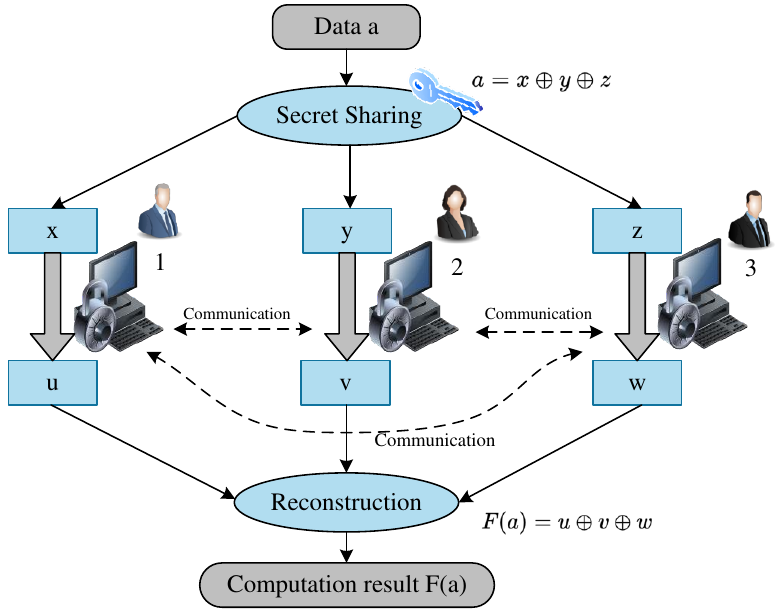}}
     \caption{Secure Multi-Party Computation \cite{SMPC}}
    \label{Secure Multi-Party Computation}	
 \end{figure}

In highly complex, heterogeneous, and dynamically evolving network environments, it is imperative to ensure data security while achieving efficient collaboration.
As network technology continues to evolve, so do network attack methods, becoming increasingly sophisticated. Conventional security mechanisms are often inadequate to counter these emerging threats. As shown in Figure \ref{Secure Multi-Party Computation}, secure multi-party computation (SMPC) \cite{yao1982protocols} is an encryption protocol that allows multiple parties to jointly compute a function while keeping their input data private. Each party obtains the correct computation result without knowing the other parties' private inputs. SMPC has characteristics such as decentralization, input data security, and accurate computation results. It enables AI to handle complex network environments, protect user privacy and data security, support emerging application scenarios, and promote data sharing and utilization.

The main supporting technologies of SMPC include Garbled Circuits \cite{shamir1979share}, Oblivious Transfer \cite{haque2022garbled}, Secret Sharing \cite{lindell2009proof}, and Homomorphic Encryption \cite{acar2018survey}.

(1) Garbled Circuits technique involves compiling computational logic into a circuit and encrypting each gate (e.g., AND, XOR, etc.) within the circuit. Participants can complete the computation by interacting with the encrypted information without knowing the specific logic of the circuit.

(2) Oblivious Transfer is a protocol that allows one party (e.g., the sender) to send one of many pieces of information to another party (the receiver), where the receiver can choose only one piece of information. The sender does not know which piece of information the receiver has chosen.

(3) Secret Sharing involves splitting a secret into multiple shares and distributing them among participants. The original secret can only be reconstructed when a sufficient number of shares are combined.

(4) Homomorphic Encryption is a special encryption method that allows specific computations (such as addition and multiplication) to be performed on encrypted data. The result of these computations, when decrypted, corresponds to the result of the same computations performed on the original data.

Recent research has leveraged SMPC to achieve secure model aggregation in FL, which involves training ML models on decentralized devices while keeping the training data localized. 
To counter reverse attacks on data, a privacy-preserving data aggregation mechanism based on secret sharing techniques in SMPC can efficiently protect FL against such attacks \cite{wang2022privacy}. To address high communication overhead and poor scalability in conventional SMPC, a two-phase SMPC-based FL framework has been proposed \cite{song2022eppda}. This framework enables multiple clients to collaboratively train AI models while protecting their data privacy. {Furthermore, a hierarchical FL architecture utilizing SMPC has been developed to alleviate the communication costs and scalability issues associated with aggregating global models at only a few nodes, enhancing communication efficiency and scalability without compromising privacy \cite{zhu2020privacy}.}
A novel hybrid FL architecture has been proposed, which combines FL with Trusted Execution Environments, SMPC, and Beidou satellites. This architecture achieves secure key distribution, encryption, and decryption, and provides verification mechanisms for each participant to ensure the security of local data \cite{huang2021starfl}. Additionally, SMPC can combine with blockchain technology to track and mitigate malicious behavior, achieving a secure and trustworthy collaborative edge learning framework \cite{tang2022secure}. 

{
SMPC protects data privacy while enabling collaborative computation across multiple parties, making it a valuable asset for 6G applications. It can contribute to FL, distributed resource allocation and optimization, secure data fusion and analysis, distributed identity verification and key management, and anti-fraud billing systems. By leveraging encrypted computation, SMPC ensures the secure processing of sensitive information. 
While SMPC enhances data sharing privacy and security in 6G networks, challenges remain, including computational and communication overhead, protocol standardization, and finding a balance between privacy protection and efficiency.}

\subsubsection{Semantic Communication}
\begin{figure}[htbp]
    \centerline{ \includegraphics[width=0.85\textwidth]{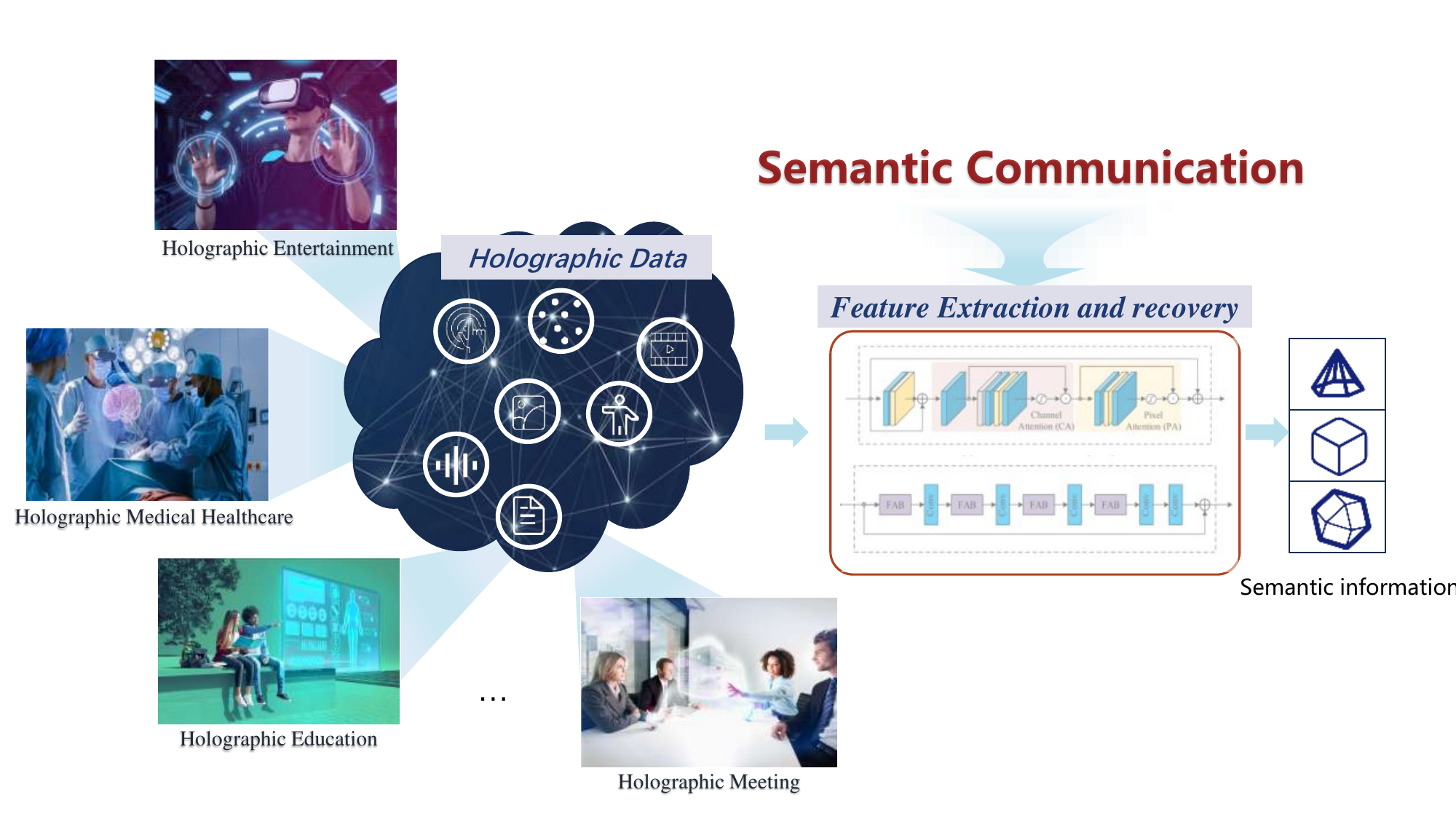}}
     \caption{{Llustration of applications that require ultra-low-latency interaction in 6G networks, and deep learning-based semantic communications can greatly benefit these applications.}}
    \label{semantic_communication}	
 \end{figure}

Semantic communication, commonly relying on DNNs, is considered a promising technology in 6G. Compared with traditional syntactic communication which focuses on the accurate transmission of data in bits, semantic communication only transmits the semantics of the message.  
It can also preserve essential semantic relationships when transmitting information, thereby aiding downstream AI tasks. These characteristics help meet the demands of new data-hungry 6G applications, e.g., holographic communication and XR, as shown in Figure \ref{semantic_communication}.

\begin{figure}[t]
    \centerline{ \includegraphics[width=1\textwidth]{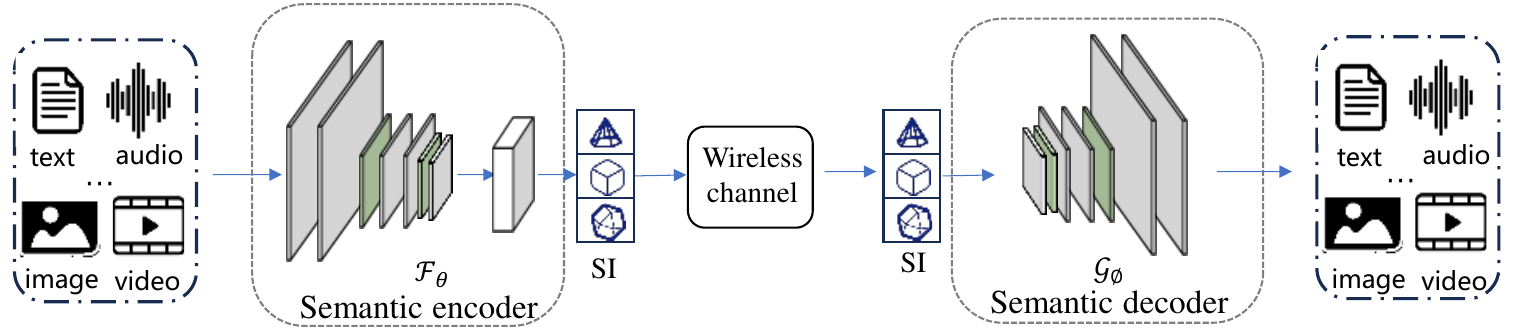}}
     \caption{A basic semantic communication model}
    \label{semantic_communication_model}	
 \end{figure}

Figure \ref{semantic_communication_model} shows a basic semantic communication model, encompassing three essential components: the semantic encoder, the channel, and the semantic decoder~\cite{10820866}. At the transmitter, the semantic encoder extracts and encodes the SI of the source data. This process involves extracting the SI and compressing or removing the irrelevant information. To achieve this, the raw data is encoded by a neural network, denoted as ${\mathcal{F}}_\theta $, which outputs the semantics that retain the critical meaning while discarding non-essential details. The compact semantics are then transmitted over a noisy physical channel, such as a wireless channel. 
At the receiver, the semantic decoder decodes the received data. 
This process involves “understanding” and inferring the semantic content sent by the transmitter. To accomplish this, the semantic decoder, denoted as ${\mathcal{G}}_\phi $, processes the received data with the goal of mapping it back to the source data. 

The authors of \cite{niu2024mathematical} provided a pioneering work in SI theory, establishing the mathematical framework for SI theory. The concept of modulation division multiple access technology was proposed in \cite{zhang2023model}, which efficiently distinguishes users based on semantic features from the information dimension, thereby improving resource utilization in multi-user semantic communication systems. Semantic extraction and recovery, also called semantic encoding and decoding, are the core components of semantic transceivers with four dominant advanced semantic extraction techniques \cite{yang2022semantic}: DL-based semantic extraction, RL-based semantic extraction, KB-assisted semantic extraction, and native semantic extraction. Moreover, most semantic communication studies focus on text transmission, audio transmission, and recognition, image transmission and recognition, and video transmission and recognition from the perspective of data modality and task classification.

\textbf{(1) Text Transmission.}
The authors of \cite{xie2021deep} developed a DL-based text semantic communication system named DeepSC, which performs joint semantic-channel coding to achieve superior performance gains compared to existing technologies, particularly at low SNRs. Recognizing that most semantic metrics are non-differentiable, an RL-based optimization paradigm for large-scale and complex text semantic transmission was introduced in \cite{lu2021reinforcement}. This technique uses self-critical random iterative updates to train decoupled semantic transceivers, addressing the non-differentiable semantic channel optimization issue. The authors of \cite{jiang2022deep} proposed a text-oriented semantic communication technique named semantic coding Reed Solomon hybrid automatic repeat request (HARQ) by combining HARQ and Reed-Solomon channel coding with semantic coding to tackle inefficiency and inflexibility with varying sentence lengths.

\textbf{(2) Audio Transmission and Recognition.}
A DL-based audio semantic communication system named DeepSC-S was developed, as noted in \cite{weng2021semantic, weng2021semantic1}. The audio semantic communication scheme employs a joint semantic encoder/decoder and channel encoder/decoder to mitigate the semantic distortion caused by the noises and interference from the wireless channels. The authors of \cite{tong2021federated} explored FL-based audio semantic communication, developing a wav2vec-based autoencoder composed of CNNs. This system effectively encodes, transmits, and decodes audio SI, reducing communication overhead. 

\textbf{(3) Image Transmission and Recognition.}
The authors of \cite{bourtsoulatze2019deep} investigated a DeepJSCC image semantic communication architecture, where the encoder and decoder functions are parameterized by CNNs and jointly trained on the same dataset to minimize the mean squared error of the reconstructed images. Joint training enables DeepJSCC to avoid the cliff effect and adapt to the varying SNR.

\textbf{(4) Video Transmission and Recognition.}
The authors of \cite{wang2022wireless} developed a variable-length DeepJSCC system that utilizes nonlinear transformations and conditional coding architectures to adaptively extract the semantic features of video frames. This system outperforms the traditional wireless video coding transmission on recognition accuracy. 

Semantic communication has unique advantages in semantic understanding, driving advancements in applications such as the tactile internet, holographic communication, XR, and human-machine symbiotic networks, fostering an increasingly intelligent and efficient network.

\section{Wireless Network Large Model}
{
LAMs exhibit remarkable capabilities and great potential in academia and industries. However, such a trend presents new challenges for the underlying network infrastructure. Wireless networks play a vital role in the infrastructure and are directly associated with LAMs' performance (e.g., data transmission, processing speed, etc).} Therefore, this section aims to explore the critical role of wireless networks in supporting the operation of these LAMs and conduct a systematic analysis and reflection on the relevant issues related to wireless networks in the context of constructing LAMs.

\subsection{Comparisons Between LLM and Wireless Network Large Model}
\begin{figure}[H]
    \centerline{ \includegraphics[width=0.8\textwidth]{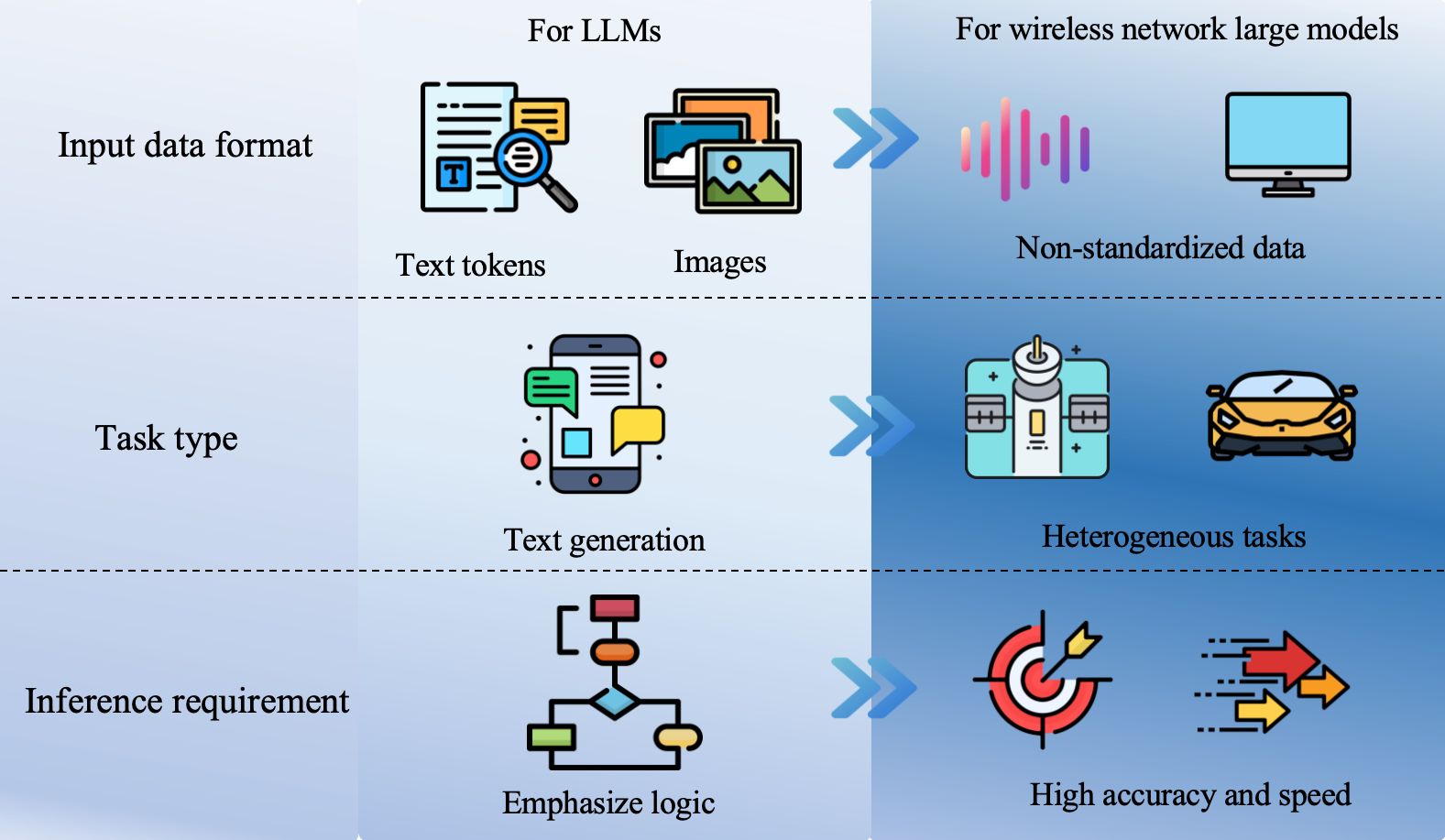}}
     \caption{Differences between LLMs and wireless network large models}
    \label{Comparisons between LLMs and wireless network large model}	
 \end{figure}

{ LLMs\cite{10500411} have demonstrated their promise in text and image tasks. It is challenging to directly apply traditional LLMs to wireless communication systems, as the data in the system consists of various protocol data in access networks, CN, and application data from the upper layer. Furthermore, these data are either structured or non-structured and may be heterogeneous with temporal and relational features.}
Specifically, the air interface involves wireless channels' transmission and scheduling management, encompassing diverse information generated by various wireless communication standards, such as LTE and 5G. The CN handles routing, authentication, and service control of user data, which includes sensitive content such as user behavior and location information, demanding high levels of privacy and security. Operational management data covers status monitoring, fault diagnosis, and performance optimization of network equipment. These datasets are typically large-scale, high-dimensional, lacking effective designs for unifying data across different standards. These characteristics necessitate the consideration of complex factors such as spatiotemporal relationships, real-time requirements, and system stability when processing and analyzing wireless communication data, presenting significant differences and challenges compared to conventional LLM data processing approaches, as illustrated in Figure \ref{Comparisons between LLMs and wireless network large model}.

Moreover, wireless communication systems constitute a highly complex ecosystem\cite{9144301} that employs multiple technological paradigms to meet diverse technical requirements and environmental scenarios. 
Unlike the relatively lenient requirements for inference speed in LLMs used in text and image processing, the demands placed on model inference speed in wireless communication tasks are critically important. Insufficient inference speed in a wireless network large model can lead to outdated results due to significant changes in channel conditions and signal characteristics during the inference process. Therefore, ensuring that models in wireless communication systems can quickly obtain inference results is a KPI for deployment effectiveness.

On the other hand, the requirements for inference accuracy in wireless network large models far exceed those of LLMs. In language models, different combinations of word arrangements have minimal impact on semantic expression; however, in wireless communication systems, which rely on exact mathematical models, sensitivity to inference errors is markedly high, with repercussions that are difficult to quantify. Signal modulation, a critical process encoding information onto carrier signals, exemplifies this sensitivity. Minor deviations in predicting modulation parameters by the model, such as errors in frequency or phase, can prevent the receiver from correctly demodulating signals, resulting in data loss or errors. In distributed wireless networks, even slight errors in predicting clock offsets or sync signals by the model can lead to synchronization failures among nodes, thereby compromising network stability. Therefore, precise control over the inference accuracy of the wireless network large model is a critical research challenge.

Although LLMs have significantly advanced text and image processing, directly applying them to construct a wireless network large model is impractical. In practical applications, integrating domain expertise with advanced data processing techniques is essential for effectively handling and analyzing wireless communication data, addressing its complexities and dynamic challenges.

\subsection{Potential Approaches to Constructing a Wireless Network Large Model}
Wireless network large model\cite{tong2023issuesnetgpt} refers to an intelligent and efficient solution to communication network operations, achieved through integrating knowledge and technologies from the communication domain. A wireless network large model provides network management, and operations fault detection and diagnostic services. The failures in networks may lead to performance degradation and service interruptions. By analyzing massive network data, a wireless network large model can monitor network status in a real-time fashion, quickly identify and locate potential reasons, and provide accurate diagnosis strategies. 

A wireless network large model optimizes task orchestration and scheduling. Modern communication networks involve complex and diverse tasks and business processes, requiring efficient scheduling and management to ensure network operations. Through an in-depth analysis of network topology and business requirements, a wireless network large model can potentially optimize task execution sequences and resource allocations, enhancing network resource utilization and business processing efficiency.

In terms of performance optimization and resource allocation \cite{shao2024wirelessllm}, the wireless network large model employs intelligent strategies and algorithms. By continuously monitoring network load and traffic conditions and integrating predictive analytics and dynamic adjustment mechanisms \cite{zou2024telecomgptframeworkbuildtelecomspecfic}, it achieves real-time optimization of network performance and rational allocation of resources, ensuring network stability and reliability. 
The prevailing industry viewpoint suggests that the wireless network large model should adopt a three-layer model structure, involving L0, L1, and L2 layers. 
\begin{itemize}
    \item 
The L2 layer can provide customized solutions tailored to specific business scenarios in this framework. For instance, the L2 layer model can optimize energy utilization efficiency in energy conservation within networks by analyzing data traffic and device loads. Regarding traffic prediction and fault monitoring, the L2 layer model can employ DL techniques to identify anomalies and issue early warnings, thereby enhancing network stability and reliability.
    \item 
The L1 layer involves domain-specific grand models. The design of the L1 layer emphasizes modeling and optimization for specific domain challenges and issues. For example, the L1 layer model can analyze wireless channel conditions and user behavior in the air interface to improve data transmission rates and coverage. The L1 layer model can utilize network topology and device status information for fault diagnosis and optimization scheduling in the CN and operations. 
    \item 
The L0 layer represents the universal grand model for the entire network. The design of this layer needs to consider the overall architecture and operational mechanisms of communication systems to ensure it possesses strong generalization capabilities. Through comprehensive data analysis and intelligent decision-making, it aims to provide optimization and management services for all communication tasks.
    
\end{itemize}

\begin{figure}[H]
    \centerline{ \includegraphics[width=0.85\textwidth]{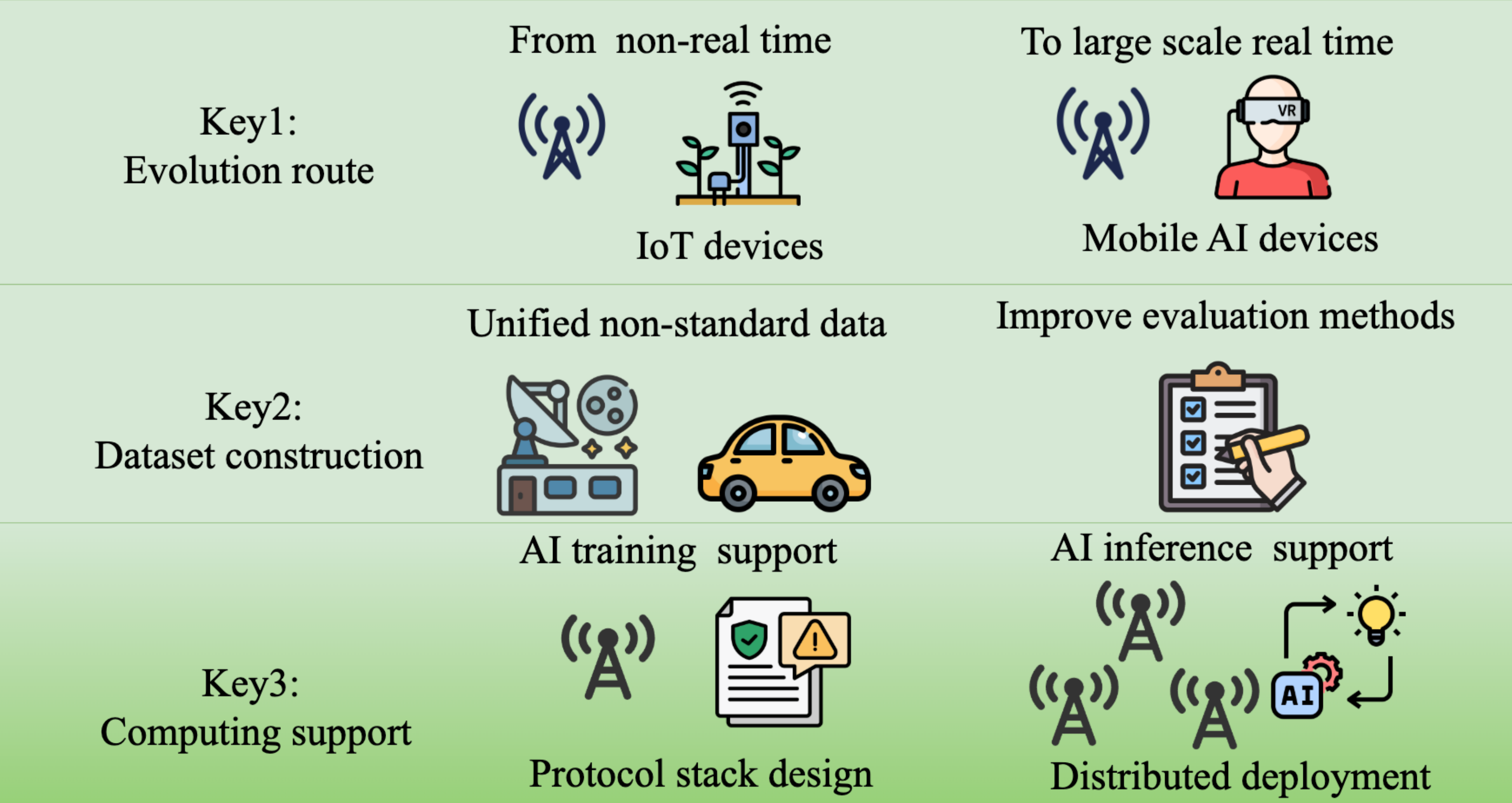}}
     \caption{Construction philosophy of wireless network large model}
    \label{Potential Approaches}	
\end{figure}

Figure \ref{Potential Approaches} shows the construction philosophy for constructing a wireless network large model in communication networks, encompassing evolution route, dataset construction, and computational support, which complement each other to form the fundamental framework of wireless network large models.

\textbf{Key 1: Evolution Route.} Constructing a universally applicable large model for the entire network faces many challenges. On the one hand, data within communication networks exhibit highly structured characteristics, encompassing extensive information from various devices and sensors~\cite{2019The}, including network traffic, device status, and user behavior. The diversity and complexity of these data present significant challenges for model 
optimization. On the other hand, tasks within communication networks can be complex and interdependent.
For instance, network traffic characteristics may vary substantially across different business scenarios, rendering conventional unified modeling approaches less feasible. 
A phased approach to model construction becomes important to address the aforementioned challenges effectively. 
The initial focus can be small-scale, offline L2-level models. Through the practical validation of critical business scenarios such as network energy efficiency, traffic prediction, and fault monitoring, efficient utilization of network resources can be achieved.
In designing L1-level models, emphasis should be placed on addressing domain-specific challenges, employing targeted modeling and optimization approaches to achieve domain generalization capabilities across air interface, operations, and CN sides. To construct a universally applicable L0-level model encompassing the entire network, carefully considering the communication system's overall architecture and operational mechanisms is essential to ensure the model's ability to support services and strategies for all communication tasks.

\textbf{Key 2: Dataset Construction. }Dataset Construction is crucial for large-scale network models\cite{8304385} directly impacting AI model training efficacy and application. Challenges revolve around data acquisition and quality difficulties \cite{10579546}. 
To tackle the difficulties in data acquisition, multiple avenues should be explored, including proactive collaboration with industry partners to promote data sharing and openness\cite{8932161}, involving telecommunications operators, equipment manufacturers, and others to expand the coverage of datasets. 
Furthermore, establishing industry standards and guidelines can guide stakeholders in adhering to unified data collection and processing standards, facilitating data sharing and exchange to overcome data acquisition challenges.
A comprehensive data quality assessment framework is also indispensable \cite{2019Big}.
Utilizing data quality metrics such as accuracy, completeness, and consistency, data should be assessed and monitored to detect any data quality anomalies and take corresponding measures for improvement and optimization. Employing data quality management tools and technologies can automate quality checks and validations.
During the data collection process, it is crucial to enhance the monitoring and management of data sources to ensure the comprehensiveness and reliability of data acquisition. Advanced data cleaning and preprocessing techniques should be implemented to denoise and handle missing values in raw data.
Moreover, establishing robust data standardization procedures to unify data formats reduces dataset heterogeneity and inconsistencies.

\textbf{Key 3: Computational Support. }Regarding the computational demands of 6G communication to support large-scale network models, optimization efforts are needed in both AI training and deployment. This aims to achieve efficient, reliable, secure, and stable data transmission and utilization of computing resources. In AI training services, key factors include efficient and reliable training data collection and transmission processes, proprietary protocol design, support from network elements, and customized traffic scheduling algorithms\cite{10283537}. When designing core and access networks, proprietary protocols tailored specifically for AI task transmission should be developed. These protocols need to account for the unique characteristics of AI tasks, such as large data volumes and computational intensity, to ensure optimal transmission efficiency and performance. 
When designing network elements to support AI task transmission, considerations should be made regarding hardware architecture and software functionality to process and transmit AI training tasks effectively. 
In traffic scheduling, selecting appropriate strategies is crucial for supporting the training of LAMs. Conventional traffic scheduling strategies may not suffice for AI training tasks, necessitating the design of customized traffic scheduling algorithms tailored to the characteristics of AI tasks. These algorithms should factor in task priorities, network load conditions, and device resource availability to ensure that AI training tasks receive sufficient network and computing resources promptly.

In AI inference services, the high demands of LAMs on storage space and computational capabilities often exceed what a single BS can provide. A viable approach is designing an effective architecture utilizing distributed node collaboration \cite{10466747} to collectively handle model storage and inference tasks. Alongside the design of distributed node collaboration architecture, effective management and scheduling of inference tasks are crucial to maximize the utilization of each node's computational resources while ensuring system efficiency and stability \cite{2020When}. A key consideration is the design of task allocation and scheduling algorithms. 
This approach allows for the rational allocation and scheduling of inference tasks.
Moreover, attention should be given to enhancing system monitoring and automated management. Automated management tools and mechanisms, such as automated configuration and deployment, fault diagnosis, and recovery, can be introduced to detect and promptly address faults and anomalies. This reduces management and maintenance costs and improves system maintainability and manageability. These optimization measures collectively provide better technical support and assurance for AI large model computation capabilities in 6G networks.


{\subsection{Challenges of Wireless Network Large Model Development}

\textbullet\textbf{ \emph {Data Heterogeneity.} }In wireless networks, the diversity and complexity of data present significant challenges in constructing comprehensive large-scale training and testing corpora. The network contains various types of structured and unstructured data, including uplink and downlink channel information, measurement reports, drive test data minimization, KPI monitoring data, network topology performance indicators, and cost and energy consumption logs. These data sources are highly heterogeneous, and the data from different origins often exhibit notable discrepancies in granularity and time resolution. It is necessary to consider how to effectively align these heterogeneous data to build a consistent and representative dataset. Exploring techniques such as data normalization \cite{8936546}, time-series alignment \cite{10679601}, and data fusion \cite{10399795} can contribute to constructing a high-quality training corpus. Additionally, attention should be given to transforming the processed data into input formats suitable for inference algorithms. This may involve feature extraction and selection, data dimensionality reduction, and the creation of appropriate input representations to enable the model to perform efficient inference and decision-making.

\textbullet\textbf{ \emph {Real-Time Inference.} } Wireless network large model can be applied to real-time decision-making tasks such as dynamic spectrum allocation, power control, and user access management. However, excessive inference time can degrade overall network performance and user experience. This is particularly critical in network resource scheduling, where strict time constraints demand that model inference be completed within milliseconds to effectively manage and allocate communication network resources. This directly impacts the real-time responsiveness and service quality of the network. To meet these requirements, the design and deployment of large-scale models in wireless networks must prioritize optimizing inference speed, which includes hardware acceleration, algorithm optimization, and network architecture adjustments\cite{10192095}. Furthermore, achieving a balance between model accuracy and computational efficiency is crucial for efficient and reliable network resource management and scheduling .

\textbullet\textbf{ \emph {Ambiguous Route.} } Building a universal large-scale model framework for wireless communication is a challenging task. Current industry practices tend to leverage large-scale pre-trained language models as foundational components, which are then fine-tuned for specific industry applications. However, the complexity and dynamics of wireless networks require models with higher adaptability and flexibility. Scholars have made some progress in exploring models based on semantic communication, which has advanced the understanding and processing of data flows in wireless networks \cite{qiao2024latency}. However, the feasibility and effectiveness of these models in practical deployments, including their stability and performance under varying environments and conditions, remain unresolved and require further empirical validation.

\textbullet\textbf{ \emph {Protocol Compatibility.} }The integration and deployment of the Wireless Network Large Model inevitably rely on the communication network's protocol stack. However, since large-scale model technologies have only emerged in recent years, compatibility issues with existing communication protocol stacks remain to be addressed. Whether the current protocol stack architecture can meet the future demands of large-scale models, and how to enhance and optimize the existing protocol stack to better support the application of these models in wireless networks, pose significant technical challenges. This requires an analysis of the existing protocol stack to identify potential barriers to large-scale model integration and the exploration of new protocol design principles to ensure the efficient and stable operation of large-scale models in wireless networks.

\textbullet\textbf{ \emph {Security Threats.} } Wireless network large model faces multiple security challenges\cite{1580513}. Firstly, it may encounter privacy attacks. Data sets from 6G networks may involve sensitive information, such as user location, mobility patterns, and communication content. Without adequate protection, this data is vulnerable to privacy threats, which can lead to a loss of user trust and legal issues. Secondly, data tampering is another significant threat to large-scale models. Attackers may influence the model's learning process and decision-making outcomes by injecting misleading data or manipulating data sources. For instance, incorrect input data could lead to misjudgments of user demands or network congestion conditions. Additionally, network attacks targeting large-scale wireless network models are also hazardous. Malicious actors may disrupt model operations or regular requests through flooding attacks or resource depletion attacks, severely affecting service availability and real-time decision-making capabilities. Finally, ensuring the security and integrity of large-scale model architectures is critical. Reverse engineering and model parameter leakage could allow attackers to gain sensitive information about intellectual property or operational mechanisms, thereby jeopardizing the stability and reliability of services.}

\section{AI as a Service}
\begin{figure}[H]
    \centerline{\includegraphics[width=1\linewidth]{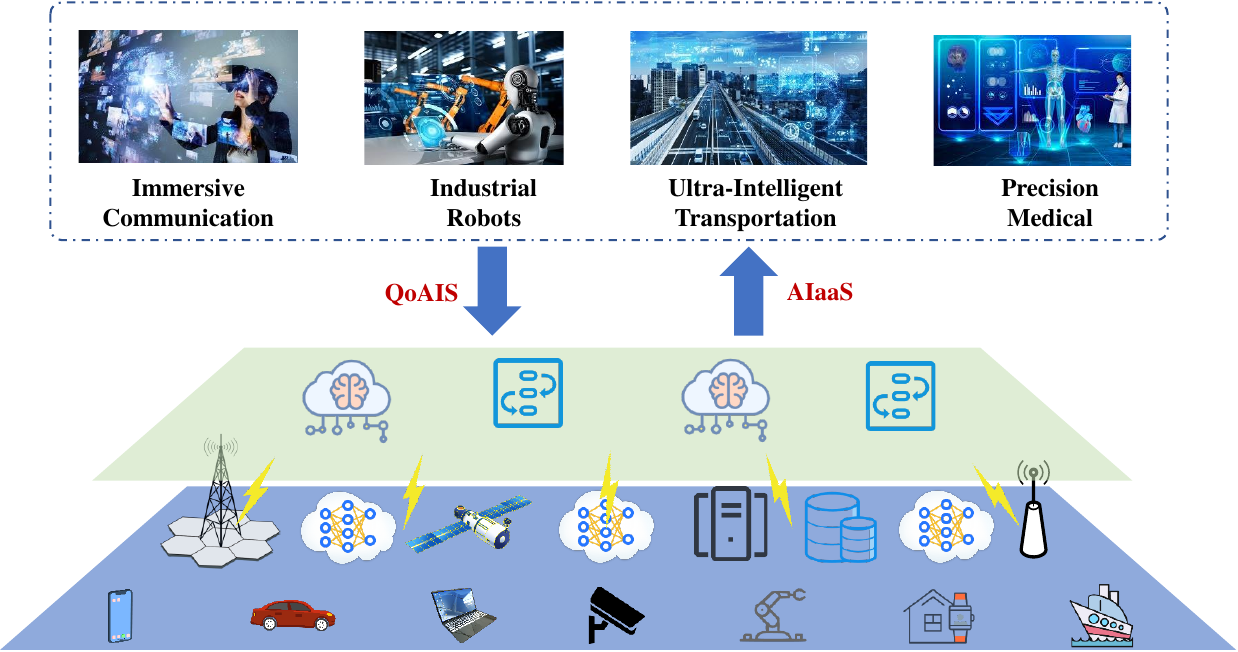}}
     \caption{AI as a Service}
    \label{AI as a Service}
 \end{figure}

To more effectively support ``native intelligence" and achieve ubiquitously ``universal intelligence", future 6G networks will treat AIaaS, giving rise to the concept of 6G AIaaS \cite{AIaaS}, which is presented in Figure \ref{AI as a Service}. 6G will provide efficient end-to-end support for AI-related businesses and applications, intelligently connecting distributed agents for large-scale AI deployment across various industries. 6G AI services will address needs for high real-time performance, high security, and privacy, or low overall energy consumption by conducting AI training or inference within the network, offering intelligent capabilities tailored to different application scenarios.
\label{Section_AIaaS}

6G networks, as inherently intelligent architectures, will be able to efficiently train or infer large-scale distributed AI through resources and functionalities within the network, including communication, computation, datasets, and foundational models \cite{AIaaS_6G}. This transition from AI4NET in 5G to networks for AI in 6G will ultimately provide intelligence that goes beyond data, directly targeting users. This transformation will redefine the ecosystem of edge clouds and create new business models through 6G mobile networks, transitioning from connectivity-focused to service-oriented networks and ultimately achieving ubiquitous intelligence.

\subsection{Typical Scenarios of AIaaS}

Typical scenarios for 6G AIaaS are contexts and situations where AI in 6G networks comes into play, including but not limited to industries and fields such as agriculture, education, and healthcare \cite{CUHK-6g}.

\subsubsection{Immersive Communication}
In future communication, immersive technologies like VR, AR, and Measurement Reports will enhance interactive experiences \cite{partarakis2024review}. AIaaS will support these technologies with real-time language translation, emotion analysis, and personalized suggestions in virtual environments. Trained AI models will be validated to ensure realistic and credible interactions.
6G AIaaS will offer comprehensive AI training, collecting user data for multimodal learning and enabling rapid model training and updates  \cite{10411981}. It will support multi-device linkage, allowing various devices to interact with virtual environments. 6G's extended coverage will seamlessly connect virtual and real worlds, facilitating cross-system information sharing.
DT technology will enhance AI validation, generating richer sample data than physical environments and improving model stability and performance. 
As the number of intelligent terminals increases, 6G's ultra-large connectivity and low latency will be essential for AI training and validation, supporting precise positioning and massive data collection.
6G AIaaS must deliver high-quality autonomous services to ensure secure interactions in virtual environments, with options ranging from fully autonomous to human-controlled services. This support will lead to more realistic and efficient immersive communication, advancing remote collaboration and communication.

\subsubsection{Intelligent Industrial Robots}
Industrial robots will be widely utilized in future industrial manufacturing and transportation scenarios \cite{benotsmane2019economic}. 6G AIaaS provides AI training services for robots, such as data collection through robots, applying model training for multi-agent learning, and distributing the trained models to the robots. The extreme transmission capabilities of 6G will support rapid model training and parameter exchange between robots and the network. 6G AIaaS also offers multi-system linkage capabilities, allowing various terminal devices to network with robots flexibly and enabling remote control via computers, smartphones, and VR devices~\cite{10002890}.
Robots can connect and interact with the environment through 6G AIaaS's sensing technology. They can even engage in cross-environment and cross-system learning with robots from different departments, sharing experiences. Additionally, the 6G network can leverage DT networks to provide AI verification services, iteratively training AI models with greater robustness and better performance through performance pre-validation. 6G AIaaS offers AI verification services for trained models, requiring the DT network to generate sample data in more scenarios than the physical environment, reducing the overhead and performance impact of data collection from the physical network.

\subsubsection{Precision Medical}
Smart healthcare will encompass various aspects of disease prevention, prediction, diagnosis, reasoning, monitoring, clinical surgery, patient care, and vaccine and drug development throughout the entire lifecycle \cite{santos2020online}. The new 6G system will better support the massive information transmission and synchronization required for smart healthcare and directly empower the processing and decision-making of medical information. By leveraging AI over 6G networks, geographically dispersed medical institutions, and individual practitioners can connect more extensively, aggregate more AI models and case data samples, and rapidly transmit and synchronize information between doctors and patients. This enhances FL and group learning, continually improving the accuracy, reliability, and real-time performance of predictive diagnostics and treatment actions. Smart healthcare can also achieve more rational planning and optimized allocation of various medical resources through 6G network AI, significantly reducing the physical and psychological stress on healthcare practitioners and patients.

\subsubsection{Ultra-Intelligent Transportation}
Ultra-intelligent transportation will involve various stages across the temporal dimension, including high-definition map downloads, vehicle environment perception, environmental prediction, and route planning \cite{8466351}. 
Taking autonomous driving as an example, the 6G system can provide AI-based data services, AI computation offloading services, and AI-based environment prediction or path planning. AI-based data services refer to the 6G system providing autonomous vehicles with AI-driven environmental perception results. For instance, when an autonomous vehicle encounters a perception blind spot, the 6G system can collect perception data through sensors or wireless signals, then use AI models to infer the environment within the blind spot and feed the results back to the vehicle, enhancing driving safety. AI computation offloading refers to the process where autonomous vehicles offload part of their AI model inference or training computations to the 6G system. For example, when a vehicle drives through a complex road segment, the computational power required for AI model inference can increase significantly, resulting in unacceptable latency. In such cases, the vehicle can offload some of the AI computation tasks to the 6G system, thereby reducing the processing burden on the vehicle's chips.

With the 6G network's DL and multi-sensor fusion capabilities, an integrated perception and decision-making architecture is established for vehicles, roadside infrastructure, and cloud platforms \cite{10700687}. This architecture advances a new era of intelligent transportation characterized by intelligent travel, ubiquitous services, and comprehensive management and control.

\subsection{QoAIS}
Quality of AI service (QoAIS) refers to a comprehensive metrics framework for evaluating AI services within a network. It encompasses key dimensions such as performance, connectivity, computation, data, security, and orchestration \cite{10273257, AIaaS_6G, AI4Net, AIaaS}. Notably, the performance metrics in QoAIS differ from traditional model KPIs used in ML. While model KPIs primarily assess the internal evaluation metrics of a model—serving purposes like optimization and fine-tuning—QoAIS extends this scope. It incorporates factors such as the generalizability and robustness of models or algorithms, thereby offering a holistic view of system-wide performance. The metrics for other dimensions within QoAIS are novel, as they specifically account for the unique characteristics and requirements of 6G networks.
As shown in Table \ref{QoAIS indicator system of AI services}, the QoAIS system should include:

\begin{table}[H]
    \centering
    \caption{QoAIS indicator system of AI services}
    \begin{tabularx}{\textwidth}{m{0.25\textwidth}X}
    \toprule
        \textbf{Indicator Dimension} & \textbf{QoAIS Indicators} \\ \toprule
         { \textbf{Performance}} & {Boundary of performance metrics, training time, generalization, reusability, robustness, interpretability, consistency between loss function and optimization objectives, fairness, etc.} \\ 
             { \textbf{Connection}} & {Bandwidth and jitter, link delay and jitter, bit error rate and jitter, reliability, etc.} \\ 
             { \textbf{Data}} & {Feature redundancy, completeness, data accuracy, time-consuming data preparation, sample space balance, sample distribution dynamics, etc.} \\ 
             {\textbf{Computation}} & {Computing accuracy, duration, efficiency, etc.} \\ 
             {\textbf{Security}} & {Information confidentiality, data/algorithm privacy levels, data authenticity, traceability, etc.} \\ 
             { \textbf{Orchestration}} & {Full autonomy, partially controllable by humans, fully controllable by humans, etc.} \\
        \bottomrule
    \end{tabularx}
    \label{QoAIS indicator system of AI services}
\end{table}

\textbullet \textbf{ \textit{Performance-related indicators.}} 
Aspects critical to evaluating and improving AI models are encompassed, including the boundary of performance metrics, training time, generalization, reusability, robustness, interpretability, and consistency between the loss function and optimization objectives. 
\textit{Boundary of performance metrics} defines the upper and lower limits for assessing the quality of model performance indicators, such as error rate, precision, and recall. 
\textit{Generalization} reflects a model's ability to make accurate predictions on new, unseen data, while \textit{reusability} highlights its capability to remain effective across diverse scenarios. \textit{Robustness} ensures the model maintains consistent performance despite perturbations, adversarial attacks, or input uncertainties. \textit{Interpretability} focuses on the extent to which the model’s internal workings and outputs can be understood and explained, fostering transparency and trust. \textit{Consistency} between the loss function and optimization objectives ensures alignment between the design of the loss function during training and the AI system's goals, accounting for all relevant variables to achieve the desired outcomes.

\textbullet \textbf{ \textit{Connection-related indicators.}} Reliable and efficient AI services depend heavily on network characteristics, including bandwidth, jitter, link delay, bit error rate, and overall connection reliability. \textit{Bandwidth} represents the data transmission capacity of the network, directly impacting the speed and quality of real-time AI applications, such as video analysis or interactive systems.
\textit{Jitter} refers to the variability in packet arrival times, which can lead to inconsistencies in service performance, particularly in latency-sensitive tasks.
\textit{Link delay} quantifies the time taken for data to travel across the network, with lower delays contributing to smoother and more responsive AI systems.
\textit{Bit error rate} measures the rate at which errors occur during data transmission, emphasizing the importance of accurate and reliable data exchange. These factors collectively define the stability and dependability of the connection, directly influencing the QoAISs.

\textbullet \textbf{ \textit{Data-related indicators.}} The quality and characteristics of data play a vital role in ensuring the success of AI systems, focusing on feature redundancy, completeness, accuracy, and preparation time. \textit{Feature redundancy} highlights the presence of overlapping or irrelevant features, which can complicate model training and reduce efficiency.
\textit{Completeness} ensures that all necessary data attributes are available to achieve comprehensive insights and predictions.
\textit{Data accuracy} emphasizes the importance of reliable, error-free data to maintain the validity of AI outputs.
\textit{Time-consuming data preparation} reflects the challenges involved in cleaning, transforming, and organizing data for training purposes.
\textit{Sample space balance} ensures equitable representation of different classes or categories within the dataset, critical for minimizing bias.
\textit{Sample distribution dynamics} track changes in data distributions over time, enabling models to adapt and remain effective in evolving environments. Together, these aspects underpin the foundation for building robust and trustworthy AI models.

\textbullet \textbf{ \textit{Computation-related indicators.}} Efficiency and accuracy in computation are crucial for ensuring reliable and high-performing AI systems, encompassing factors such as computing accuracy, duration, and efficiency. \textit{Computing accuracy} reflects the precision of the computational results, ensuring the reliability of outputs across tasks. This dimension emphasizes the efficiency and accuracy of AI computations, which include computing accuracy, duration, and efficiency.
\textit{Duration} measures the time taken to complete computational processes, with faster computations enabling real-time or near-real-time performance in AI systems.
\textit{Efficiency} evaluates the resource utilization of the computation, including memory, processing power, and energy consumption. Optimizing these aspects ensures that AI services are not only accurate but also cost-effective and sustainable.

\textbullet \textbf{ \textit{Security-related indicators.}} Protecting data, algorithms, and systems is a fundamental requirement for AI services, with key considerations including confidentiality, privacy, authenticity, and traceability. \textit{Information confidentiality} ensures that sensitive data remains inaccessible to unauthorized entities, safeguarding user trust and compliance with regulations.
\textit{Data/algorithm privacy levels} highlight the measures taken to protect proprietary algorithms and personal data, reducing the risk of exposure or misuse.
\textit{Data authenticity} guarantees that the data used in AI processes is genuine and unaltered, ensuring reliable outputs.
\textit{Traceability} refers to the ability to track the origin, flow, and modifications of data and algorithms, enabling accountability and transparency in AI systems. Collectively, these factors contribute to the safe and ethical deployment of AI services.

\textbullet \textbf{ \textit{Orchestration-related indicators.}} The level of automation and human involvement in AI services plays a key role in determining how workflows are managed and controlled. \textit{Autonomy} in AI services defines the degree of automation and the extent of human involvement required across data, training, validation, and inference workflows. It reflects user expectations regarding the level of automation and is categorized into three levels. At the highest level, \textit{full autonomy} enables AI services to operate independently without human intervention. In contrast,\textit{ partial human control} involves workflows where some stages are automated, but others require human assistance. Finally, \textit{full human control} necessitates human involvement at every stage of the workflow, with no reliance on automation.

{{
\subsection{Key Capabilities for Supporting AIaaS}
The concept of AIaaS envisions a network-wide platform that seamlessly delivers AI services to users and devices connected. Achieving this vision hinges on several critical capabilities, which can be grouped into three primary aspects:

\subsubsection{Data Support and Model Accessibility}
A fundamental requirement for an AIaaS platform is robust data support. This can be achieved by utilizing the network's sensing capabilities and the data collection functions of IoT devices and user terminals to enable real-time data acquisition and storage. Ensuring data privacy while maintaining a direct link between users and their data is crucial, as this connection underpins model training and inference services. Moreover, AI services also rely on diverse models tailored to different tasks. The platform should allow users to access pre-existing models or create custom models to suit their needs. A dedicated interface connecting users to these models would facilitate customization and enhance the flexibility of AI services.

\subsubsection{Distributed Computing and Resource Optimization}
The distributed computing power inherent in 6G networks—spanning CPUs, GPUs, and cache resources across CN nodes, BSs, user devices, and third-party systems—forms the backbone of AIaaS. By leveraging software defined network, these resources can be discovered, tagged, and virtualized for unified management~\cite{10819488}. Protocols like distributed hash tables can be employed to store, retrieve, and update resource information efficiently. With resource visibility in place, computing, caching, and bandwidth resources can be allocated dynamically to meet service demands. Techniques such as DRL, game theory, and model predictive control enable adaptive resource allocation and service scheduling, ensuring efficient operation even under varying workloads.

\subsubsection{Security and Lifecycle Safeguards}
The distribution of cloud-edge computing nodes introduces new challenges in securing dynamic and cross-domain network connections. Comprehensive security measures are needed to protect AI services throughout their lifecycle. These include protocols for secure data acquisition and device authentication to ensure the integrity of inputs. Secure data fusion algorithms can validate and clean data from multiple devices, while homomorphic encryption offers privacy-preserving data aggregation. Safeguarding models, which represent significant intellectual property, is equally important. Measures like access control, differential privacy, FL, and real-time monitoring of data flow can protect against unauthorized use, model inversion, and other emerging threats.

Integrating these capabilities into AIaaS platforms requires a cohesive approach that addresses data management, computational scheduling, and security concerns simultaneously. The multidimensional service experience provided by these platforms should align with user-specific needs. With the evolution of 6G networks, traditional network limitations will be overcome, enabling AIaaS to thrive within an intelligent, adaptive, and efficient network architecture.

}}

 \begin{figure}[H]
    \centerline{ \includegraphics[width=0.85\textwidth]{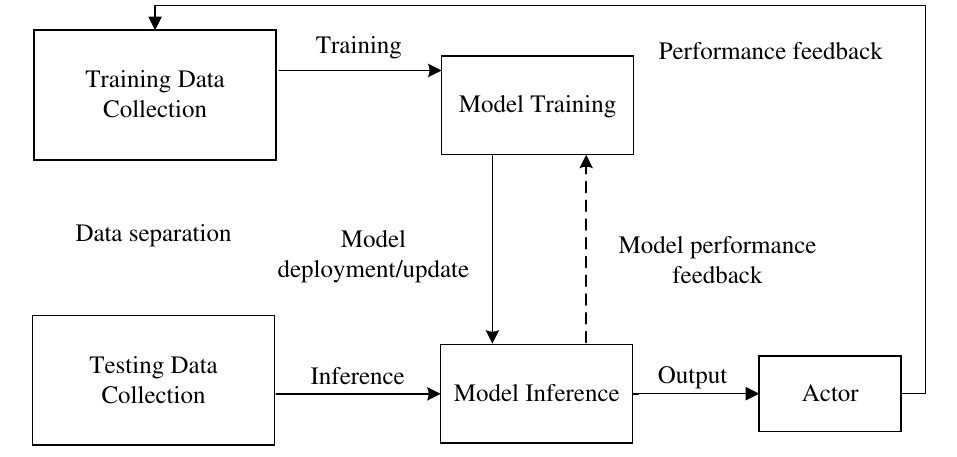}}
     \caption{This is the flowchart when applying the AI model to the communication process. The training task of the AI model needs to collect a large amount of data information in advance. The trained model must also be deployed in the corresponding position, and the model inference performance determines whether the AI model still has application value.}
    \label{Application process of AI model}	
 \end{figure}
\section{Standardization Process}
\label{Section_standardization}
As wireless communication and AI integration is a focus of the industry's research direction, important standardization organizations and industry alliances at home and abroad have launched research on integrating mobile communication networks and AI since 2017. These include the 3GPP, ITU, European Telecommunication Standardization Institute (ETSI), China Communication Standardization Association, IMT-2020 (5G) promotion group, IMT-2030 (6G) promotion group, etc. Currently, 3GPP has three technical specification groups (TSG): the CN, RAN, and service and system aspects (SA). Each TSG has 4 to 6 WG. The 3GPP SA WG2  (responsible for developing the overall 3GPP system architecture and services, including user equipment, access network, CN and terminals (CT), and internet protocol (IP) multimedia subsystem system architecture and services) and the 3GPP RAN WG3 (responsible for the overall radio network architecture and the specification of protocols for the related network interfaces) already accomplished some specification work on data collection, data analytic, and associated procedures at CN and radio network. 3GPP SA WG5 has progressed in standardizing network management data analytics services for managing AI/ML Models and functions. ETSI's zero-touch network and service management WG focuses on the research and analysis of technologies related to automated closed-loop operations, aiming to achieve a high degree of automation in network and service management. ETSI's experiential networked intelligence (ENI) WG is focused on defining an experiential and aware network management architecture that uses AI to improve operators' experience in network deployment and operations. Its goal is to use AI technology to improve the efficiency and automation of network management by defining a perception-adaptive-decision-execute control model. Figure \ref{Application process of AI model} displays the application process of the AI model. 

\subsection{3GPP}
3GPP has conducted relevant research on network intelligence and automation and introduced NWDAF as a CN AI element since Release-15. NWDAF can collect data from any network functions in 5GC and third-party application functions.  Due to time limitations, only slice-specific network data analytics is supported in Rel-15. Various network data and third-party application data from AF and 5GC network functions could be used as input to NWDAF, and output analytic results from NWDAF could be used by other NFs to guide network optimization, such as the following use cases been introduced in the Release-16: customized mobile management, 5G QoS enhancement, UPF selection, network functions load balancing, slicing SLA guarantee, etc. Rel-17 allows multiple NWDAFs to be deployed in a distributed architecture. Another enhancement is decomposing the monolithic tWDAF into a Model Training Logic  Function, Analytics Logical Function, and Analytics Data Repository Function. Further, the data collection coordination function enhances the data collection framework. Rel-18, the model performance monitoring, multi-vendor ML Model sharing, Horizontal FL, and six more use cases have been standardized.  

3GPP has identified typical application scenarios for self-organizing networks (SON) and has applied ML techniques, successfully verifying its usefulness in optimizing network performance. In 2018, 3GPP SA5 Group set up a research project, ``A study on SON for 5G", aimed at the integration of SON and 5G network technology, studying to promote the integration of SON and 5G network infrastructure by strengthening the collection and application of wireless big data. 

{{
The RAN1 WG initiated the study item ``Study on AI/ML for NR air interface" as part of the Release-18 phase \cite{38.843}. The study item explores how AI/ML algorithms can enhance air interface performance (throughput, robustness, accuracy, or reliability) and reduce overhead. The main application scenarios of this research include AI-based CSI feedback, AI-based beam management, AI-based positioning enhancement, and life cycle management of AI/ML model/functionality for the air interface.
Similarly, the 3GPP RAN2 WG launched the ``Study on AI/ML for mobility in NR" research project in December 2023 \cite{38.744}. The study project evaluates the potential benefits and specification impacts of AI/ML-aided mobility for network-triggered L3-based handoff. The project addresses several critical aspects, including AI/ML-based radio resource management measurement and event prediction for L3 Mobility, handoff/radio link failure prediction, and Measurement events prediction.

The 3GPP RAN3 WG identified the study item ``Study on further enhancement for data collection" in July 2020 \cite{37.817}. This study item aimed to enable an intelligent functional architecture for RAN. By leveraging use cases, it targeted to enhance data collection and identify potential standardization implications for  NG-RAN nodes and interfaces. In February 2022, the study item was completed, and the intelligent typical functional architecture on the RAN side, input, output, and feedback. The potential standard impact of three high-priority use cases (e.g., network energy saving, load balancing, and mobility optimization) was incorporated into TR 37.817. The intelligent universal functional architecture on the RAN side includes functions, such as data collection, model training, model inference, and action execution.
Recently, the 3GPP RAN3 WG has launched the ``AI/ML for NG-RAN" work item \cite{38.743}. The WI aims to incorporate data collection enhancements and signaling support specified in existing NG-RAN interfaces and architectures to enable AI/ML-based network energy saving, load balancing, and mobility optimization. 

The TSG SA (responsible for the overall architecture and service capabilities of 3GPP systems and the cross-3GPP TSG coordination) launched the study item, “3GPP AI/ML Consistency Alignment,” in June 2024 \cite{22.850}. The study item investigates ongoing AI/ML work in TSG CT, TSG RAN, and TSG SA WGs, and identifies the instances of any potential misalignment and/or inconsistencies. TSG SA leads the close collaboration and takes inputs from TSG CT and TSG RAN. The technical report of this study item will be finalized by June 2025.}}

\subsection{5G-MoNAch}
The 5G-mobile network architecture (5G-MoNArch) project in Europe introduced RAN-DAF, an independent AI analysis function explicitly designed for control unit plane in 5G NR, for data analysis and decision-making \cite{5G-MoNArch}. As a network element for AI and data analysis on the RAN side, RAN-DAF can monitor and collect UE and RAN data, and AI can prioritize local processing based on these real-time data to solve the need for rapid response of operations such as wireless resource management. MoNArch recommends that the RAN-DAF pass information to the controller on the RAN side to collaborate with the RCA to optimize wireless side performance, such as slice-sensitive selection, live wireless resource control, cross-slice resource management, and so on.

\subsection{ITU-T}
ITU-T study group 13 established the ITU-T focus group on ML for future networks and 5G in November 2017. The group has drafted several technical reports and specifications for ML in future networks, including network architectures, interfaces, protocols, algorithms, etc. There are three WG under the focus group where WG1 studies potential ML use cases in future networks and identifies the needs of use cases, and WG2 mainly works to classify ML in mobile communication networks and define the required data formats and related mechanisms to protect security and privacy. WG3 primarily studies the requirements of ML on network architecture in mobile communication networks, including network functions, interfaces, resources, etc. 

\subsection{ETSI}
In 2017, ETSI established the ENI WG dedicated to applying AI technologies to network operations. Improve the experience of operator network deployment and operations with a closed-loop AI mechanism based on context-aware and data-driven strategies. Since its inception, ENI has published several versions of specifications and reports, including system architectures with context-aware policy management, data processing mechanisms, and hierarchical evaluation methods for network AI. In 2019, ETSI ENI published the first edition of this research report and standard and launched the second phase of the closed-loop control study for real-time networks.

 \begin{figure}[h]
    \centerline{ \includegraphics[width=1\textwidth]{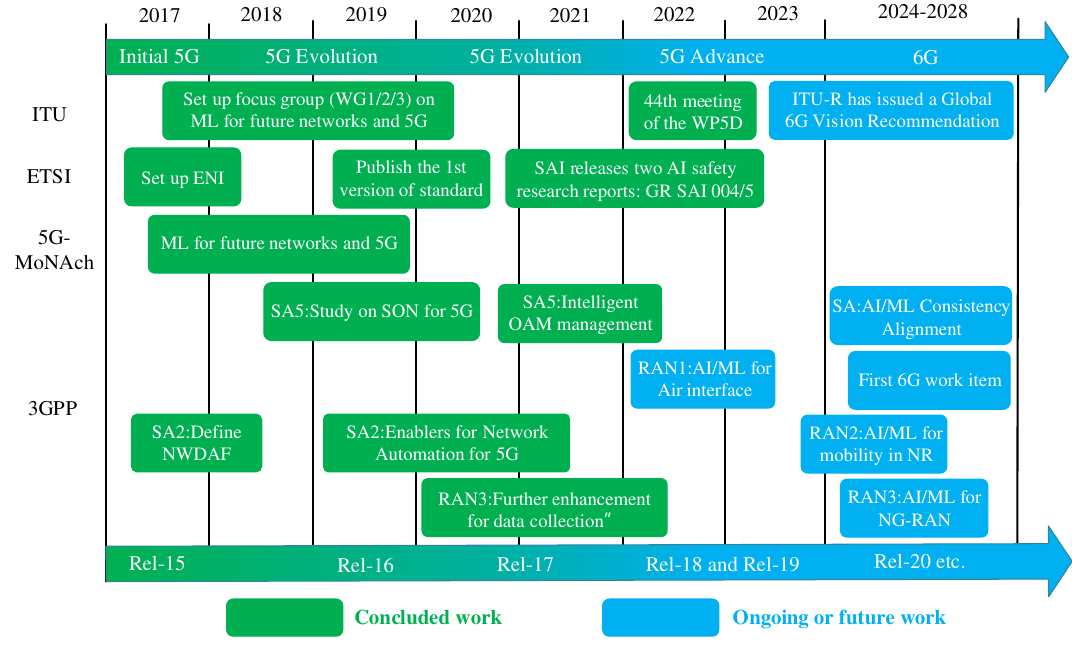}}
     \caption{The standardization process of AI/ML for wireless networks}
    \label{AI standardization process}	
 \end{figure}
\subsection{Collaborative Efforts in 6G}
In recent years, standardization organizations have also focused on the research and preparation of 6G. In June 2019, the China IMT-2030 (6G) promotion group was established to gather the strength of Chinese industry, universities, and research institutes, promoting China's 6G mobile communication technology research and carrying out international exchanges and cooperation and successively releasing several white papers and research reports such as the data format and model proposal of mobile communication and AI integration. In June 2023, the 44th meeting of the ITU-R WG 5D was held in Geneva, Switzerland, and the recommendation on the framework and overall objectives of the IMT for 2030 and Beyond was completed. This proposal aims to set a framework and overall goals for the development of IMT in 2030 and beyond, involving a global 6G vision consensus, 6G goals and trends, typical 6G scenarios, and capability indicators system, including the convergence of AI and communication scenarios, and define AI-related indicators. On September 12, 2024, at the 105th Plenary meeting held in Melbourne, Australia, 3GPP officially launched the first 6G study item -- ``6G use cases and Service Requirements". This initiative formally marks the 6G technology from the pre-research stage into the substantive standardization stage.
The vigorous development of AI standardization will significantly promote the integration of communication and AI, as shown in Figure \ref{AI standardization process}. AI still has a way to go from standardization to landing, which is also the main research direction of many standardization organizations in the future.

\section{Challenges of  AI and communication for 6G}
\label{Section_challenges}
This section delves into multiple challenges confronted by AI and communication in 6G. Herein, AI4NET and NET4AI represent different perspectives of AI empowering networks and networks supporting AI\cite{6gdistributed}, respectively. 
Sustainability 
cannot be overlooked in the integration process of 6G and AI. The demands for real-time AI, security and privacy, and human-AI collaboration also bring challenges to the development of AI and communication in the 6G era.

\subsection{Challenges of AI4NET}
{
\noindent \textbullet  \textbf{ \textit{Reliability of AI}}

The reliability of AI in networks poses a significant challenge due to the dynamic, heterogeneous, and often unpredictable nature of 6G communication systems \cite{10506539}. AI models, which rely heavily on data-driven learning, can struggle to maintain consistent performance when exposed to situations that deviate from their training environments. For instance, unexpected network anomalies, rare edge cases, or sudden shifts in user behavior patterns can lead to errors or even system failures \cite{9925609}. Additionally, the distributed and decentralized nature of 6G networks introduces further complexity, as AI models must operate across diverse nodes with varying resource constraints and environmental conditions. This interconnectedness increases the risk of cascading failures, where an error in part of the system can propagate and disrupt the entire network. 
An approach based on model-driven engineering principles was proposed in \cite{9643788}, which allows ML experts to schedule the execution of drift-detecting algorithms on a computing cluster. The author of \cite{9145835} proposed a distributed parallel modeling and monitoring framework for plant-wide processes with big data, where the multilevel monitoring indexes and fault contribution indexes are established based on the Bayesian fusion algorithm. 

\vspace{3 mm}
\noindent \textbullet  \textbf{ \textit{Stability of AI}}

Stability refers to an AI system's ability to produce reliable outputs when exposed to variations in input data, environmental conditions, or system parameters. The stability of AI in networks is a critical challenge, as it directly impacts the consistency and predictability of AI-driven decisions in dynamic and complex environments \cite{10159517}. 
This is particularly challenging in networked systems due to the heterogeneity and unpredictability of real-world conditions \cite{10375527}. For instance, fluctuating network traffic, sudden surges in user demand, or unexpected hardware failures can introduce variations destabilizing AI models. 
The simulation results of \cite{9868259} indicate that reducing the number of model parameters can enhance the model's robustness against maximum adversarial attacks, while increasing the number of model parameters can improve the model's robustness against minimum adversarial attacks. The authors of \cite{8826332} studied the robustness of mainstream link prediction methods under various network attacks and discovered that the link prediction method with high performance probably has low attack robustness.

\vspace{3 mm}
\noindent \textbullet  \textbf{ \textit{Generalization of AI models}}

The generalization problem is widely prevalent in integrating AI with communication systems \cite{2024A}. 
The application of AI in wireless communications imposes even higher demands on model generalization~\cite{9954418}. In wireless environments, users are situated in varying contexts, with 
their relative position to the BSs and mobility playing significant roles. This variability means that pre-trained AI models may lose effectiveness when confronted with new data that deviate from their training sets. In the dynamic and complex wireless environments of 6G networks, models with poor generalization capabilities will lack the potential for deployment and widespread use. Data enhancement or meta-learning can potentially improve model generalization. The scenario-adaptive meta-learning for mmWave beam alignment method was proposed to improve the domain adaption capabilities of beam alignment problem in \cite{xu2023samba}. A fast adaptive channel prediction meta-learning technique combined with a denoising process was proposed in \cite{28} to improve the generalization of the predictor.

\vspace{3 mm}
\noindent \textbullet  \textbf{ \textit{Interpretability of AI}}

DL is often treated as a black box due to its highly nonlinear nature. 
Humans can hardly understand the meaning, importance, and fluctuation range of parameters in the neural network through basic statistical assumptions like linear regression parameters, leading to unexplainability. 
Improving the explainability of AI helps humans understand the decision-making process, working principle, and potential bias of AI, and build trustworthy AI~\cite{10839451,10663726}. 
However, millions or even billions of parameters in DNNs and their nonlinear activation function form a highly complex decision boundary, hindering the interpretability of a model's decision-making process.
Through the gradient derivation and visualization of the model training process, the interpretability of the model can be effectively improved. The authors of \cite{18} deduced the gradient expression of the meta-model to improve the interpretability in the process of backpropagation and promoted the application of meta-learning. t-SNE visualization was proposed in \cite{van2008visualizing} to visualize the representations of data points. t-SNE is an embedded model that can map data from a high-dimensional space to a low-dimensional space and retain the local characteristics of datasets.
}

\subsection{Challenges of NET4AI}
{
\noindent \textbullet  \textbf{ \textit{Dynamics of Networks}}

Constructing an AI execution environment in wireless networks inherently entails significant dynamic variations. The interference and loss in the wireless connections between terminals and BSs are impacted by the surrounding environment, changes in the operational states of other BSs, and the mobility of terminals.
These variations contribute to the dynamic nature of mobile wireless networks, resulting in slow model training convergence and adversely affecting model inference performance, and 
even causing model divergence.
A mitigation strategy for unreliable communication nodes was designed in \cite{9745426} to prevent unreliable nodes from contributing to stable convergence during the AI training process. In response to the transmission errors caused by the dynamics of networks, the authors of \cite{10551685} modified the global aggregation method to counteract the model drift resulting from data packet errors. The theoretical relationship between dynamic unreliable communication channels and collaborative learning was also established in \cite{10253642}. 

\vspace{3 mm}
\noindent \textbullet  \textbf{ \textit{Heterogeneity of Networks}}

There is a myriad of heterogeneities in wireless networks: communication heterogeneity arising from the geographical locations of nodes and the time-varying nature of communication links; computational heterogeneity due to differences in the number, manufacturing processes, and architectures of node chips; data heterogeneity in the volume and distribution of data possessed by nodes; and model heterogeneity stemming from varying task objectives. 
Data heterogeneity results in disparities among local gradients computed by AI models.
Differences in computational power and communication capabilities lead to variations in model training and transmission efficiency, giving rise to the straggler effect. 
In \cite{10750008},  the sampling of training nodes in collaborative learning was optimized from the dimensions of communication, data, and computation, achieving efficient training on heterogeneous edge networks. For the heterogeneity of different models, the authors of \cite{10354479} provided approximate rules for the decision boundaries of each model to bridge the gap caused by model heterogeneity, achieving more accurate model precision.


\vspace{3 mm}
\noindent \textbullet  \textbf{ \textit{Complexity of Networks}}

The large scale and complexity of 6G networks significantly affect the deployment and performance of AI systems. 
In some cases, AI models must 
train and infer in real time to make effective decisions, which is a daunting task due to the high dimension and dynamic nature of the network data. The deployment of AI in 6G networks also necessitates decentralized architectures, complicating data aggregation and synchronization. Furthermore, the intricate interplay between various network components, such as edge and CN, requires AI models to handle multifaceted interactions and dependencies.
To address these challenges, there is a need to advance research in scalable AI techniques.
The authors of \cite{10184998} proposed a hierarchical RL framework based on human prior knowledge. Through process decomposition, stage transformation, key feature selection, and a policy gradient with a parameter-based exploration method, it provides insights 
in solving complex operations. In response to the differentiated data distribution, a clustering collaborative learning framework was proposed in \cite{10106044}, where the server dynamically determines the optimal number of clusters by iteratively performing incremental clustering.
Another focus could be on implementing scalable AI solutions that can dynamically adapt to the network complexity and scale. The authors of \cite{10.1145/3589303} used MARL to learn an adaptive fully distributed collaboration strategy for each collaborative node in complex wireless networks. 

\vspace{3 mm}
\noindent \textbullet  \textbf{ \textit{Resource Scarcity of Networks}}

The localized model training and transmission may not always succeed due to constraints on mobile devices' power and computational resources, bandwidth limitations, and wireless channel impairments, directly compromising the performance of FL regarding model accuracy and convergence speed. Consequently, evaluating the trade-off relationship between cognitive performance and multi-dimensional resource allocation becomes imperative upon introducing distributed learning.
For FL-enabled edge intelligence,
the authors of \cite{liu2022resource} derived mathematical relationships between FL performance, consumed computational resources, and communication resources, and provided a theoretical foundation for the design of intelligent access networks. 
To meet the low-latency requirements of multimedia MEC services, a task offloading and resource allocation algorithm based on the double DQN was proposed in \cite{10417030}, which assigns appropriate task volumes to different devices and optimizes their communication and computing resource settings via BS.
Moreover, the authors of \cite{10376360} 
introduced a distributed resource allocation method. Leveraging meta-federated RL, each device can autonomously optimize its transmission power and channel usage according to wireless environments. 
}

\subsection{Consideration of Sustainability}
{
While the advanced capabilities of AI are crucial for managing complex 6G operations to achieve unprecedented speeds, low latency, and high reliability, the computational power required comes at a high energy cost, which, if unchecked, could increase carbon emissions. 
The energy consumption of AI pertains to communication networks, cloud computing centers, and edge computing devices. For example, the computational requirements of LLM (e.g., GPT-4) and complex AI tasks ( e.g., RL/DRL) have increased substantially \cite{10347516}. The training stage may consume several hundreds to thousands of MWh of energy, while the energy consumption in the inference stage is directly related to the invocation frequency, model scale, and real-time requirements \cite{10363447}. 
For the energy consumption assessment of cloud computing centers and edge, a more comprehensive perspective is needed. For instance, the energy efficiency of a data center can be evaluated by the power usage effectiveness, which is the ratio of the total energy consumption of the data center to the energy consumption of IT equipment. The cost of leasing these computing resources is affected by the duration of resource usage, the required level of computing power, and market price fluctuations, and is ultimately passed on to end - users in the form of service pricing, subscription fees, etc.

Techniques such as pruning, quantization, and using neural architecture search can help create models that maintain high accuracy while reducing computational load \cite{Muhammad}. 
Energy-aware AI models, which are capable of dynamically adjusting their energy consumption in response to network demands, can also contribute to sustainable operations \cite{Energy-Aware}. Additionally, optimizing data processing and transmission within 6G networks represents another crucial aspect \cite{Intelligent-Traffic}.
Integrating renewable energy sources into 6G infrastructure can also contribute to the green 6G vision. 
}

\subsection{Requirement of Real-Time AI}
{The 6G aims to achieve uRLLC in support of real-time applications, including autonomous driving, telemedicine, and industrial automation. These 
applications require data transmission and processing to be completed within milliseconds or even microseconds. However, the state-of-the-art AI models, characterized by large parameter spaces and intensive computation, typically employ distributed deployment and incur additional latency \cite{9739684}. 
In scenarios such as sensor data fusion for industrial automation and traffic monitoring and analysis for intelligent transportation systems, algorithms are required to process data and make accurate fusion decisions within an extremely short time \cite{9385927}.

DL algorithms construct DNNs to extract deep feature representations from vast and complex data, unlike traditional methods that rely on manually designed feature extraction rules. This significantly reduces errors and limitations caused by human factors, thereby enhancing the accuracy and real-time performance of data fusion.
RL/DRL dynamically adjusts fusion weights and methods based on real-time feedback information from different data sources, adapting to the constantly changing data environment and task requirements. 
Reducing the computational requirements of models without significantly compromising accuracy is possible through designing more efficient network architectures, e.g., lightweight versions of CNN and RNN \cite{RNN-IOT}. Adopting edge computing can reduce latency by shifting data processing tasks from the central cloud to network edge \cite{Edge-Computing}. Innovations in network architecture, e.g., network slicing~\cite{XIAO2022109279}, can also provide customized network resources for different services and applications, ensuring critical tasks receive the necessary bandwidth and latency guarantees. Additionally, 
dedicated AI processors, such as GPU, tensor processing units, and field-programmable gate arrays, can expedite model inference, helping meet the stringent real-time requirements of 6G networks.}

\subsection{Security and Privacy}
{
AI models are susceptible to adversarial attacks, and imperceptible perturbations in the input data can lead to incorrect outputs~\cite{10263803}.
Moreover, the security and privacy risks faced by AI algorithms in the process of data fusion are formidable. Specifically, the data generally originate from diverse data sources, which frequently encompass sensitive information, such as personal identities, medical records, and financial transactions. This gives rise to risks on all aspects of data fusion, ranging from storage and transmission to processing \cite{8428412,10753492}. For instance, the server storing the fused data might be subject to hacker attacks.
During the data preprocessing of AI model training, data are susceptible to theft if appropriate encryption protection measures are not in place.

One defense approach is developing robust AI through adversarial training, which can enhance the resilience of AI systems 
\cite{Adversarial}. Another approach is 
FL and/or MARL, where AI models are trained across multiple 
devices without exchanging the raw data, thereby preserving privacy \cite{Federated-Learning,10763434}. 
For example, the authors of \cite{10045665} proposed an adaptive feature-correlation region segmentation mechanism to offer privacy protection for DL models under a given moderate privacy budget.
SMPC and homomorphic encryption can also be performed to
ensure data privacy even when some processing needs to be conducted at third parties \cite{Lagendijk}. 
Moreover, regulatory frameworks and policies must be established to define clear data protection and cybersecurity standards in the context of AI-6G integration. This includes setting guidelines for responsible AI usage, ensuring compliance with privacy regulations, e.g., general data protection regulation, and fostering international collaboration to combat cyber threats.}

\subsection{Human-AI Interactive Collaboration} 
The core challenges faced by human-AI interactive collaboration 
include trust, transparency, fairness, and ethics. The ``black-box'' nature of AI makes it difficult for humans to understand their decision-making logic, resulting in insufficient trust or overreliance. For instance, the complexity of DL models often makes it impossible for users to judge their reliability, thus affecting the effectiveness of collaboration \cite{10605530}. Meanwhile, typically dynamic human-AI interactive collaboration requires high adaptability in task allocation and role switching. However, the existing collaboration frameworks are 
prone to 
unreasonable task allocation and delayed role switching \cite{9590721}. Moreover, AI ethics and responsibility also need to be addressed urgently. For example, when an AI-driven decision is incorrect, it becomes crucial to establish a clear framework for assigning responsibility between humans and AI. 

Through natural language explanations, visualization tools, and interactive interfaces, users can better understand the decision-making logic of AI  and establish trust \cite{10646349}. 
For example, 
the division of labor between humans and machines can be adjusted in real time according to task complexity and AI capabilities \cite{9283461}. 
Coupled with an intuitive interface design, it can effectively reduce the difficulty for users to adapt to AI systems. 
For reducing bias and enhancing fairness, training with diverse and high-quality datasets and the application of fairness algorithms are crucial. Real-time bias-correction mechanisms can monitor and correct unfairness in AI-made decisions \cite{10529943}. Furthermore, establishing a global ethical and legal framework will provide institutional guarantees for the development and application of AI. The boundaries of AI responsibility can be clarified through legislation and interdisciplinary ethical review mechanisms 
ensure the compliance of AI technology with social values and legal norms \cite{9940606}.

\section{Future Works}
\label{Section_future}
This section delves into directions that revolve around 6G, with the aim of providing robust support for its innovations. In-depth research on these areas will help lay a foundation for the 
development of 6G.

\vspace{3 mm}
\noindent \textbullet  \textbf{ \textit{Adaptive Learning Mechanisms}}

Adaptive learning mechanisms, leveraging AI, are set to become the cornerstone of smart network management. These mechanisms must handle the complexity and dynamism of 6G characterized by heterogeneous devices, variable traffic patterns, and the need for real-time processing.
These models should be scalable to manage the exponential increase in connected devices and robust to ensure consistent performance under varying network states. 
One promising approach is DRL, which can dynamically optimize network decisions~\cite{10192095}. 
Integrating transfer learning techniques can enable AI models to apply knowledge gained in one domain to different yet relevant problems. 
Transfer learning can reduce the need for retraining models, saving computational resources and time. Another approach can be self-healing networks using AI. By incorporating predictive analytics, networks can anticipate failures and automatically reroute traffic or adjust parameters.
This proactive approach to fault management will be critical for maintaining the high reliability and availability required by 6G applications.

\vspace{3 mm}
\noindent \textbullet  \textbf{ \textit{Network Architecture Design}}

With their promise of higher bandwidths and lower latencies, 6G networks provide an ideal platform for AI applications but pose challenges in architectural design.
Future research can delve into creating network architectures that are inherently flexible and can dynamically adapt to the needs of AI applications. 
On the other hand, AI can significantly enhance network capability to manage these requirements by predicting traffic patterns and adjusting resources. A key component of such architectures is edge computing, which brings computational resources closer to the data sources, which is crucial for latency-sensitive AI applications.
Research is expected to develop edge computing solutions that can seamlessly integrate with central cloud resources.
Furthermore, deploying AI at the network edge introduces new challenges in distributed learning and data privacy. FL emerges as a viable solution, allowing AI models to be trained locally at the edge, with only model updates being shared to the central server. 

\vspace{3 mm}
\noindent \textbullet  \textbf{ \textit{Green AI and 6G Networking}}

The sustainability of 6G networks is critical
environmentally and economically.
AI has the potential to drive advancements in this area by optimizing network operations to reduce energy consumption. 
Future research can be devoted to AI models that can accurately predict the energy consumption of network components and optimize their operation to minimize energy use. This includes the dynamic adjustment of network infrastructure, such as BSs and data centers, based on user demand and traffic patterns. AI can enable these components to enter low-power states when demand is low and quickly ramp up when the demand increases. 
Future research can also 
explore the design of new energy-efficient hardware that can support AI operations at the network edge. This includes the development of specialized processors that can run AI algorithms with high efficiency and low power consumption.
Moreover, 
AI can manage the integration of renewable energy sources, e.g., solar and wind, and wireless power transfer into the network's power supply~\cite{10057423}. This involves predicting energy generation patterns and adjusting network operations to match the availability of renewable energy. 

\vspace{3 mm}
\noindent \textbullet  \textbf{ \textit{AI for Ultra-Low Latency Communications}}

Ultra-low latency communication is a key aspect of 6G networks~\cite{8636206}.
AI is poised to play a pivotal role in achieving this goal by providing real-time data processing and decision-making capabilities.
Future research is expected to develop AI systems that can operate at the network edge, providing near-instantaneous response time for critical applications. This includes creating lightweight AI models that can be deployed on edge devices with limited computational resources. One challenge is the need for AI algorithms that can make decisions based on incomplete or uncertain data. The use of techniques, such as probabilistic modeling and approximate computing, is a promising direction to enable AI systems to operate effectively under these conditions. Another challenge is the development of AI-driven network orchestration tools that can allocate computational resources in real-time. These tools are expected to predict the computational needs of applications and allocate resources accordingly to ensure latency.

\vspace{3 mm}
\noindent \textbullet  \textbf{ \textit{AI-Enhanced Cybersecurity in 6G}}

AI has the potential to enhance the cybersecurity of 6G networks by providing advanced threat detection and response capabilities~\cite{10706120}. Future research can look into developing AI-based security frameworks that can identify and respond to cyber threats in real-time. This involves using ML algorithms to analyze network traffic and detect anomalies that may indicate a security breach.
One challenge is the need for AI systems that can adapt to the constantly evolving landscape of cyber threats. Efforts can be directed to explore adaptive learning algorithms that can update their models in response to new types of attacks. Another challenge is developing AI-driven security protocols that can automatically patch vulnerabilities and respond to incidents without human intervention~\cite{10707443}. This requires the creation of intelligent systems that can understand the context of security incidents and take appropriate actions. 

\vspace{3 mm}
\noindent \textbullet  \textbf{ \textit{AI-Driven 6G Innovations}}

The incorporation of AI within 6G networks 
can potentially drive technological innovations, 
including intelligent reflecting surface (IRS) \cite{WuQIRS}, ISAC, O-RAN,  non-terrestrial network (NTN), and near-field communication \cite{2024Intelligent}. 
An IRS can dynamically augment wireless environments to optimize signal propagation, where AI can facilitate high-dimensional channel estimation and beamforming.
ISAC merge sensing and communication functions to improve spectrum utilization and provide environmental context for network optimization, where AI can contribute to joint waveform design and resource allocation for
seamless ISAC integration \cite{GuangxuAI} and AI-based real-time data fusion can be essential to extract actionable insights from ISAC systems. 
ORAN can benefit from the AI-based network management of its software-defined architecture, enhancing adaptability and resilience. NTN, including satellite systems, can leverage AI for efficient resource management and fault detection.
Interesting research directions can be AI models tailored to NTNs and addressing challenges like latency and dynamic network topology. 
By advancing AI applications in these novel areas, researchers can unlock new levels of performance and efficiency in 6G networks.

\vspace{3 mm}
{\noindent \textbullet  \textbf{ \textit{Ubiquitous Computing}}

With the proliferation of intelligent devices and the advancement of IoT technologies, the demand for intelligent services from users and devices continues to grow across wide areas. The integration of 6G and AI aims to provide AI services to users and devices worldwide, offering ubiquitous computing and meeting diverse intelligent service needs. By deploying edge servers on unmanned aerial vehicles (UAV)~\cite{10246260} or LEO satellites, 
multi-layered and heterogeneous computing services can be delivered,
satisfying their differentiated QoS requirements. The primary challenge lies in designing rational, efficient, and dynamic resource allocation schemes for heterogeneous networks that integrate terrestrial BSs, UAVs, and LEO satellites. Compared to classical theories, such as convex optimization and Lyapunov optimization, particular emphasis needs to be placed on advanced algorithms like DRL to explore computation offloading strategies and resource allocation designs.}

\vspace{3 mm}
{{\noindent \textbullet  \textbf{ \textit{LLM-Driven Cognitive Networking for 6G}}

The rise of low-cost open-source LLMs, such as DeepSeek \cite{DeepSeekV3}, is set to drastically lower barriers to AI-driven innovation. 
These accessible frameworks enable global researchers and organizations to rapidly prototype and deploy lightweight LLMs for 6G’s mission-critical applications. 
The integration of LLMs with 6G networks is poised to revolutionize cognitive networking by enabling adaptive resource allocation~\cite{10742580}, intent-based automation~\cite{10285423}, embodied AI~\cite{9687596}, and semantic communication~\cite{10559618}. Future research aims to embed LLMs into edge devices through lightweight architectures for real-time network optimization, while addressing challenges such as computational overhead and privacy risks. Key directions include federated learning for distributed LLM training, cross-modal semantic encoding to reduce bandwidth consumption, and DT-aided network simulation. Innovations in sparsity-aware inference and hybrid quantum-classical frameworks may unlock scalable deployment across 6G infrastructures. However, balancing model compression with performance retention and ensuring robust decision-making under dynamic conditions remain critical hurdles.}}

\section{Conclusions}\label{Section_conclusion}
In this all-encompassing review, we have initially highlighted that 6G network evolution towards an integrated wireless infrastructure platform amalgamating communication, sensing, computation, intelligence, and storage hinges on the profound integration of AI and communications. 
Concurrently, the exposition of the three-stage integration of AI within 6G networks, namely “AI4NET,” “NET4AI,” and “AIaaS,” furnishes a lucid blueprint for understanding the technological development trajectory. These three consecutive stages present the strategic shift from utilizing AI to optimize the network, through the network's facilitation of AI support to the network offering AIaaS, thereby underlining the interdependent and coordinated relationship between communication and AI. The 6G networks will introduce entirely new service paradigms, seamlessly integrating AI capabilities into various application scenarios, thereby serving as a powerful platform to support the advancement of AI.
Wireless network large models are poised to serve as pivotal drivers in 6G systems, effectively integrating AI with communication technologies to advance intelligent, autonomous, and highly efficient network operations.
By analyzing the standardization process of AI for wireless networks, we have highlighted the crucial milestones and ongoing efforts. 
We have further explored the challenges faced by the integration of AI and communications in the 6G era, including sustainability considerations, real-time AI demands, security and privacy issues, as well as human-AI interaction and collaboration. 
Finally, we have outlined promising future research opportunities that are expected to advance the development and optimization of AI and 6G communications. 
We are convinced that this review holds profound significance for the research on the integration of AI and communication for the 6G network, and it can provide deep insights for academic researchers and industry experts alike.

\Supplements{Appendix A.}
\bibliographystyle{scis_references}
\bibliography{references}

\begin{thebibliography}{100}
\expandafter\ifx\csname url\endcsname\relax
  \def\url#1{{\tt #1}}\fi
\expandafter\ifx\csname urlprefix\endcsname\relax\def\urlprefix{URL }\fi
\expandafter\ifx\csname eprint\endcsname\relax\def\eprint#1{\url{#1}}\fi

\bibitem{7414384}
Agiwal M, Roy A, Saxena N.
\newblock Next generation {5G} wireless networks: A comprehensive survey.
\newblock {IEEE} Commun. Surv. Tut., 2016, 18: 1617--1655

\bibitem{9335927}
Bhat J~R, Alqahtani S~A.
\newblock {6G} ecosystem: Current status and future perspective.
\newblock IEEE Access, 2021, 9: 43134--43167

\bibitem{9237460}
Yang H, Alphones A, Xiong Z, et~al.
\newblock Artificial-intelligence-enabled intelligent {6G} networks.
\newblock {IEEE} Netw., 2020, 34: 272--280

\bibitem{2017A}
Behera R, Das K.
\newblock A survey on machine learning: Concept, algorithms and applications.
\newblock International Journal of Innovative Research in Computer and Communication Engineering, 2017, 2: 2

\bibitem{GUO201627}
Guo Y, Liu Y, Oerlemans A, et~al.
\newblock Deep learning for visual understanding: A review.
\newblock Neurocomputing, 2016, 187: 27--48

\bibitem{chowdhary2020natural}
Morgan D~P, Scofield C~L.
\newblock Natural language processing, Springer US, Boston, MA.
\newblock 1991, 245--288

\bibitem{9023918}
Wang C~X, Renzo M~D, Stanczak S, et~al.
\newblock Artificial intelligence enabled wireless networking for {5G} and beyond: Recent advances and future challenges.
\newblock IEEE Wireless Commun., 2020, 27: 16--23

\bibitem{10208153}
Li K, Lau B~P~L, Yuan X, et~al.
\newblock Toward ubiquitous semantic metaverse: Challenges, approaches, and opportunities.
\newblock {IEEE} Internet Things J., 2023, 10: 21855--21872

\bibitem{10118940}
Zheng J, Li K, Mhaisen N, et~al.
\newblock Federated learning for online resource allocation in mobile edge computing: A deep reinforcement learning approach.
\newblock In: Proceedings of IEEE Wirel. Commun. Netw. Conf. (WCNC), Glasgow, United Kingdom, 2023. 1--6

\bibitem{wong2024real}
Wong M~L, Arjunan T.
\newblock Real-time detection of network traffic anomalies in big data environments using deep learning models.
\newblock Emerging Trends in Machine Intelligence and Big Data, 2024, 16: 1--11

\bibitem{9023920}
Cui Q, Ni W, Li S, et~al.
\newblock Learning-assisted clustered access of {5G/B5G} networks to unlicensed spectrum.
\newblock {IEEE} Commun. Mag., 2020, 27: 31--37

\bibitem{10269761}
Guan X, Xu Z, Liu Y, et~al.
\newblock Reduction in energy consumption of the 5{G} communication system and beyond through collaborative optimization for bs site operation: Challenges, efforts and the new approach.
\newblock {IEEE} Trans. Ind. Inform., 2024, 20: 3948--3963

\bibitem{9652477}
Zhang J, Cui Q, Zhang X, et~al.
\newblock Online optimization of energy-efficient user association and workload offloading for mobile edge computing.
\newblock {IEEE} Trans. Veh. Technol., 2022, 71: 1974--1988

\bibitem{9059015}
Cui Q, Zhang J, Zhang X, et~al.
\newblock Online anticipatory proactive network association in mobile edge computing for {IoT}.
\newblock {IEEE} Trans. Wireless Commun., 2020, 19: 4519--4534

\bibitem{8663994}
Cui Q, Gong Z, Ni W, et~al.
\newblock Stochastic online learning for mobile edge computing: Learning from changes.
\newblock {IEEE} Commun. Mag., 2019, 57: 63--69

\bibitem{wang2022high}
Wang J, Jiang C, Kuang L.
\newblock High-mobility satellite-uav communications: Challenges, solutions, and future research trends.
\newblock {IEEE} Commun. Mag., 2022, 60: 38--43

\bibitem{Security}
Van Den~Broek F, Verdult R, De~Ruiter J.
\newblock Defeating {IMSI} catchers.
\newblock In: Proceedings of 22Nd ACM SIGSAC Conference on Computer and Communications Security, New York, NY, USA, 2015. 340--351

\bibitem{9999559}
Saleem R, Ni W, Ikram M, et~al.
\newblock Deep-reinforcement-learning-driven secrecy design for intelligent-reflecting-surface-based {6G-IoT} networks.
\newblock {IEEE} Internet Things J., 2023, 10: 8812--8824

\bibitem{CustomizedSecurity}
Dutta A, Hammad E.
\newblock {5G} security challenges and opportunities: A system approach.
\newblock In: Proceedings of IEEE 3rd 5G world forum (5GWF), Bangalore, India, 2020. 109--114

\bibitem{9239911}
Lyu X, Ren C, Ni W, et~al.
\newblock Online learning of optimal proactive schedule based on outdated knowledge for energy harvesting powered internet-of-things.
\newblock {IEEE} Trans. Wireless Commun., 2021, 20: 1248--1262

\bibitem{ITU-R}
{ITU-Rec M2160, Int Telecommun Union, Geneva}.
\newblock Framework and overall objectives of the future development of {IMT} for 2030 and beyond.
\newblock 2023.
\newblock \eprint{https://www.itu.int/rec/R-REC-M.2160-0-202311-I/en}

\bibitem{10540053}
Gong Z, Hashash O, Wang Y, et~al.
\newblock {UAV}-aided lifelong learning for {AoI} and energy optimization in nonstationary {IoT} networks.
\newblock {IEEE} Internet Things J., 2024, 11: 39206--39224

\bibitem{10001188}
Li X, Cui Q, Xue Q, et~al.
\newblock A new batch access scheme with global {QoS} optimization for satellite-terrestrial networks.
\newblock In: Proceedings of IEEE Global Commun. Conf. (GLOBECOM), Rio de Janeiro, Brazil, 2022. 3929--3934

\bibitem{CuiBigData}
Cui Q, Zhang X, Ni W, et~al.
\newblock Big data analytics for intelligent management of autonomous vehicles in smart cities, The Institution of Engineering and Technology, chap.~9.
\newblock 2021, 201--229.
\newblock \eprint{https://digital-library.theiet.org/doi/pdf/10.1049/PBTE090E_ch9}

\bibitem{8473689}
Cui Q, Wang Y, Chen K~C, et~al.
\newblock Big data analytics and network calculus enabling intelligent management of autonomous vehicles in a smart city.
\newblock {IEEE} Internet Things J., 2019, 6: 2021--2034

\bibitem{reinsel2018data}
Reinsel D, Gantz J, Rydning J.
\newblock Data age 2025: the digitization of the world from edge to core.
\newblock IDC white paper, 2018.
\newblock \eprint{https://www.seagate.com/files/www-content/our-story/trends/files/idc-seagate-dataage-whitepaper.pdf?gid=164649}

\bibitem{zhinengfenlei}
Yue L, Chen T.
\newblock {AI} large model and {6G} network.
\newblock In: Proceedings of IEEE Global Commun. Conf. Workshops (GC Wkshps), Kuala Lumpur, Malaysia, 2023. 2049--2054

\bibitem{tong2022nine}
Tong W, Li G~Y.
\newblock Nine challenges in artificial intelligence and wireless communications for {6G}.
\newblock {IEEE} Wireless Commun., 2022, 29: 140--145

\bibitem{tao2023wireless}
Tao Z, Xu W, Huang Y, et~al.
\newblock Wireless network digital twin for {6G}: Generative {AI} as a key enabler.
\newblock IEEE Wireless Commun., 2024, 31: 24--31

\bibitem{IMT2030-6gnetwork}
{IMT-2030 (6G) Promotion Group}.
\newblock 6{G} network architecture outlook white paper.
\newblock 2023.
\newblock \eprint{https://www.imt2030.org.cn/html/default/zhongwen/chengguofabu/baipishu/index.html?index=2}

\bibitem{yang2022kubeedge}
Yang T, Ning J, Lan D, et~al.
\newblock Kubeedge wireless for integrated communication and computing services everywhere.
\newblock {IEEE} Wireless Commun., 2022, 29: 140--145

\bibitem{UserExperience}
Chen X, Guo Z, Wang X, et~al.
\newblock Foundation model based native {AI} framework in {6G} with cloud-edge-end collaboration.
\newblock 2023.
\newblock \eprint{https://arxiv.org/pdf/2310.17471}

\bibitem{ZJU-6g}
Wu J, Li R, An X, et~al.
\newblock Toward native artificial intelligence in 6{G} networks: System design, architectures, and paradigms.
\newblock 2021.
\newblock \eprint{https://arxiv.org/pdf/2103.02823}

\bibitem{LLMFanhua}
Liu G, Deng J, Zheng Q, et~al.
\newblock {6G} native intelligence: Technical challenges architecture and key features.
\newblock Mobile Communications, 2021, 45: 68--78

\bibitem{cao2023comprehensive}
Cao Y, Li S, Liu Y, et~al.
\newblock A comprehensive survey of {AI}-generated content {(AIGC)}: A history of generative {AI} from gan to chatgpt.
\newblock 2023.
\newblock \eprint{https://arxiv.org/pdf/2303.04226}

\bibitem{6GLLM}
Maatouk A, Piovesan N, Ayed F, et~al.
\newblock Large language models for telecom: Forthcoming impact on the industry.
\newblock 2023.
\newblock \eprint{https://arxiv.org/abs/2308.06013}

\bibitem{10558819}
Jiang F, Peng Y, Dong L, et~al.
\newblock Large {AI} model-based semantic communications.
\newblock {IEEE} Wireless Commun., 2024, 31: 68--75

\bibitem{deepseekai2025}
DeepSeek-AI, Guo D, Yang D, et~al.
\newblock Deepseek-{R1}: Incentivizing reasoning capability in {LLMs} via reinforcement learning.
\newblock 2025.
\newblock \eprint{https://arxiv.org/abs/2501.12948}

\bibitem{AI4Net}
{6G Alliance of Network AI (6GANA)}.
\newblock {6G} network {AI} concept and terminology white paper.
\newblock 2022.
\newblock \eprint{https://www.6g-ana.com/upload/file/20220523/6378893017578935554735081.pdf}

\bibitem{10247266}
Soury H, Smida B, Aliabadi S~M.
\newblock Accurate {MMSE} expressions for short-packet pilot-based channel estimation.
\newblock {IEEE} Wireless Commun. Lett., 2023, 12: 2188--2192

\bibitem{8308193}
Farzamnia A, Hlaing N~W, Haldar M~K, et~al.
\newblock Channel estimation for sparse channel {OFDM} systems using least square and minimum mean square error techniques.
\newblock In: Proceedings of International Conference on Engineering and Technology (ICET), Antalya, Turkey, 2017. 1--5

\bibitem{8934725}
Li J, Zhang Q, Xin X, et~al.
\newblock Deep learning-based massive {MIMO CSI} feedback.
\newblock In: Proceedings of 18th International Conference on Optical Communications and Networks (ICOCN), Huangshan, China, 2019. 1--3

\bibitem{Huawei}
{Huawei Technologies Co, Ltd}.
\newblock {R1-2304653: Evaluation on AI/ML for CSI feedback enhancement}.
\newblock 2023.
\newblock \eprint{https://www.3gpp.org/ftp/tsg_ran/WG1_RL1/TSGR1_113/Docs}

\bibitem{8972904}
Guo J, Wen C~K, Jin S, et~al.
\newblock Convolutional neural network-based multiple-rate compressive sensing for massive {MIMO CSI} feedback: Design, simulation, and analysis.
\newblock {IEEE} Trans. Wireless Commun., 2020, 19: 2827--2840

\bibitem{9466243}
Zhang Y, Zhang X, Liu Y.
\newblock Deep learning based {CSI} compression and quantization with high compression ratios in {FDD} massive {MIMO} systems.
\newblock {IEEE} Wireless Commun. Lett., 2021, 10: 2101--2105

\bibitem{9625585}
Wang Y, Sun J, Wang J, et~al.
\newblock Multi-rate compression for downlink {CSI} based on transfer learning in {FDD} massive {MIMO} systems.
\newblock In: Proceedings of IEEE 94th Vehicular Technology Conference (VTC2021-Fall), Norman, OK, USA, 2021. 1--5

\bibitem{9178295}
Yu X, Li X, Wu H, et~al.
\newblock {DS-NLCsiNet}: Exploiting non-local neural networks for massive {MIMO CSI} feedback.
\newblock {IEEE} Commun. Lett., 2020, 24: 2790--2794

\bibitem{9149229}
Lu Z, Wang J, Song J.
\newblock Multi-resolution {CSI} feedback with deep learning in massive {MIMO} system.
\newblock In: Proceedings of IEEE International Commun. Conf. (ICC), Dublin, Ireland, 2020. 1--6

\bibitem{8509622}
Gao X, Jin S, Wen C~K, et~al.
\newblock {ComNet}: Combination of deep learning and expert knowledge in {OFDM} receivers.
\newblock {IEEE} Commun. Lett., 2018, 22: 2627--2630

\bibitem{9723316}
Pihlajasalo J, Korpi D, Honkala M, et~al.
\newblock Deep learning based {OFDM} physical-layer receiver for extreme mobility.
\newblock In: Proceedings of 55th Asilomar Conference on Signals, Systems, and Computers, Pacific Grove, CA, USA, 2021. 395--399

\bibitem{9287725}
Mendonça M~O~K, Diniz P~S~R.
\newblock {OFDM} receiver using deep learning: Redundancy issues.
\newblock In: Proceedings of 28th European Signal Processing Conference (EUSIPCO), Amsterdam, Netherlands, 2021. 1687--1691

\bibitem{9446039}
Chen W, Tang Z.
\newblock Research on improved receiver of {NOMA-OFDM} signal based on deep learning.
\newblock In: Proceedings of International Conference on Communications, Information System and Computer Engineering (CISCE), Beijing, China, 2021. 173--177

\bibitem{8663458}
Balevi E, Andrews J~G.
\newblock One-bit {OFDM} receivers via deep learning.
\newblock {IEEE} Trans. Commun., 2019, 67: 4326--4336

\bibitem{9637770}
Wang B, Xu K, Song P, et~al.
\newblock A deep learning-based intelligent receiver for {OFDM}.
\newblock In: Proceedings of IEEE 18th International Conference on Mobile Ad Hoc and Smart Systems (MASS), Denver, CO, USA, 2021. 562--563

\bibitem{9526047}
An Y, Wu Z, Chunyang T.
\newblock Performance of {OFDM} system receiver based on deep learning.
\newblock In: Proceedings of International Conference on Intelligent Transportation, Big Data \& Smart City (ICITBS), Xi'an, China, 2021. 571--574

\bibitem{8023460}
Hussain M, Michelusi N.
\newblock Throughput optimal beam alignment in millimeter wave networks.
\newblock In: Proceedings of Information Theory and Applications Workshop (ITA), San Diego, CA, USA, 2017. 1--6

\bibitem{8444984}
Zhu D, Choi J, Cheng Q, et~al.
\newblock High-resolution angle tracking for mobile wideband millimeter-wave systems with antenna array calibration.
\newblock {IEEE} Trans. Wireless Commun., 2018, 17: 7173--7189

\bibitem{7460513}
Giordani M, Mezzavilla M, Barati C~N, et~al.
\newblock Comparative analysis of initial access techniques in {5G} mmwave cellular networks.
\newblock In: Proceedings of Annual Conference on Information Science and Systems (CISS), Princeton, NJ, USA, 2016. 268--273

\bibitem{9665388}
Cui Q, Zhang Z, Shi Y, et~al.
\newblock Dynamic multichannel access based on deep reinforcement learning in distributed wireless networks.
\newblock {IEEE} Systems J., 2022, 16: 5831--5834

\bibitem{heng2021machine}
Heng Y, Andrews J~G.
\newblock Machine learning-assisted beam alignment for mm{W}ave systems.
\newblock {IEEE} Trans. Cogn. Commun. Netw., 2021, 7: 1142--1155

\bibitem{yang2023hierarchical}
Yang J, Zhu W, Tao M, et~al.
\newblock Hierarchical beam alignment for millimeter-wave communication systems: A deep learning approach.
\newblock {IEEE} Trans. Wireless Commun., 2024, 23: 3541--3556

\bibitem{9690703}
Heng Y, Mo J, Andrews J~G.
\newblock Learning site-specific probing beams for fast mmwave beam alignment.
\newblock {IEEE} Trans. Wireless Commun., 2022, 21: 5785--5800

\bibitem{sayed2003wireless}
Sayed A~H, Yousef N~R.
\newblock Wireless location.
\newblock New York: Wiley, 2003

\bibitem{luan20226g}
Luan N, Xiong K, Zhang Y, et~al.
\newblock {6G}: Typical applications, key technologies and challenges.
\newblock Chin. J. Internet Things, 2022, 6: 29--42

\bibitem{maung2020enhanced}
Maung N~A~M, Lwi B~Y, Thida S.
\newblock An enhanced {RSS} fingerprinting-based wireless indoor positioning using random forest classifier.
\newblock In: Proceedings of International Conference on Advanced Information Technologies (ICAIT), Yangon, Myanmar, 2020. 59--63

\bibitem{shan2018outdoor}
Shan L, et~al.
\newblock Outdoor positioning technology based on telecom data.
\newblock Computer Engineering \& Science, 2018, 40: 649

\bibitem{zhang2022indoor}
Zhang L, Wang B, Xu Y, et~al.
\newblock Indoor location method based on {RSSI} probability distribution and {CSI} modified model.
\newblock In: Proceedings of Global Conference on Robotics, Artificial Intelligence and Information Technology (GCRAIT), Chicago, IL, USA, 2022. 429--433

\bibitem{10505907}
Wang Q, Zou S, Sun Y, et~al.
\newblock Toward intelligent and adaptive task scheduling for {6G}: An intent-driven framework.
\newblock {IEEE} Trans. Cogn. Commun. Netw., 2024, 10: 1975--1988

\bibitem{2020A}
Saad W, Bennis M, Chen M.
\newblock A vision of 6{G} wireless systems: Applications, trends, technologies, and open research problems.
\newblock {IEEE} Netw., 2020, 34: 134--142

\bibitem{2017Spatiotemporal}
Wang J, Tang J, Xu Z, et~al.
\newblock Spatiotemporal modeling and prediction in cellular networks: A big data enabled deep learning approach.
\newblock In: Proceedings of IEEE Conference on Computer Communications, Atlanta, GA, USA, 2017. 1--9

\bibitem{2017Network}
Nie L, Jiang D, Yu S, et~al.
\newblock Network traffic prediction based on deep belief network in wireless mesh backbone networks.
\newblock In: Proceedings of IEEE Wireless Communications and Networking Conference (WCNC), San Francisco, CA, USA, 2017. 1--5

\bibitem{2020Cellular}
Jaffry S.
\newblock Cellular traffic prediction with recurrent neural network.
\newblock 2020.
\newblock \eprint{https://arxiv.org/pdf/2003.02807}

\bibitem{zhang2021dual}
Zhang C, Dang S, Shihada B, et~al.
\newblock Dual attention-based federated learning for wireless traffic prediction, vancouver, bc, canada. 2021.
\newblock In: Proceedings of IEEE Conference on Computer Communications. 1--10

\bibitem{zhang2022efficient}
Zhang L, Zhang C, Shihada B.
\newblock Efficient wireless traffic prediction at the edge: A federated meta-learning approach.
\newblock {IEEE} Commun. Lett., 2022, 26: 1573--1577

\bibitem{2019Deep}
Zhang C, Zhang H, Qiao J, et~al.
\newblock Deep transfer learning for intelligent cellular traffic prediction based on cross-domain big data.
\newblock {IEEE} J. Sel. Areas Commun., 2019, 37: 1389--1401

\bibitem{zhao2018federated}
Zhao Y, Li M, Lai L, et~al.
\newblock Federated learning with {Non-IID} data.
\newblock 2018.
\newblock \eprint{https://arxiv.org/pdf/1806.00582}

\bibitem{2017Model}
Finn C, Abbeel P, Levine S.
\newblock Model-agnostic meta-learning for fast adaptation of deep networks.
\newblock In: Proceedings of International Conference on Machine Learning, Sydney, NSW, Australia, 2017. 1126–1135

\bibitem{9215362}
Inamdar M~A, Kumaraswamy H~V.
\newblock Energy efficient 5{G} networks: Techniques and challenges.
\newblock In: Proceedings of International Conference on Smart Electronics and Communication (ICOSEC), Trichy, India, 2020. 1317--1322

\bibitem{Jinlin2014Stochastic}
Jinlin, Peng, Peilin, et~al.
\newblock Stochastic analysis of optimal base station energy saving in cellular networks with sleep mode.
\newblock {IEEE} Commun. Lett., 2014, 18: 612--615

\bibitem{2015AE}
Debaillie B, Desset C, Louagie F.
\newblock A flexible and future-proof power model for cellular base stations.
\newblock In: Proceedings of IEEE Vehicular Technology Conference (VTC Spring), Glasgow, UK, 2015. 1--7

\bibitem{li2023carbon}
Li T, Yu L, Ma Y, et~al.
\newblock Carbon emissions of 5{G} mobile networks in china.
\newblock Nature Sustainability, 2023, 6: 1620--1631

\bibitem{2019Reinforcement}
Salem F~E, Altman Z, Gati A, et~al.
\newblock Reinforcement learning approach for advanced sleep modes management in 5{G} networks.
\newblock In: Proceedings of IEEE Vehicular Technology Conference (VTC-Fall), Chicago, IL, USA, 2018. 1--5

\bibitem{masoudi2020reinforcement}
Masoudi M, Khafagy M~G, Soroush E, et~al.
\newblock Reinforcement learning for traffic-adaptive sleep mode management in 5{G} networks.
\newblock In: Proceedings of IEEE 31st Annual International Symposium on Personal, Indoor and Mobile Radio Communications, London, UK, 2020. 1--6

\bibitem{lin2022dades}
Lin S, Qiu C, Tan J, et~al.
\newblock {DADEs}: 5{G} dual-adaptive delay-aware and energy-saving system with tandem learning.
\newblock In: Proceedings of IEEE Global Commun. Conf. (GLOBECOM), Rio de Janeiro, Brazil. 2022. 1--6

\bibitem{Mchenry2007XG}
McHenry M, Livsics E, Nguyen T, et~al.
\newblock {XG} dynamic spectrum access field test results [topics in radio communications].
\newblock {IEEE} Commun. Mag., 2007, 45: 51--57

\bibitem{kaur2022comprehensive}
Kaur A, Kumar K.
\newblock A comprehensive survey on machine learning approaches for dynamic spectrum access in cognitive radio networks.
\newblock Journal of Experimental \& Theoretical Artificial Intelligence, 2022, 34: 1--40

\bibitem{alabi2023artificial}
Alabi C~A, Imoize A~L, Giwa M~A, et~al.
\newblock Artificial intelligence in spectrum management: Policy and regulatory considerations.
\newblock In: Proceedings of International Conference on Multidisciplinary Engineering and Applied Science (ICMEAS), Abuja, Nigeria, 2023. vol.~1, 1--6

\bibitem{data2020machine}
Liu Y, Bi S, Shi Z, et~al.
\newblock When machine learning meets big data: A wireless communication perspective.
\newblock {IEEE} Veh. Technol. Mag., 2020, 15: 63--72

\bibitem{6666502}
Chernogorov F, Puttonen J.
\newblock User satisfaction classification for minimization of drive tests {QoS} verification.
\newblock In: Proceedings of IEEE 24th Annual International Symposium on Personal, Indoor, and Mobile Radio Communications (PIMRC), London, UK, 2013. 2165--2169

\bibitem{10155734}
Luo Z~Q, Zheng X, López-Pérez D, et~al.
\newblock {SRCON}: A data-driven network performance simulator for real-world wireless networks.
\newblock {IEEE} Commun. Mag., 2023, 61: 96--102

\bibitem{8382166}
Mao Q, Hu F, Hao Q.
\newblock Deep learning for intelligent wireless networks: A comprehensive survey.
\newblock {IEEE} Commun. Surv. Tut., 2018, 20: 2595--2621

\bibitem{you2014development}
You X, Pan Z, Gao X, et~al.
\newblock Development trend and some key technologies of {5G} mobile communication.
\newblock Sci. China Inf. Sci., 2014, 44: 551--563

\bibitem{10332666}
Tu Y~H, Ma Y~W, Li Z~X, et~al.
\newblock Applying deep reinforcement learning for self-organizing network architecture.
\newblock In: Proceedings of IEEE 6th International Conference on Knowledge Innovation and Invention (ICKII), Sapporo, Japan, 2023. 16--20

\bibitem{wang2021research}
Wang W, Duan X, Sun W, et~al.
\newblock Research on mobility prediction in 5{G} and beyond for vertical industries.
\newblock In: Proceedings of IEEE/CIC International Conference on Communications in China (ICCC Workshops), Xiamen, China, 2021. 379--383

\bibitem{shubyn2020deep}
Shubyn B, Lutsiv N, Syrotynskyi O, et~al.
\newblock Deep learning based adaptive handover optimization for ultra-dense 5{G} mobile networks.
\newblock In: Proceedings of International Conference on Advanced Trends in Radioelectronics, Telecommunications and Computer Engineering (TCSET), Lviv-Slavske, Ukraine, 2020. 869--872

\bibitem{liu2020proactive}
Liu Q, Chuai G, Wang J, et~al.
\newblock Proactive mobility management with trajectory prediction based on virtual cells in ultra-dense networks.
\newblock {IEEE} Trans. Veh. Technol., 2020, 69: 8832--8842

\bibitem{prado2023enabling}
Prado A, St{\"o}ckeler F, Mehmeti F, et~al.
\newblock Enabling proportionally-fair mobility management with reinforcement learning in 5{G} networks.
\newblock {IEEE} J. Sel. Areas Commun., 2023, 41: 1845--1858

\bibitem{yan2019machine}
Yan L, Ding H, Zhang L, et~al.
\newblock Machine learning-based handovers for sub-6 {G}{H}z and mmwave integrated vehicular networks.
\newblock {IEEE} Trans. Wireless Commun., 2019, 18: 4873--4885

\bibitem{10121779}
Awad~Abdellatif A, Abo-Eleneen A, Mohamed A, et~al.
\newblock Intelligent-slicing: An {AI}-assisted network slicing framework for {5G}-and-beyond networks.
\newblock {IEEE} Trans. Netw. Service Manag., 2023, 20: 1024--1039

\bibitem{9998551}
Qin X, Li Y, Song X, et~al.
\newblock Timeliness of information for computation-intensive status updates in task-oriented communications.
\newblock {IEEE} J. Sel. Areas Commun., 2023, 41: 623--638

\bibitem{10003251}
Wang Y, Chen K~C, Gong Z, et~al.
\newblock Reliability-guaranteed uplink resource management in proactive mobile network for minimal latency communications.
\newblock {IEEE} Trans. Wireless Commun., 2023, 22: 5018--5030

\bibitem{9858867}
Zhang P, Su Y, Wang J, et~al.
\newblock Reinforcement learning assisted bandwidth aware virtual network resource allocation.
\newblock {IEEE} Trans. Netw. Service Manag., 2022, 19: 4111--4123

\bibitem{10323248}
Zhang P, Chen N, Xu G, et~al.
\newblock Multi-target-aware dynamic resource scheduling for cloud-fog-edge multi-tier computing network.
\newblock {IEEE} Trans. Intell. Transp. Syst., 2024, 25: 3885--3897

\bibitem{10330565}
Nguyen V~D, Vu T~X, Nguyen N~T, et~al.
\newblock Network-aided intelligent traffic steering in {6G O-RAN}: A multi-layer optimization framework.
\newblock {IEEE} J. Sel. Areas Commun., 2024, 42: 389--405

\bibitem{9380677}
Sami H, Otrok H, Bentahar J, et~al.
\newblock {AI}-based resource provisioning of {IoE} services in {6G}: A deep reinforcement learning approach.
\newblock {IEEE} Trans. Netw. Service Manag., 2021, 18: 3527--3540

\bibitem{9655323}
Dong T, Zhuang Z, Qi Q, et~al.
\newblock Intelligent joint network slicing and routing via {GCN}-powered multi-task deep reinforcement learning.
\newblock {IEEE} Trans. Cogn. Commun. Netw., 2022, 8: 1269--1286

\bibitem{9459763}
Mei J, Wang X, Zheng K, et~al.
\newblock Intelligent radio access network slicing for service provisioning in {6G}: A hierarchical deep reinforcement learning approach.
\newblock {IEEE} Trans. Commun., 2021, 69: 6063--6078

\bibitem{9810019}
Bhattacharya P, Patel F, Alabdulatif A, et~al.
\newblock A deep-{Q} learning scheme for secure spectrum allocation and resource management in {6G} environment.
\newblock {IEEE} Trans. Netw. Service Manag., 2022, 19: 4989--5005

\bibitem{9075574}
Fan J, Mu D, Liu Y.
\newblock Research on network traffic prediction model based on neural network.
\newblock In: Proceedings of 2nd International Conference on Information Systems and Computer Aided Education (ICISCAE), Dalian, China, 2019. 554--557

\bibitem{7587350}
Wu J, Ota K, Dong M, et~al.
\newblock Big data analysis-based security situational awareness for smart grid.
\newblock {IEEE} Trans. Big Data, 2018, 4: 408--417

\bibitem{9775698}
Yin K, Yang Y, Yao C, et~al.
\newblock Long-term prediction of network security situation through the use of the transformer-based model.
\newblock {IEEE} Access, 2022, 10: 56145--56157

\bibitem{8330210}
Xu Z, Liu J, Luo X, et~al.
\newblock Cross-version defect prediction via hybrid active learning with kernel principal component analysis.
\newblock In: Proceedings of IEEE 25th International Conference on Software Analysis, Evolution and Reengineering (SANER), Campobasso, Italy, 2018. 209--220

\bibitem{10541561}
Zhang Z, Li J.
\newblock Application of active learning strategies in fault detection and diagnosis within communication networks.
\newblock In: Proceedings of 7th World Conference on Computing and Communication Technologies (WCCCT), Chengdu, China, 2024. 186--193

\bibitem{10543904}
Iswarya P, Manikandan K.
\newblock Algorithms for fault detection and diagnosis in wireless sensor networks using deep learning and machine learning - an overview.
\newblock In: Proceedings of 10th International Conference on Communication and Signal Processing (ICCSP), Melmaruvathur, India, 2024. 1404--1409

\bibitem{ChinaMobile-6g}
Liu G, Zhang H, Tong Z, et~al.
\newblock {6G} mobile information network architecture: migrate from communication to {XaaS} (in {Chinese}).
\newblock Sci. Sin. Inform., 2024, 54: 1236--1266

\bibitem{ChinaTelecom-6g}
{China Telecom Research Institute}.
\newblock 6{G} vision and technology white paper.
\newblock 2022.

\bibitem{ChinaUnicom-6g}
{China Unicom Research Institute}.
\newblock China unicom 6{G} white paper.
\newblock 2021.

\bibitem{Huawei-task}
Wu J, Deng J, Peng C, et~al.
\newblock Task-centered {6G} network {AI} architecture.
\newblock Radio Commun. Tech., 2022, 48: 599--613

\bibitem{CICT-6gvision}
CICT, DTmobile.
\newblock Full coverage, scene intelligence-6{G} vision and technology trends white paper.
\newblock 2020.

\bibitem{CICT-6gnetwork}
CICT, {CICT Mobile}.
\newblock Full coverage, scene intelligence-{6G} network architecture white paper.
\newblock 2022.

\bibitem{OPPO-aicube}
{OPPO Research Institute}.
\newblock {6G} {AI}-cube intelligent network (white paper).
\newblock 2021.
\newblock \eprint{https://www.oppo.com/content/dam/oppo/cn/mkt/newsroom/press/455/whitepaper.pdf}

\bibitem{Ericsson-6g}
Ericsson.
\newblock 6{G}–connecting a cyber-physical world (white paper).
\newblock 2022.
\newblock \eprint{https://www.ericsson.com/en/reports-and-papers/white-papers/a-research-outlook-towards-6g}

\bibitem{Nokia-6g}
Hoydis J, Aoudia F~A, Valcarce A, et~al.
\newblock Toward a {6G AI}-native air interface.
\newblock {IEEE} Commun. Mag., 2021, 59: 76--81

\bibitem{B5GPC-6g}
Consortium B~G~P.
\newblock Beyond {5G} white paper - version 1.0.
\newblock 2022.
\newblock \eprint{https://b5g.jp/w/wp-content/uploads/pdf/whitepaper_en_1-0.pdf}

\bibitem{Zhang-6g}
Zhang P, Niu K, Tian H, et~al.
\newblock Technology prospect of 6{G} mobile communications.
\newblock J. Commun., 2019, 40: 141--148

\bibitem{Cui-6gran}
{Beijing University of Posts and Telecommunications}.
\newblock 6{G} green wireless access network white paper for multi-dimensional stereo and full scenarios.
\newblock 2023.

\bibitem{Southeast-6g}
{Southeast University}, {Purple Mountain Laboratories}.
\newblock 6{G} research white paper.
\newblock 2020.

\bibitem{UESTC-6g}
Zhang L, Liang Y~C, Niyato D.
\newblock 6{G} visions: Mobile ultra-broadband, super internet-of-things, and artificial intelligence.
\newblock China Commun., 2019, 16: 1--14

\bibitem{THU-6g}
Luo H, Zhang T, Zhao C, et~al.
\newblock Integrated sensing and communications framework for 6{G} networks.
\newblock 2024.
\newblock \eprint{https://arxiv.org/pdf/2405.19925}

\bibitem{SJTU-6g}
Tao M, Zhou Y, Shi Y, et~al.
\newblock Federated edge learning for {6G}: Foundations, methodologies, and applications.
\newblock Proceedings of the IEEE, Early Access, 2024.

\bibitem{CUHK-6g}
Yang Y, Ma M, Wu H, et~al.
\newblock {6G} network {AI} architecture for everyone-centric customized services.
\newblock {IEEE} Netw., 2022, 37: 71--80

\bibitem{Oslo-6g}
Lu Y, Maharjan S, Zhang Y.
\newblock Adaptive edge association for wireless digital twin networks in {6G}.
\newblock {IEEE} Internet Things J., 2021, 8: 16219--16230

\bibitem{Oulu-6g}
Taleb T, Aguiar R~L, Grida Ben~Yahia I, et~al.
\newblock White paper on 6{G} networking.
\newblock 2020.
\newblock \eprint{http://urn.fi/ urn:isbn:9789526226842}

\bibitem{liu2021-6gnative}
Liu G, Deng J, Zheng Q, et~al.
\newblock {6G} native intelligence: Technical challenges, architecture and key features.
\newblock Mobile Commun., 2021, 45: 68--78

\bibitem{liu2022-nativeai}
Liu G, Deng J, Li N, et~al.
\newblock Native {AI} and service based architecture for {6G} wireless network.
\newblock Radio Commun. Tech., 2022, 48: 562--573

\bibitem{22.261}
{3rd Generation Partership Project (3GPP)}.
\newblock Service requirements for the {5G} system.
\newblock TS 22.261.
\newblock \eprint{https://www.3gpp.org/ftp/Specs/archive/22_series/22.261}

\bibitem{22.874}
{3rd Generation Partership Project (3GPP)}.
\newblock Study on traffic characteristics and performance requirements for {AI}/{ML} model transfer in {5GS}.
\newblock TR 22.874.
\newblock \eprint{https://www.3gpp.org/ftp/Specs/archive/22_series/22.874}

\bibitem{6gdata}
{6G Alliance of Network AI (6GANA)}.
\newblock 6{G} data service concept and requirements white paper.
\newblock 2022.
\newblock \eprint{https://www.6g-ana.com/upload/file/20220523/6378893028172687311971492.pdf}

\bibitem{alkhateeb2019deepmimo}
Alkhateeb A.
\newblock Deep{MIMO}: A generic deep learning dataset for millimeter wave and massive {MIMO} applications.
\newblock 2019.
\newblock \eprint{https://arxiv.org/pdf/1902.06435}

\bibitem{oshea2016}
O'Shea T~J, Corgan J, Clancy T~C.
\newblock Convolutional radio modulation recognition networks.
\newblock 2016.
\newblock \eprint{https://arxiv.org/pdf/1602.04105}

\bibitem{AI/MLITU}
{International Telecommunication Union (ITU)}.
\newblock {AI/ML in 5G} challenge.
\newblock 2023.
\newblock \eprint{https://www.itu.int/en/ITU-T/AI/challenge}

\bibitem{9050836}
Lyu X, Ren C, Ni W, et~al.
\newblock Distributed online learning of cooperative caching in edge cloud.
\newblock {IEEE} Trans. Mobile Comput., 2021, 20: 2550--2562

\bibitem{security5}
Liu Y, Zhang W, Li L, et~al.
\newblock Toward autonomous trusted networks-from digital twin perspective.
\newblock {IEEE} Netw., 2024, 38: 84--91

\bibitem{security6}
{NGMN}.
\newblock 6{G} requirements and design considerations (white paper).
\newblock 2023.
\newblock \eprint{https://www.ngmn.org/wp-content/uploads/NGMN_6G_Requirements_and_Design_Considerations.pdf}

\bibitem{cui2021edge}
Cui Q, Zhu Z, Ni W, et~al.
\newblock Edge-intelligence-empowered, unified authentication and trust evaluation for heterogeneous beyond 5{G} systems.
\newblock {IEEE} Wireless Commun., 2021, 28: 78--85

\bibitem{li2022lightweight}
Li K, Cui Q, Zhu Z, et~al.
\newblock Lightweight, privacy-preserving handover authentication for integrated terrestrial-satellite networks.
\newblock In: Proceedings of IEEE International Commun. Conf. {(ICC)}, Seoul, South Korea, 2022. 25--31

\bibitem{9446488}
Hu S, Chen X, Ni W, et~al.
\newblock Distributed machine learning for wireless communication networks: Techniques, architectures, and applications.
\newblock {IEEE} Commun. Surv. Tut., 2021, 23: 1458--1493

\bibitem{6gorchestration}
{6G Alliance of Network AI (6GANA)}.
\newblock White paper on native {AI} technology requirements of {6G} network.
\newblock 2022.
\newblock \eprint{https://www.6g-ana.com/upload/file/20220523/6378893017730497434706068.pdf}

\bibitem{10073536}
Yuan X, Ni W, Ding M, et~al.
\newblock Amplitude-varying perturbation for balancing privacy and utility in federated learning.
\newblock IEEE Trans. Info. Forensics Secur., 2023, 18: 1884--1897

\bibitem{10838599}
Zhao X, Cui Q, Li W, et~al.
\newblock Convergence-privacy-fairness trade-off in personalized federated learning.
\newblock IEEE Trans. Mach. Learn. Commun. Netw., 2025, 3: 246--262

\bibitem{XIAO2024}
Xiao B, Yu X, Ni W, et~al.
\newblock Over-the-air federated learning: Status quo, open challenges, and future directions.
\newblock Fundamental Research, 2024

\bibitem{10145450}
Yu X, Xiao B, Ni W, et~al.
\newblock Optimal adaptive power control for over-the-air federated edge learning under fading channels.
\newblock {IEEE} Trans. Commun., 2023, 71: 5199--5213

\bibitem{10649032}
Sun P, Liu E, Ni W, et~al.
\newblock Reconfigurable intelligent surface-assisted wireless federated learning with imperfect aggregation.
\newblock IEEE Trans. Commun., 2024, to appear: 1--1

\bibitem{10542235}
Li W, Lv T, Ni W, et~al.
\newblock Decentralized federated learning over imperfect communication channels.
\newblock IEEE Trans. Commun., 2024, 72: 6973--6991

\bibitem{li2019convergence}
Li X, Huang K, Yang W, et~al.
\newblock On the convergence of fedavg on non-{IID} data.
\newblock 2019.
\newblock \eprint{https://arxiv.org/abs/1907.02189}

\bibitem{9796935}
Luo B, Xiao W, Wang S, et~al.
\newblock Tackling system and statistical heterogeneity for federated learning with adaptive client sampling.
\newblock In: Proceedings of IEEE Conference on Computer Communications, London, United Kingdom, 2022. 1739--1748

\bibitem{10229070}
Wang S, Perazzone J, Ji M, et~al.
\newblock Federated learning with flexible control.
\newblock In: Proceedings of IEEE Conference on Computer Communications, New York City, NY, USA, 2023. 1--10

\bibitem{9745426}
Zhao B, Cui Q, Liang S, et~al.
\newblock Green concerns in federated learning over 6{G}.
\newblock China Commun., 2022, 19: 50--69

\bibitem{10750008}
Liang S, Cui Q, Huang X, et~al.
\newblock Efficient hierarchical federated services for heterogeneous mobile edge.
\newblock {IEEE} Trans. Services Comput., Early Access, 2024.

\bibitem{fallah2020personalized}
Fallah A, Mokhtari A, Ozdaglar A.
\newblock Personalized federated learning with theoretical guarantees: A model-agnostic meta-learning approach.
\newblock Adv. Neural Inf. Process. Syst., 2020, 33: 3557--3568

\bibitem{tan2022towards}
Tan A~Z, Yu H, Cui L, et~al.
\newblock Towards personalized federated learning.
\newblock {IEEE} Trans. Neural Netw. Learn. Syst., 2023, 34: 9587--9603

\bibitem{passerat2020blockchain}
Passerat-Palmbach J, Farnan T, McCoy M, et~al.
\newblock Blockchain-orchestrated machine learning for privacy preserving federated learning in electronic health data.
\newblock In: Proceedings of IEEE International Conference on Blockchain, Rhodes, Greece, 2020. 550--555

\bibitem{polap2021agent}
Po{\l}ap D, Srivastava G, Yu K.
\newblock Agent architecture of an intelligent medical system based on federated learning and blockchain technology.
\newblock J. Inf. Secur. Appl., 2021, 58: 102748

\bibitem{kevin2021federated}
Kevin I, Wang K, Zhou X, et~al.
\newblock Federated transfer learning based cross-domain prediction for smart manufacturing.
\newblock {IEEE} Trans. Ind. Inform., 2021, 18: 4088--4096

\bibitem{10423783}
Wan Y, Qu Y, Ni W, et~al.
\newblock Data and model poisoning backdoor attacks on wireless federated learning, and the defense mechanisms: A comprehensive survey.
\newblock IEEE Commun. Surv. Tut., 2024, 26: 1861--1897

\bibitem{10419367}
Li K, Zheng J, Yuan X, et~al.
\newblock Data-agnostic model poisoning against federated learning: A graph autoencoder approach.
\newblock IEEE Trans. Info.n Forensics Secur., 2024, 19: 3465--3480

\bibitem{liu2024vertical}
Liu Y, Kang Y, Zou T, et~al.
\newblock Vertical federated learning: Concepts, advances, and challenges.
\newblock {IEEE} Trans. Knowl. Data Eng., 2024, 36: 3615--3634

\bibitem{zhang2021vertical}
Zhang J, Jiang Y.
\newblock A vertical federation recommendation method based on clustering and latent factor model.
\newblock In: Proceedings of International Conference on Electronic Information Engineering and Computer Science (EIECS), Changchun, China, 2021. 362--366

\bibitem{tang2023ihvfl}
Tang F, Liang S, Ling G, et~al.
\newblock {IHVFL}: A privacy-enhanced intention-hiding vertical federated learning framework for medical data.
\newblock Cybersecurity, 2023, 6: 37

\bibitem{yuan2022fedstn}
Yuan X, Chen J, Yang J, et~al.
\newblock Fedstn: Graph representation driven federated learning for edge computing enabled urban traffic flow prediction.
\newblock {IEEE} Trans. Intell. Transp. Syst., 2022, 24: 8738--8748

\bibitem{chu2019multi}
Chu T, Wang J, Codec{\`a} L, et~al.
\newblock Multi-agent deep reinforcement learning for large-scale traffic signal control.
\newblock {IEEE} Trans. Intell. Transp. Syst., 2019, 21: 1086--1095

\bibitem{wu2020multi}
Wu T, Zhou P, Liu K, et~al.
\newblock Multi-agent deep reinforcement learning for urban traffic light control in vehicular networks.
\newblock {IEEE} Trans. Veh. Technol., 2020, 69: 8243--8256

\bibitem{palanisamy2020multi}
Palanisamy P.
\newblock Multi-agent connected autonomous driving using deep reinforcement learning.
\newblock In: Proceedings of International Joint Conference on Neural Networks (IJCNN), Glasgow, UK, 2020. 2020, 1--7

\bibitem{10754633}
Yu K, Cui Q, Lyu X, et~al.
\newblock Efficient collaborative computing for multi-layer {LEO} satellites with spatiotemporal dynamics: A long-term continuous timescale optimization.
\newblock {IEEE} Internet Things J., Early Access, 2024.

\bibitem{zhou2021multi}
Zhou T, Tang D, Zhu H, et~al.
\newblock Multi-agent reinforcement learning for online scheduling in smart factories.
\newblock Robotics and computer-integrated Manufacturing, 2021, 72: 102202

\bibitem{cao2020multiagent}
Cao Z, Zhou P, Li R, et~al.
\newblock Multiagent deep reinforcement learning for joint multichannel access and task offloading of mobile-edge computing in industry 4.0.
\newblock {IEEE} Internet Things J., 2020, 7: 6201--6213

\bibitem{duan2022combined}
Duan Q, Hu S, Deng R, et~al.
\newblock Combined federated and split learning in edge computing for ubiquitous intelligence in internet of things: State-of-the-art and future directions.
\newblock Sensors, 2022, 22: 5983

\bibitem{10646623}
Lyu X, Li Y, He Y, et~al.
\newblock Objective-driven differentiable optimization of traffic prediction and resource allocation for split {AI} inference edge networks.
\newblock IEEE Trans. Machine Learn. Commun. Netw., 2024, 2: 1178--1192

\bibitem{10529950}
Lin Z, Qu G, Chen X, et~al.
\newblock Split learning in {6G} edge networks.
\newblock {IEEE} Wireless Commun., 2024, 31: 170--176

\bibitem{vepakomma2018split}
Vepakomma P, Gupta O, Swedish T, et~al.
\newblock Split learning for health: Distributed deep learning without sharing raw patient data.
\newblock 2018.
\newblock \eprint{https://arxiv.org/abs/1812.00564}

\bibitem{poirot2019split}
Poirot M~G, Vepakomma P, Chang K, et~al.
\newblock Split learning for collaborative deep learning in healthcare.
\newblock 2019.
\newblock \eprint{https://arxiv.org/abs/1912.12115}

\bibitem{li2024split}
Li Z, Yan C, Zhang X, et~al.
\newblock Split learning for distributed collaborative training of deep learning models in health informatics.
\newblock In: AMIA Annu Symp Proc., 2024. 1047--1056

\bibitem{9016486}
Jeon J, Kim J.
\newblock Privacy-sensitive parallel split learning.
\newblock In: Proceedings of International Conference on Information Networking (ICOIN), Barcelona, Spain, 2020. 7--9

\bibitem{10040976}
Wu W, Li M, Qu K, et~al.
\newblock Split learning over wireless networks: Parallel design and resource management.
\newblock {IEEE} J. Sel. Areas Commun., 2023, 41: 1051--1066

\bibitem{10637271}
Mao Y, Yu X, Huang K, et~al.
\newblock Green edge {AI}: A contemporary survey.
\newblock Proceedings of the IEEE, 2024, 112: 880--911

\bibitem{nguyen2021digital}
Nguyen H~X, Trestian R, To D, et~al.
\newblock Digital twin for 5{G} and beyond.
\newblock {IEEE} Commun. Mag., 2021, 59: 10--15

\bibitem{grieves2005product}
Grieves M~W.
\newblock Product lifecycle management: the new paradigm for enterprises.
\newblock Int. J. Product Develop., 2005, 2: 71--84

\bibitem{cao2024channeltwinningenablernextgeneration}
Cao Y, Dai L, Tan J, et~al.
\newblock Advancing ubiquitous wireless connectivity through channel twinnin.
\newblock IEEE Wirel. Commun., 2024, to appear

\bibitem{tao2018digital}
Tao F, Zhang H, Liu A, et~al.
\newblock Digital twin in industry: State-of-the-art.
\newblock {IEEE} Trans. Ind. Inform., 2018, 15: 2405--2415

\bibitem{10234427}
Cui Y, Lv T, Ni W, et~al.
\newblock Digital twin-aided learning for managing reconfigurable intelligent surface-assisted, uplink, user-centric cell-free systems.
\newblock IEEE J. Select. Areas Commun., 2023, 41: 3175--3190

\bibitem{barricelli2019survey}
Barricelli B~R, Casiraghi E, Fogli D.
\newblock A survey on digital twin: Definitions, characteristics, applications, and design implications.
\newblock {IEEE} Access, 2019, 7: 167653--167671

\bibitem{li2022novel}
Li Z, Duan M, Xiao B, et~al.
\newblock A novel anomaly detection method for digital twin data using deconvolution operation with attention mechanism.
\newblock {IEEE} Trans. Ind. Inform., 2022, 19: 7278--7286

\bibitem{qin2024machine}
Qin B, Pan H, Dai Y, et~al.
\newblock Machine and deep learning for digital twin networks: A survey.
\newblock {IEEE} Internet Things J., 2024

\bibitem{wu2022digital}
Wu X, Lian W, Zhou M, et~al.
\newblock A digital twin-based fault diagnosis framework for bogies of high-speed trains.
\newblock IEEE J. Radio Freq. Identification, 2022, 7: 203--207

\bibitem{cui2023human}
Cui C, Ma Y, Cao X, et~al.
\newblock Human-autonomy teaming on autonomous vehicles with large language model-enabled human digital twins.
\newblock In: Proceedings of IEEE/ACM Symposium on Edge Computing (SEC), Wilmington, DE, USA, 2023. 319--324

\bibitem{liu2021digital}
Liu T, Tang L, Wang W, et~al.
\newblock Digital-twin-assisted task offloading based on edge collaboration in the digital twin edge network.
\newblock {IEEE} Internet Things J., 2021, 9: 1427--1444

\bibitem{huang2023collective}
Huang Z, Li D, Cai J, et~al.
\newblock Collective reinforcement learning based resource allocation for digital twin service in 6{G} networks.
\newblock J. Netw. Comput. Appl., 2023, 217: 103697

\bibitem{wang2023towards}
Wang J, Zhang J, Zhang Y, et~al.
\newblock Towards 6{G} digital twin channel using radio environment knowledge pool.
\newblock 2023.
\newblock \eprint{https://arxiv.org/abs/2312.10287}

\bibitem{wang2024digital}
Wang H, Zhang J, Nie G, et~al.
\newblock Digital twin channel for 6{G}: Concepts, architectures and potential applications.
\newblock 2024.
\newblock \eprint{https://arxiv.org/abs/2403.12467}

\bibitem{nie2022predictive}
Nie G, Zhang J, Zhang Y, et~al.
\newblock A predictive 6{G} network with environment sensing enhancement: From radio wave propagation perspective.
\newblock China Commun., 2022, 19: 105--122

\bibitem{miao2023demo}
Miao Y, Zhang Y, Zhang J, et~al.
\newblock Demo abstract: Predictive radio environment for digital twin communication platform via enhanced sensing.
\newblock In: Proceedings of IEEE Conference on Computer Communications Workshops (INFOCOM WKSHPS), New York, USA, 2023. 1--2

\bibitem{zhang2016interdisciplinary}
Zhang J.
\newblock The interdisciplinary research of big data and wireless channel: A cluster-nuclei based channel model.
\newblock China Commun., 2016, 13: 14--26

\bibitem{wang2024electromagnetic}
Wang J, Zhang J, Sun Y, et~al.
\newblock Electromagnetic wave property inspired radio environment knowledge construction and {AI}-based verification for 6{G} digital twin channel.
\newblock 2024.
\newblock \eprint{https://arxiv.org/abs/2406.00690}

\bibitem{ITU-T-DT}
{International Telecommunication Union (ITU)}.
\newblock Digital twin network – requirements and architecture.
\newblock 2022.
\newblock \eprint{https://www.itu.int/rec/dologin_pub.asp?lang=e&id=T-REC-Y.3090-202202-I!!PDF-E&type=items}

\bibitem{10.1145/3659099}
Jiang Y, Ma B, Wang X, et~al.
\newblock Blockchained federated learning for internet of things: A comprehensive survey.
\newblock ACM Comput. Surv., 2024, 56: 37

\bibitem{van2022urllc}
Van~Huynh D, Nguyen V~D, Khosravirad S~R, et~al.
\newblock {URLLC} edge networks with joint optimal user association, task offloading and resource allocation: A digital twin approach.
\newblock {IEEE} Trans. Commun., 2022, 70: 7669--7682

\bibitem{khan2022digital}
Khan L~U, Han Z, Saad W, et~al.
\newblock Digital twin of wireless systems: Overview, taxonomy, challenges, and opportunities.
\newblock {IEEE} Commun. Surv. Tut., 2022, 24: 2230--2254

\bibitem{lu2020communication}
Lu Y, Huang X, Zhang K, et~al.
\newblock Communication-efficient federated learning and permissioned blockchain for digital twin edge networks.
\newblock {IEEE} Internet Things J., 2020, 8: 2276--2288

\bibitem{zhang2021adaptive}
Zhang K, Cao J, Zhang Y.
\newblock Adaptive digital twin and multiagent deep reinforcement learning for vehicular edge computing and networks.
\newblock {IEEE} Trans. Ind. Inform., 2021, 18: 1405--1413

\bibitem{hu2021digital}
Hu C, Fan W, Zeng E, et~al.
\newblock Digital twin-assisted real-time traffic data prediction method for 5{G}-enabled internet of vehicles.
\newblock {IEEE} Trans. Ind. Inform., 2021, 18: 2811--2819

\bibitem{chen2022artificial}
Chen D, Lv Z.
\newblock Artificial intelligence enabled digital twins for training autonomous cars.
\newblock Internet of Things and Cyber-Physical Systems, 2022, 2: 31--41

\bibitem{liu2023cooperative}
Liu K, Xu X, Dai P, et~al.
\newblock Cooperative sensing and uploading for quality-cost tradeoff of digital twins in {VEC}.
\newblock IEEE Trans. on Consum. Electron., 2024, 70: 3614--3625

\bibitem{IMT2030-6gkeytechnology}
{IMT-2030 (6G) Promotion Group}.
\newblock {6G} network architecture vision and key technology outlook white paper.
\newblock 2021.
\newblock \eprint{https://www.imt2030.org.cn/}

\bibitem{yukun2024computing}
Yukun S, Bo L, Juniin L, et~al.
\newblock Computing power network: A survey.
\newblock China Commun., 2024, 21: 109--145

\bibitem{qingmin2022design}
Qingmin J, Kai G, Xiaomao Z, et~al.
\newblock Design and discussion for new computing power network architecture.
\newblock Information and Communication Technology and Policy, 2022, 48: 18

\bibitem{yao2019trend}
Yao H, Geng L.
\newblock Trend of next generation network architecture: computing and networking convergence evolution.
\newblock Telecommun. Sci., 2019, 35: 38--43

\bibitem{tang2021computing}
Tang X, Cao C, Wang Y, et~al.
\newblock Computing power network: The architecture of convergence of computing and networking towards 6{G} requirement.
\newblock China Commun., 2021, 18: 175--185

\bibitem{10146517}
Xiao D, Zhang J~A, Liu X, et~al.
\newblock A two-stage {GCN}-based deep reinforcement learning framework for {SFC} embedding in multi-datacenter networks.
\newblock IEEE Trans. Netw. Serv. Manag., 2023, 20: 4297--4312

\bibitem{wang2020net}
Wang X, Ren X, Qiu C, et~al.
\newblock Net-in-{AI}: A computing-power networking framework with adaptability, flexibility, and profitability for ubiquitous {AI}.
\newblock {IEEE} Netw., 2020, 35: 280--288

\bibitem{tokusashi2019case}
Tokusashi Y, Dang H~T, Pedone F, et~al.
\newblock The case for in-network computing on demand.
\newblock In: Proceedings of the Fourteenth EuroSys Conference, Dresden, Germany, 2019. 1--16

\bibitem{ren2022ai}
Ren X, Qiu C, Wang X, et~al.
\newblock {AI}-bazaar: A cloud-edge computing power trading framework for ubiquitous {AI} services.
\newblock IEEE Trans. Cloud Comput., 2022, 11: 2337--2348

\bibitem{10194241}
Duan L, Sun Y, Ni W, et~al.
\newblock Attacks against cross-chain systems and defense approaches: A contemporary survey.
\newblock IEEE/CAA J. Automatica Sinica, 2023, 10: 1647--1667

\bibitem{10130487}
Duan L, Yang L, Liu C, et~al.
\newblock A new smart contract anomaly detection method by fusing opcode and source code features for blockchain services.
\newblock IEEE Trans. Netw. Serv. Manag., 2023, 20: 4354--4368

\bibitem{liu2022computing}
Liu J, Sun Y, Su J, et~al.
\newblock Computing power network: A testbed and applications with edge intelligence.
\newblock In: Proceedings of IEEE Conference on Computer Communications Workshops (INFOCOM WKSHPS), New York, USA, 2022. 1--2

\bibitem{mingxuan2020research}
Mingxuan L, Chang C, Xiongyan T, et~al.
\newblock Research on edge resource scheduling solutions for computing power network.
\newblock Frontiers of Data and Comput., 2020, 2: 80--91

\bibitem{chen2024two}
Chen Q, Yang C, Lan S, et~al.
\newblock Two-stage evolutionary search for efficient task offloading in edge computing power networks.
\newblock {IEEE} Internet Things J., 2024, 11: 30787--30799

\bibitem{sodhro2020ai}
Sodhro A~H, Sodhro G~H, Guizani M, et~al.
\newblock {AI}-enabled reliable channel modeling architecture for fog computing vehicular networks.
\newblock {IEEE} Wireless Commun., 2020, 27: 14--21

\bibitem{deng2022uav}
Deng C, Fang X, Wang X.
\newblock Uav-enabled mobile-edge computing for {AI} applications: Joint model decision, resource allocation, and trajectory optimization.
\newblock {IEEE} Internet Things J., 2022, 10: 5662--5675

\bibitem{yao1982protocols}
Yao A~C.
\newblock Protocols for secure computations.
\newblock In: Proceedings of IEEE 23rd Annual Symposium on Foundations of Computer Science (SFCS), Chicago, Illinois, 1982. 160--164

\bibitem{SMPC}
Okamura T, Teranishi I.
\newblock Enhancing fintech security with secure multi-party computation technology.
\newblock NEC Tech. J., 2017, 11: 46--50.
\newblock \eprint{https://www.nec.com/en/global/techrep/journal/g16/n02/pdf/160211.pdf}

\bibitem{shamir1979share}
Shamir A.
\newblock How to share a secret.
\newblock Commun. of the ACM, 1979, 22: 612--613

\bibitem{haque2022garbled}
Haque A, Heath D, Kolesnikov V, et~al.
\newblock Garbled circuits with sublinear evaluator.
\newblock In: Procddings of Springer 41st Annual International Conference on the Theory and Application of Cryptographic Technology. 2022, 37--64

\bibitem{lindell2009proof}
Lindell Y, Pinkas B.
\newblock A proof of security of yao’s protocol for two-party computation.
\newblock Springer J. cryptology, 2009, 22: 161--188

\bibitem{acar2018survey}
Acar A, Aksu H, Uluagac A~S, et~al.
\newblock A survey on homomorphic encryption schemes: Theory and implementation.
\newblock ACM Comput. Surv., 2018, 51: 1--35

\bibitem{wang2022privacy}
Wang R, Lai J, Zhang Z, et~al.
\newblock Privacy-preserving federated learning for internet of medical things under edge computing.
\newblock IEEE J. Biomed. and Health Inform., 2022, 27: 854--865

\bibitem{song2022eppda}
Song J, Wang W, Gadekallu T~R, et~al.
\newblock Eppda: An efficient privacy-preserving data aggregation federated learning scheme.
\newblock IEEE Trans. Netw. Sci. and Eng., 2022, 10: 3047--3057

\bibitem{zhu2020privacy}
Zhu H, Goh R~S~M, Ng W~K.
\newblock Privacy-preserving weighted federated learning within the secret sharing framework.
\newblock IEEE Access, 2020, 8: 198275--198284

\bibitem{huang2021starfl}
Huang A, Liu Y, Chen T, et~al.
\newblock Starfl: Hybrid federated learning architecture for smart urban computing.
\newblock ACM Transactions on Intelligent Systems and Technology (TIST), 2021, 12: 1--23

\bibitem{tang2022secure}
Tang X, Zhu L, Shen M, et~al.
\newblock Secure and trusted collaborative learning based on blockchain for artificial intelligence of things.
\newblock {IEEE} Wireless Commun., 2022, 29: 14--22

\bibitem{10820866}
Yu X, Lv T, Li W, et~al.
\newblock Multi-task semantic communication with graph attention-based feature correlation extraction.
\newblock IEEE Trans. Mobile Comput., 2025: 1--18

\bibitem{niu2024mathematical}
Niu K, Zhang P.
\newblock A mathematical theory of semantic communication.
\newblock 2024.
\newblock \eprint{https://arxiv.org/abs/2401.13387}

\bibitem{zhang2023model}
Zhang P, Xu X, Dong C, et~al.
\newblock Model division multiple access for semantic communications.
\newblock Frontiers of Information Technology \& Electronic Engineering, 2023, 24: 801--812

\bibitem{yang2022semantic}
Yang W, Du H, Liew Z~Q, et~al.
\newblock Semantic communications for future internet: Fundamentals, applications, and challenges.
\newblock {IEEE} Commun. Surv. Tut., 2022, 25: 213--250

\bibitem{xie2021deep}
Xie H, Qin Z, Li G~Y, et~al.
\newblock Deep learning enabled semantic communication systems.
\newblock {IEEE} Trans. Signal Process., 2021, 69: 2663--2675

\bibitem{lu2021reinforcement}
Lu K, Li R, Chen X, et~al.
\newblock Reinforcement learning-powered semantic communication via semantic similarity.
\newblock 2021.
\newblock \eprint{https://arxiv.org/abs/2108.12121}

\bibitem{jiang2022deep}
Jiang P, Wen C~K, Jin S, et~al.
\newblock Deep source-channel coding for sentence semantic transmission with {HARQ}.
\newblock {IEEE} Trans. Commun., 2022, 70: 5225--5240

\bibitem{weng2021semantic}
Weng Z, Qin Z.
\newblock Semantic communication systems for speech transmission.
\newblock {IEEE} J. Sel. Areas Commun., 2021, 39: 2434--2444

\bibitem{weng2021semantic1}
Weng Z, Qin Z, Li G~Y.
\newblock Semantic communications for speech signals.
\newblock In: Proceedings of IEEE International Commun. Conf. (ICC), Montreal, QC, Canada, 2021. 1--6

\bibitem{tong2021federated}
Tong H, Yang Z, Wang S, et~al.
\newblock Federated learning for audio semantic communication.
\newblock Frontiers in communications and networks, 2021, 2: 734402

\bibitem{bourtsoulatze2019deep}
Bourtsoulatze E, Kurka D~B, G{\"u}nd{\"u}z D.
\newblock Deep joint source-channel coding for wireless image transmission.
\newblock {IEEE} Trans. Cogn. Commun. Netw., 2019, 5: 567--579

\bibitem{wang2022wireless}
Wang S, Dai J, Liang Z, et~al.
\newblock Wireless deep video semantic transmission.
\newblock {IEEE} J. Sel. Areas Commun., 2022, 41: 214--229

\bibitem{10500411}
Yenduri G, Ramalingam M, Selvi G~C, et~al.
\newblock {GPT} (generative pre-trained transformer)— a comprehensive review on enabling technologies, potential applications, emerging challenges, and future directions.
\newblock {IEEE} Access, 2024, 12: 54608--54649

\bibitem{9144301}
Chowdhury M~Z, Shahjalal M, Ahmed S, et~al.
\newblock 6{G} wireless communication systems: Applications, requirements, technologies, challenges, and research directions.
\newblock {IEEE} Open J. Commun. Soc., 2020, 1: 957--975

\bibitem{tong2023issuesnetgpt}
Tong W, Peng C, Yang T, et~al.
\newblock Ten issues of {NetGPT}, 2023.
\newblock \eprint{https://arxiv.org/abs/2311.13106}

\bibitem{shao2024wirelessllm}
Shao J, Tong J, Wu Q, et~al.
\newblock {WirelessLLM}: Empowering large language models towards wireless intelligence.
\newblock Journal of Communications and Information Networks, 2024, 9: 99--112

\bibitem{zou2024telecomgptframeworkbuildtelecomspecfic}
Zou H, Zhao Q, Tian Y, et~al.
\newblock Telecomgpt: A framework to build telecom-specfic large language models, 2024.
\newblock \eprint{https://arxiv.org/abs/2407.09424}

\bibitem{2019The}
Letaief K~B, Chen W, Shi Y, et~al.
\newblock The roadmap to 6{G} -- {AI} empowered wireless networks.
\newblock {IEEE} Commun. Mag., 2019, 57: 84--90

\bibitem{8304385}
Zhang N, Yang P, Ren J, et~al.
\newblock Synergy of big data and 5{G} wireless networks: Opportunities, approaches, and challenges.
\newblock {IEEE} Wireless Commun., 2018, 25: 12--18

\bibitem{10579546}
Chen Z, Zhang Z, Yang Z.
\newblock Big {AI} models for 6{G} wireless networks: Opportunities, challenges, and research directions.
\newblock {IEEE} Wireless Commun., 2024: 1--9

\bibitem{8932161}
Qian L, Zhu J, Zhang S.
\newblock Survey of wireless big data.
\newblock Journal of Communications and Information Networks, 2017, 2: 1--18

\bibitem{2019Big}
Dai H~N, Wong C~W, Wang H, et~al.
\newblock Big data analytics for large scale wireless networks: Challenges and opportunities.
\newblock 2019.
\newblock \eprint{https://arxiv.org/pdf/1909.08069}

\bibitem{10283537}
Chen T, Tang Q, Liu G.
\newblock Efficient task scheduling and resource allocation for {AI} training services in native {AI} wireless networks.
\newblock In: Proceedings of IEEE International Commun. Conf. Workshops (ICC Workshops), Rome, Italy, 2023. 637--642

\bibitem{10466747}
Chen Y, Li R, Zhao Z, et~al.
\newblock Netgpt: An {AI}-native network architecture for provisioning beyond personalized generative services.
\newblock {IEEE} Netw., 2024: 1--1

\bibitem{2020When}
Challita U, Ryden H, Tullberg H.
\newblock When machine learning meets wireless cellular networks: Deployment, challenges, and applications.
\newblock {IEEE} Commun. Mag., 2020, 58: 12--18

\bibitem{8936546}
Passalis N, Tefas A, Kanniainen J, et~al.
\newblock Deep adaptive input normalization for time series forecasting.
\newblock {IEEE} Trans. Neural Netw. Learn. Syst., 2020, 31: 3760--3765

\bibitem{10679601}
Zhang H, Zhang Y~F, Zhang Z, et~al.
\newblock {LogoRA}: Local-global representation alignment for robust time series classification.
\newblock {IEEE} Trans. Knowl. Data Eng., 2024, 36: 8718--8729

\bibitem{10399795}
Xu C, Du X, Fan X, et~al.
\newblock {FastVSDF}: An efficient spatiotemporal data fusion method for seamless data cube.
\newblock {IEEE} Trans. Geosci. Remote Sens., 2024, 62: 1--22

\bibitem{10192095}
Fang C, Hu Z, Meng X, et~al.
\newblock {DRL}-driven joint task offloading and resource allocation for energy-efficient content delivery in cloud-edge cooperation networks.
\newblock {IEEE} Trans. Veh. Technol., 2023, 72: 16195--16207

\bibitem{qiao2024latency}
Qiao L, Mashhadi M~B, Gao Z, et~al.
\newblock Latency-aware generative semantic communications with pre-trained diffusion models.
\newblock {IEEE} Wireless Commun. Lett., 2024, 13: 2652--2656

\bibitem{1580513}
Shin M, Ma J, Mishra A, et~al.
\newblock Wireless network security and interworking.
\newblock Proceedings of the IEEE, 2006, 94: 455--466

\bibitem{AIaaS}
IMT-2030.
\newblock 6{G} {AlaaS} requirement research (white paper).
\newblock 2023.
\newblock \eprint{https://www.itu.int/rec/R-REC-M.2160-0-202311-I/en}

\bibitem{AIaaS_6G}
{6GANA TG1}.
\newblock {6G} {AlaaS} requirements whitepaper.
\newblock 2023.
\newblock \eprint{https://6g-ana.com/}

\bibitem{partarakis2024review}
Partarakis N, Zabulis X.
\newblock A review of immersive technologies, knowledge representation, and {AI} for human-centered digital experiences.
\newblock Electronics, 2024, 13: 269

\bibitem{10411981}
Guo J, Chen H, Song B, et~al.
\newblock Distributed task-oriented communication networks with multimodal semantic relay and edge intelligence.
\newblock {IEEE} Commun. Mag., 2024, 62: 82--89

\bibitem{benotsmane2019economic}
Benotsmane R, Kov{\'a}cs G, Dud{\'a}s L.
\newblock Economic, social impacts and operation of smart factories in industry 4.0 focusing on simulation and artificial intelligence of collaborating robots.
\newblock Social Sciences, 2019, 8: 143

\bibitem{10002890}
Cui Q, Zhao X, Ni W, et~al.
\newblock Multi-agent deep reinforcement learning-based interdependent computing for mobile edge computing-assisted robot teams.
\newblock IEEE Trans. Veh. Technol., 2023, 72: 6599--6610

\bibitem{santos2020online}
Santos M~A, Munoz R, Olivares R, et~al.
\newblock Online heart monitoring systems on the internet of health things environments: A survey, a reference model and an outlook.
\newblock Information Fusion, 2020, 53: 222--239

\bibitem{8466351}
Zhang S, Chen J, Lyu F, et~al.
\newblock Vehicular communication networks in the automated driving era.
\newblock {IEEE} Commun. Mag., 2018, 56: 26--32

\bibitem{10700687}
Gao B, Liu J, Zou H, et~al.
\newblock Vehicle-road-cloud collaborative perception framework and key technologies: A review.
\newblock IEEE Trans. Intell. Transp. Syst., 2024, 25: 19295--19318

\bibitem{10273257}
Yang Y, Wu J, Chen T, et~al.
\newblock Task-oriented {6G} native-{AI} network architecture.
\newblock {IEEE} Netw., 2024, 38: 219--227

\bibitem{10819488}
Shen J, Wu B, Xiang W, et~al.
\newblock Novel bandwidth-aware network coding for fast cloud-of-clouds disaster backup.
\newblock IEEE Trans. Netw. Serv. Manag., Early Access, 2024.

\bibitem{38.843}
{3rd Generation Partership Project (3GPP)}.
\newblock Study on artificial intelligence {(AI)}/machine learning {(ML)} for {NR} air interface.
\newblock TR 38.843.
\newblock \eprint{https://www.3gpp.org/ftp/Specs/archive/38_series/38.843}

\bibitem{38.744}
{3rd Generation Partership Project (3GPP)}.
\newblock Study on artificial intelligence {(Al)}/machine learning {(ML)} for mobility in {NR}.
\newblock TR 38.744.
\newblock \eprint{https://www.3gpp.org/ftp/Specs/archive/38_series/38.744}

\bibitem{37.817}
{3rd Generation Partership Project (3GPP)}.
\newblock Study on further enhancement for data collection.
\newblock TR 37.817.
\newblock \eprint{https://www.3gpp.org/ftp/Specs/archive/37_series/37.817}

\bibitem{38.743}
{3rd Generation Partership Project (3GPP)}.
\newblock Study on enhancements for artificial intelligence {(AI)}/machine learning {(ML)} for {NG-RAN}.
\newblock TR 38.743.
\newblock \eprint{https://www.3gpp.org/ftp/Specs/archive/38_series/38.743}

\bibitem{22.850}
{3rd Generation Partership Project (3GPP)}.
\newblock {3GPP AI/ML} consistency alignment.
\newblock TR 22.850.
\newblock \eprint{https://www.3gpp.org/ftp/Specs/archive/22_series/22.850}

\bibitem{5G-MoNArch}
{5G-MoNArch}.
\newblock 5{G} mobile network architecture for diverse services, use cases, and applications in 5{G} and beyond.
\newblock 2019.
\newblock \eprint{https://5g-monarch.eu}

\bibitem{6gdistributed}
{6G Alliance of Network AI (6GANA)}.
\newblock Whitepaper on distributed learning of {6G}.
\newblock 2023.
\newblock \eprint{https://www.6g-ana.com/upload/file/20240129/6384212141255667878785630.pdf}

\bibitem{10506539}
Zhang R, Xiong K, Du H, et~al.
\newblock Generative {AI}-enabled vehicular networks: Fundamentals, framework, and case study.
\newblock {IEEE} Netw., 2024, 38: 259--267

\bibitem{9925609}
Chen L, Qi J, Su X, et~al.
\newblock {REMR}: A reliability evaluation method for dynamic edge computing network under time constraint.
\newblock {IEEE} Internet Things J., 2023, 10: 4281--4291

\bibitem{9643788}
Kourouklidis P, Kolovos D, Noppen J, et~al.
\newblock A model-driven engineering approach for monitoring machine learning models.
\newblock In: Proceedings of ACM/IEEE International Conference on Model Driven Engineering Languages and Systems Companion (MODELS-C), Fukuoka, Japan, 2021. 160--164

\bibitem{9145835}
Yao L, Ge Z.
\newblock Industrial big data modeling and monitoring framework for plant-wide processes.
\newblock {IEEE} Trans. Ind. Inform., 2021, 17: 6399--6408

\bibitem{10159517}
Nan G, Li Z, Zhai J, et~al.
\newblock Physical-layer adversarial robustness for deep learning-based semantic communications.
\newblock {IEEE} J. Sel. Areas Commun., 2023, 41: 2592--2608

\bibitem{10375527}
Cai X, Shi K, Sun Y, et~al.
\newblock Stability analysis of networked control systems under {DoS} attacks and security controller design with mini-batch machine learning supervision.
\newblock {IEEE} Trans. Inf. Forensics Secur., 2024, 19: 3857--3865

\bibitem{9868259}
Xu P, Wang K, Hassan M~M, et~al.
\newblock Adversarial robustness in graph-based neural architecture search for edge {AI} transportation systems.
\newblock {IEEE} Trans. Intell. Transp. Syst., 2023, 24: 8465--8474

\bibitem{8826332}
Pu C, Wang K, Xia Y.
\newblock Robustness of link prediction under network attacks.
\newblock IEEE Transactions on Circuits and Systems II: Express Briefs, 2020, 67: 1472--1476

\bibitem{2024A}
Dong C, Xiong X, Xue Q, et~al.
\newblock A survey on the network models applied in the industrial network optimization.
\newblock Sci. China Inf. Sci., 2024, 67: 121301

\bibitem{9954418}
Yang R, Zhang Z, Zhang X, et~al.
\newblock Meta-learning for beam prediction in a dual-band communication system.
\newblock {IEEE} Trans. Commun., 2023, 71: 145--157

\bibitem{xu2023samba}
Xu Z, Wang S, Zhang Y~J~A.
\newblock {SAMBA}: Scenario-adaptive meta-learning for mm{W}ave beam alignment.
\newblock In: Proceedings of IEEE Globecom Workshops (GC Wkshps), Kuala Lumpur, Malaysia, 2023. 1--6

\bibitem{28}
Kim H, Choi J, Love D~J.
\newblock Massive {MIMO} channel prediction via meta-learning and deep denoising: Is a small dataset enough?
\newblock {IEEE} Trans. Wireless Commun., 2023, 22: 9278--9290

\bibitem{10839451}
Wu B, Zou S, Liwang M, et~al.
\newblock Explainable application intent for zero-touch networking: An incorporation of hypergraph and transformer.
\newblock IEEE Trans. Commun., Early Access, 2025.

\bibitem{10663726}
Zou S, Liwang M, Wu B, et~al.
\newblock Intent-oriented network slicing with hypergraphs.
\newblock IEEE Netw., Early Access, 2024.

\bibitem{18}
Nichol A, Achiam J, Schulman J.
\newblock On first-order meta-learning algorithms.
\newblock 2018.
\newblock \eprint{https://arxiv.org/pdf/1803.02999}

\bibitem{van2008visualizing}
Van~der Maaten L, Hinton G.
\newblock Visualizing data using t-{SNE}.
\newblock Journal of machine learning research, 2008, 9

\bibitem{10551685}
Wang R, Yang L, Tang T, et~al.
\newblock Robust federated learning for heterogeneous clients and unreliable communications.
\newblock {IEEE} Trans. Wireless Commun., 2024, 23: 13440--13455

\bibitem{10253642}
Zheng P, Zhu Y, Hu Y, et~al.
\newblock Federated learning in heterogeneous networks with unreliable communication.
\newblock {IEEE} Trans. Wireless Commun., 2024, 23: 3823--3838

\bibitem{10354479}
Pang Y, Zhang H, Deng J~D, et~al.
\newblock Collaborative learning with heterogeneous local models: A rule-based knowledge fusion approach.
\newblock {IEEE} Trans. Knowl. Data Eng., 2024, 36: 5768--5783

\bibitem{10184998}
Liu X, Wang G, Liu Z, et~al.
\newblock Hierarchical reinforcement learning integrating with human knowledge for practical robot skill learning in complex multi-stage manipulation.
\newblock {IEEE} Trans. Autom. Sci. Eng., 2024, 21: 3852--3862

\bibitem{10106044}
Yan Y, Tong X, Wang S.
\newblock Clustered federated learning in heterogeneous environment.
\newblock {IEEE} Trans. Neural Netw. Learn. Syst., 2024, 35: 12796--12809

\bibitem{10.1145/3589303}
Wang Y, Wu Y, Chen X, et~al.
\newblock Incentive-aware decentralized data collaboration.
\newblock In: Proceedings of the ACM on Management of Data, New York, NY, USA, 2023. 1--27.
\newblock \eprint{https://doi.org/10.1145/3589303}

\bibitem{liu2022resource}
Liu Y~J, Feng G, Sun Y, et~al.
\newblock Resource consumption for supporting federated learning enabled network edge intelligence.
\newblock In: Proceedings of IEEE International Commun. Conf. Workshops (ICC Workshops), Seoul, Korea, 2022. 01--06

\bibitem{10417030}
Zhang R, Pan C, Wang Y, et~al.
\newblock Federated deep reinforcement learning for multimedia task offloading and resource allocation in mec networks.
\newblock {IEEE} Trans. Commun., 2024, E107-B: 446--457

\bibitem{10376360}
Ji Z, Qin Z, Tao X.
\newblock Meta federated reinforcement learning for distributed resource allocation.
\newblock {IEEE} Trans. Wireless Commun., 2024, 23: 7865--7876

\bibitem{10347516}
Cherif N, Jaafar W, Yanikomeroglu H, et~al.
\newblock {RL}-based cargo-{UAV} trajectory planning and cell association for minimum handoffs, disconnectivity, and energy consumption.
\newblock {IEEE} Trans. Veh. Technol., 2024, 73: 7304--7309

\bibitem{10363447}
Samsi S, Zhao D, McDonald J, et~al.
\newblock From words to watts: Benchmarking the energy costs of large language model inference.
\newblock In: 2023 IEEE High Performance Extreme Computing Conference (HPEC), Boston, MA, USA, 2023. 1--9

\bibitem{Muhammad}
Zawish M, Dharejo F~A, Khowaja S~A, et~al.
\newblock {AI} and 6{G} into the metaverse: Fundamentals, challenges and future research trends.
\newblock {IEEE} Open J. Commun. Soc., 2024, 5: 730--778

\bibitem{Energy-Aware}
Zawish M, Ashraf N, Ansari R~I, et~al.
\newblock Energy-aware {AI}-driven framework for edge-computing-based {IoT} applications.
\newblock {IEEE} Internet Things J., 2023, 10: 5013--5023

\bibitem{Intelligent-Traffic}
Li X, Zhang H, Shen Y, et~al.
\newblock Intelligent traffic data transmission and sharing based on optimal gradient adaptive optimization algorithm.
\newblock {IEEE} Trans. Intell. Transp. Syst., 2023, 24: 13330--13340

\bibitem{9739684}
Bian J, Arafat A~A, Xiong H, et~al.
\newblock Machine learning in real-time internet of things {(IoT)} systems: A survey.
\newblock {IEEE} Internet Things J., 2022, 9: 8364--8386

\bibitem{9385927}
Chu K~F, Lam A~Y~S, Li V~O~K.
\newblock Traffic signal control using end-to-end off-policy deep reinforcement learning.
\newblock {IEEE} Trans. Intell. Transp. Syst., 2022, 23: 7184--7195

\bibitem{RNN-IOT}
Woźniak M, Siłka J, Wieczorek M, et~al.
\newblock Recurrent neural network model for {IoT} and networking malware threat detection.
\newblock {IEEE} Trans. Ind. Inform., 2021, 17: 5583--5594

\bibitem{Edge-Computing}
Mao Y, You C, Zhang J, et~al.
\newblock A survey on mobile edge computing: The communication perspective.
\newblock {IEEE} Trans. Ind. Inform., 2017, 19: 2322--2358

\bibitem{XIAO2022109279}
Xiao D, Chen S, Ni W, et~al.
\newblock A sub-action aided deep reinforcement learning framework for latency-sensitive network slicing.
\newblock Computer Networks, 2022, 217: 109279

\bibitem{10263803}
Wang Y, Sun T, Li S, et~al.
\newblock Adversarial attacks and defenses in machine learning-empowered communication systems and networks: A contemporary survey.
\newblock IEEE Commun. Surv. Tut., 2023, 25: 2245--2298

\bibitem{8428412}
Jing X, Yan Z, Pedrycz W.
\newblock Security data collection and data analytics in the internet: A survey.
\newblock {IEEE} Commun. Surv. Tut., 2019, 21: 586--618

\bibitem{10753492}
Li K, Zheng J, Ni W, et~al.
\newblock Biasing federated learning with a new adversarial graph attention network.
\newblock IEEE Trans. Mobile Comput., Early Access, 2024.

\bibitem{Adversarial}
Raja A, Njilla L, Yuan J.
\newblock Adversarial attacks and defenses toward {AI}-assisted {UAV} infrastructure inspection.
\newblock {IEEE} Internet Things J., 2022, 9: 23379--23389

\bibitem{Federated-Learning}
Wei K, Li J, Ding M, et~al.
\newblock Federated learning with differential privacy: Algorithms and performance analysis.
\newblock {IEEE} Trans. Inf. Forensics Secur., 2020, 15: 3454--3469

\bibitem{10763434}
You F, Yuan X, Ni W, et~al.
\newblock Privacy-preserving multi-agent deep reinforcement learning for effective resource auction in multi-access edge computing.
\newblock IEEE Trans. Cogn. Commun. Netw., Early Access, 2024.

\bibitem{10045665}
Wang F, Xie M, Tan Z, et~al.
\newblock Preserving differential privacy in deep learning based on feature relevance region segmentation.
\newblock {IEEE} Trans. Emerg. Topics Comput., 2024, 12: 307--315

\bibitem{Lagendijk}
Lagendijk R, Erkin Z, Barni M.
\newblock Encrypted signal processing for privacy protection: Conveying the utility of homomorphic encryption and multiparty computation.
\newblock {IEEE} Signal Process. Mag., 2013, 30: 82--105

\bibitem{10605530}
Valina L, Teixeira B, Reis A, et~al.
\newblock Explainable artificial intelligence for deep synthetic data generation models.
\newblock In: Proceedings of IEEE Conference on Artificial Intelligence (CAI), Singapore, Singapore, 2024. 555--556

\bibitem{9590721}
Liu J, Zhao Y.
\newblock Role-oriented task allocation in human-machine collaboration system.
\newblock In: Proceedings of IEEE 4th International Conference on Information Systems and Computer Aided Education (ICISCAE), Dalian, China, 2021. 243--248

\bibitem{10646349}
Alam S, Khan M~F.
\newblock Enhancing {AI}-human collaborative decision-making in industry 4.0 management practices.
\newblock IEEE Access, 2024, 12: 119433--119444

\bibitem{9283461}
Dubois C, Le~Ny J.
\newblock Adaptive task allocation in human-machine teams with trust and workload cognitive models.
\newblock In: Proceedings of IEEE International Conference on Systems, Man, and Cybernetics (SMC), Toronto, ON, Canada, 2020. 3241--3246

\bibitem{10529943}
Zhang X, Ke Q, Zhao X.
\newblock Travel demand forecasting: A fair {AI} approach.
\newblock {IEEE} Trans. Intell. Transp. Syst., 2024, 25: 14611--14627

\bibitem{9940606}
Pasricha S.
\newblock {AI} ethics in smart healthcare.
\newblock IEEE Consumer Electronics Magazine, 2023, 12: 12--20

\bibitem{10057423}
Chen X, Dai W, Ni W, et~al.
\newblock Augmented deep reinforcement learning for online energy minimization of wireless powered mobile edge computing.
\newblock IEEE Trans. Commun., 2023, 71: 2698--2710

\bibitem{8636206}
Sutton G~J, Zeng J, Liu R~P, et~al.
\newblock Enabling technologies for ultra-reliable and low latency communications: From {PHY} and {MAC} layer perspectives.
\newblock IEEE Commun. Surv. Tut., 2019, 21: 2488--2524

\bibitem{10706120}
Wang Y, Sun T, Yuan X, et~al.
\newblock Minimizing adversarial training samples for robust image classifiers: Analysis and adversarial example generator design.
\newblock IEEE Trans. Info. Forensics Secur., 2024, 19: 9613--9628

\bibitem{10707443}
Huang H, Duan L, Li C, et~al.
\newblock A secure and lightweight aggregation method for blockchain-based distributed federated learning.
\newblock In: 2024 IEEE International Conference on Web Services (ICWS). 2024, 447--456

\bibitem{WuQIRS}
Wu Q, Zhang R.
\newblock Intelligent reflecting surface enhanced wireless network via joint active and passive beamforming.
\newblock {IEEE} Trans. Wireless Commun., 2019, 18: 5394--5409

\bibitem{2024Intelligent}
Zhang N, Zhang J, Xing C, et~al.
\newblock Intelligent secure near-field communication.
\newblock Sci. China Inf. Sci., 2024, 67: 199302

\bibitem{GuangxuAI}
Zhu G, Lyu Z, Jiao X, et~al.
\newblock Pushing {AI} to wireless network edge: An overview on integrated sensing, communication, and computation towards 6{G}.
\newblock Sci. China Inf. Sci., 2023, 66: 130301

\bibitem{10246260}
Kurunathan H, Huang H, Li K, et~al.
\newblock Machine learning-aided operations and communications of unmanned aerial vehicles: A contemporary survey.
\newblock {IEEE} Commun. Surv. Tut., 2024, 26: 496--533

\bibitem{DeepSeekV3}
DeepSeek-AI, Liu A, Feng B, et~al.
\newblock Deepseek-{V3} technical report.
\newblock 2024.
\newblock \eprint{https://arxiv.org/pdf/2412.19437}

\bibitem{10742580}
Al-Hraishawi H, Alsenwi M, Rehman J~u, et~al.
\newblock Digital twin for enhanced resource allocation in {6G} non-terrestrial networks.
\newblock {IEEE} Commun. Mag., Early Access, 2024.

\bibitem{10285423}
Abbas K, Nauman A, Bilal M, et~al.
\newblock {AI}-driven data analytics and intent-based networking for orchestration and control of {B5G} consumer electronics services.
\newblock IEEE Transactions on Consumer Electronics, 2024, 70: 2155--2169

\bibitem{9687596}
Duan J, Yu S, Tan H~L, et~al.
\newblock A survey of embodied {AI}: From simulators to research tasks.
\newblock IEEE Transactions on Emerging Topics in Computational Intelligence, 2022, 6: 230--244

\bibitem{10559618}
Wang Y, Ni W, Yi W, et~al.
\newblock Federated contrastive learning for personalized semantic communication.
\newblock {IEEE} Commun. Lett., 2024, 28: 1875--1879

\end{thebibliography}

\clearpage
\begin{appendix}
\section{ }
\begin{table}[H]
\centering
\caption{A list of abbreviations}
\resizebox{\textwidth}{!}{ 
\begin{tabular}{ll|ll}
\toprule
\textbf{Abbreviation} & \textbf{Definition} & \textbf{Abbreviation} & \textbf{Definition} \\
\midrule
3GPP   & 3rd generation partnership project                   & LLM    & large language model                        \\
4G     & fourth-generation                                    & LoS    & line-of-sight                               \\
5G     & fifth-generation                                     & LAM    & large-scale AI model                        \\
6G     & sixth-generation                                     & ML     & machine learning                            \\
AI     & artificial intelligence                              & MIMO   & multiple-input multiple-output              \\
AI4NET & AI for network                                       & mMTC   & massive machine type communications         \\
AIaaS  & AI as a service                                      & MEC    & mobile edge computing                       \\
AR     & augmented reality                                    & MARL   & multi-agent reinforcement learning          \\
BS     & base station                                         & NLP    & natural language processing                 \\
CSI    & channel state information                            & NLoS   & non-line-of-sight                           \\
CNN    & convolutional neural network                         & NWDAF  & network data analytics function             \\
CN     & core network                                         & NTN    & non-terrestrial network                     \\
CT     & core network and terminal                            & NR     & new radio                                   \\
CP     & cyclic prefix                                        & NET4AI & network for AI                              \\
CPU    & central processing units                             & O\&M   & operation and maintenance                   \\
DL     & deep learning                                        & OFDM   & orthogonal frequency division multiplexing  \\
DRL    & deep reinforcement learning                          & O-RAN  & open radio access networks                  \\
DT     & digital twins                                        & PFL    & personalized federation learning            \\
DTC    & digital twins channel                                & QoS    & quality of service                          \\
DQN    & deep Q-network                                       & QoAIS  & quality of AI service                       \\
DDPG   & deep deterministic policy gradient                   & RL     & reinforcement learning                      \\
DNN    & deep neural network                                  & RAN    & radio access network                        \\
eMBB   & enhanced mobile broadband                            & REK    & radio environment knowledge                 \\
ETSI   & european telecommunication standardization institute & RNN    & recurrent neural network                    \\
ENI    & experiential networked intelligence                  & SON    & self-organizing networks                    \\
FL     & federated learning                                   & SGCS   & square of the generalized cosine similarity \\
FedAvg & federated averaging                                  & SINR   & signal-to-interference-plus-noise ratio     \\
gNB    & generation node B                                    & SNR    & signal-to-noise ratio                       \\
GPU    & graphics processing units                            & SL     & split learning                              \\
HARQ   & hybrid automatic repeat request                      & SMPC   & secure multi-party computation              \\
IP     & internet protocol                                    & SI     & semantic information                        \\
ITU    & international telecommunication union                & SA     & system aspects                              \\
ICT    & information and communication technology             & TSG    & technical specification groups              \\
INC    & in-network computing                                 & uRLLC  & ultra-reliable low latency communications   \\
IMT    & international mobile communications                  & UPF    & user plane functions                        \\
IRS    & intelligent reflecting surface                       & UAV    & unmanned aerial vehicle                     \\
ISAC   & integrated sensing and communication                 & VR     & virtual reality                             \\
IoT    & internet of things                                   & WG     & working groups                              \\
KPI    & key performance indicator                            & XR     & extensive reality                           \\
LEO    & low earth orbit  \\
\bottomrule
\end{tabular}
}
\end{table}

\end{appendix}

\end{document}